\documentclass[11pt]{article}
\usepackage[T1]{fontenc}
\usepackage{lmodern,microtype}
\usepackage[margin=1in,marginparwidth=0.85in,marginparsep=0.08in]{geometry}
\usepackage{amsmath,amssymb,amsthm,mathtools}
\usepackage{enumitem,booktabs}
\usepackage{aliascnt}
\usepackage{tikz}
\usetikzlibrary{decorations.pathreplacing}
\usepackage{algorithm}
\usepackage[noend]{algpseudocode}
\usepackage[hidelinks]{hyperref}
\usepackage[nameinlink,noabbrev]{cleveref}
\usepackage{thmtools}
\usepackage{xcolor}
\setlist{itemsep=2pt,topsep=5pt}
\newtheorem{theorem}{Theorem}[section]

\newcommand{\newsiblingtheorem}[2]{%
  \newaliascnt{#1cnt}{theorem}%
  \newtheorem{#1}[#1cnt]{#2}%
  \aliascntresetthe{#1cnt}%
  \crefname{#1cnt}{#2}{#2s}\Crefname{#1cnt}{#2}{#2s}}
\newsiblingtheorem{lemma}{Lemma}
\newsiblingtheorem{corollary}{Corollary}
\newsiblingtheorem{proposition}{Proposition}
\theoremstyle{definition}
\newsiblingtheorem{definition}{Definition}
\theoremstyle{remark}
\newsiblingtheorem{remark}{Remark}
\newcommand{\E}{\mathbb E}
\newcommand{\Prb}{\mathbb P}
\newcommand{\1}{\mathbf 1}
\newcommand{\ceil}[1]{\left\lceil #1\right\rceil}
\newcommand{\floor}[1]{\left\lfloor #1\right\rfloor}
\newcommand{\defn}[1]{{\boldmath\textbf{#1}}}
\newcommand{\Fill}{\mathsf{TryInsertSubtree}}

\title{Incremental Open Addressing Has No Time-Space Tradeoff}
\author{
  Mart\'{\i}n Farach-Colton\\
  New York University\\
  \texttt{martin@farach-colton.com}
  \and
  Andrew Krapivin\\
  Carnegie Mellon University\\
  \texttt{andrew@krapivin.net}
  \and
  William Kuszmaul\\
  Carnegie Mellon University\\
  \texttt{kuszmaul@cmu.edu}
}
\date{}
\begin{document}
\maketitle
\begin{abstract}
In this paper, we show that there is no time-space tradeoff in incremental open addressing: there is an incremental open-addressed hash table that supports $O(1)$ worst-case expected time insertions, positive queries, and negative queries, at any load factor (even load factor $1$). As an intermediate result, we also revisit the problem of constructing an incremental open-addressed hash table that does not perform reordering (does not move elements around over time). In this setting, we show that it is possible to get $O(1)$ amortized expected positive query time together with $O(1)$ worst-case expected insertion time, a combination that was not previously known to be possible.

\end{abstract}

\section{Introduction}
\label{sec:intro}

Open addressing is a classic framework for hash-table design that captures many of the most widely used hash tables (e.g., linear probing, quadratic probing, double hashing, Robin Hood hashing, and cuckoo hashing). What these hash tables share in common, and what makes them examples of open addressing, is that in each case:
\begin{enumerate}
    \item The data structure itself consists of an array of $n$ slots, each of which can store at most one element, along with at most $O(1)$ words of auxiliary metadata (such as a counter of the number of elements present) stored outside the array.
    \item Each element $x$ is associated with a fixed probe sequence $h_1(x), h_2(x), \ldots \in [n]$ that is fully determined by $x$, $n$, and random bits (i.e., hash functions).
    \item The procedure for querying an element $x$ is to examine the array locations $h_1(x), h_2(x), \ldots$ in order, terminating either when $x$ is found (a \emph{positive query}) or when the query is able to conclude, based on what it has seen so far, that $x$ is not present in the table (a \emph{negative query}).\footnote{This standard model of open addressing is sometimes referred to as \emph{classical} or \emph{oblivious} open addressing~\cite{DBLP:conf/focs/BenderKZ24,DBLP:conf/stoc/BenderKZ25}, in contrast to \emph{non-oblivious} open addressing~\cite{DBLP:journals/jacm/FiatNSS92,DBLP:journals/siamcomp/FiatN93,DBLP:conf/stoc/BenderKZ25}, where a query is permitted to adaptively choose which slot to probe next based on the contents of previously inspected slots.}
\end{enumerate}
Insertions (and deletions, when supported), by contrast, are allowed to inspect slots of the array in any order, and may also rearrange (i.e., \emph{reorder}) elements that are already stored in the table.

Open addressing comes in two natural forms: the \emph{incremental} (or \emph{insert-only}) setting \cite{DBLP:journals/ibmrd/Peterson57,Knu63,KW66,DBLP:journals/jacm/Ullman72,DBLP:journals/cacm/Brent73,DBLP:journals/cj/AmbleK74,DBLP:journals/cj/Mallach77,DBLP:journals/jcss/GuibasS78,DBLP:journals/ipl/AjtaiKS78,DBLP:journals/siamcomp/GonnetM79,DBLP:journals/jacm/Yao85,DBLP:conf/focs/CelisLM85,DBLP:journals/combinatorica/LuekerM93,DBLP:conf/focs/Farach-ColtonKK24,DBLP:conf/stoc/Farach-ColtonKK26}, in which elements are inserted one after another into an initially empty array of size $n$ until reaching some target load factor $1 - \epsilon$ (meaning that a $1 - \epsilon$ fraction of the slots are occupied), and the \emph{dynamic} setting \cite{DBLP:journals/jal/PaghR04,DBLP:journals/mst/FotakisPSS05,DBLP:journals/tcs/DietzfelbingerW07,DBLP:conf/focs/BenderKK21,DBLP:conf/focs/BenderKZ24,DBLP:conf/stoc/BenderKZ25}, in which elements are both inserted and deleted over time while maintaining a load factor of at most $1 - \epsilon$.

For more than half a century, the central question in the study of open-addressed hash tables has been to determine the optimal tradeoff between time and load factor. As the load factor parameter $\epsilon$ goes to zero, it becomes more and more difficult to place additional elements into the hash table in a way that keeps both insertions and queries cheap. But, for most of the history of the problem, it has remained open what the optimal query and insertion costs are as a function of $\epsilon^{-1}$.

Recently, Bender, Kuszmaul, and Zhou~\cite{DBLP:conf/focs/BenderKZ24} gave a full resolution to this question in the \emph{dynamic} setting. They introduced a hash table that supports $O(1)$ expected-time queries with $O(\log \log \epsilon^{-1})$ expected-time insertions and deletions. They also proved a matching lower bound, showing that any dynamic open-addressed hash table must incur $\Omega(\log \log \epsilon^{-1})$ worst-case expected time  per operation.

What remains unclear is what the optimal tradeoff should be in the \emph{incremental} setting. Does the $\Omega(\log \log \epsilon^{-1})$ lower bound continue to hold, or can an alternative data structure somehow do better?

\paragraph{This paper: Optimal incremental open addressing.} The main contribution of this paper is an optimal construction for incremental open addressing. We show that there is actually \emph{no fundamental time-space tradeoff} in this setting: it is possible to achieve $O(1)$ worst-case expected time for every operation (insertions, positive queries, and negative queries) at all load factors up to and including $1$ (i.e., even when the hash table is completely full, with $\epsilon = 0$). Note that, as a convention, we define the load factor of an insertion to be the load factor after the insertion is complete.

\begin{restatable}{theorem}{thmintrofull}\label{thm:intro-full}
There is an incremental open-addressed hash table with reordering that stores no metadata besides a counter of the number of elements present and, at every load factor up to and including $1$, supports insertions, positive queries, and negative queries in $O(1)$ worst-case expected time.
\end{restatable}

Here, as is standard, \emph{worst-case expected time} means that the $O(1)$ bound applies individually to every operation (with the expectation taken solely over the random hash functions), rather than merely on average across elements or insertions. In particular, even the $n$-th insertion (which must fill the final remaining empty slot in the array), and even an arbitrary query (positive or negative) that is performed when the array is $100\%$ full must take $O(1)$ expected time.

\paragraph{Warmup: revisiting incremental open addressing \emph{without} reordering.}
Consider first
a more basic question: how to build an optimal incremental open-addressed hash table that \emph{cannot perform reordering}. In this setting, once an element is placed in a slot of the array, the hash table is not permitted to move it in subsequent operations. The only modification an insertion makes to the array is to place the new element into some empty slot in the table.

Without reordering, it is not possible to achieve $O(1)$ worst-case expected query time at high load factors: even for positive queries, known lower bounds force a worst-case expected query time of $\Omega(\log \epsilon^{-1})$~\cite{DBLP:conf/focs/Farach-ColtonKK24}. Consequently, research in this setting tends to focus on optimizing a weaker notion of query complexity known as \emph{amortized expected positive query time} (or \emph{amortized query time} for short), which is defined to be the expected time to query a uniformly random element among those currently in the table.

For most of the history of the problem, it was believed that even amortized query time had to be $\Omega(\log \epsilon^{-1})$. Indeed, in a celebrated 1985 result, Yao~\cite{DBLP:journals/jacm/Yao85} (proving a conjecture of Ullman~\cite{DBLP:journals/jacm/Ullman72}) showed that $\Omega(\log \epsilon^{-1})$ amortized query time is unavoidable for any \emph{greedy} open-addressed hash table (in which insertions follow the same probe sequence as queries, placing the element in the first empty slot). Recently, however, researchers have shown that \emph{non-greedy} schemes~\cite{DBLP:conf/focs/Farach-ColtonKK24,DBLP:conf/stoc/Farach-ColtonKK26} can actually do much better, achieving $O(1)$ amortized expected positive query time at all load factors (even load factor $1$).

The resulting hash tables~\cite{DBLP:conf/focs/Farach-ColtonKK24,DBLP:conf/stoc/Farach-ColtonKK26} still suffer from slow \emph{insertions}, however, incurring an amortized expected insertion time of $\Omega(\log \epsilon^{-1})$. This time bound is forced by a basic \emph{coupon-collecting bottleneck}: each insertion searches for an empty slot by performing random probes into (some large part of) the hash table. If, across all insertions, we must probe at least a $(1 - \epsilon)$ fraction of slots, then we cannot hope to do this in fewer than $\Omega(n \log \epsilon^{-1})$ probes.

Our first result is an incremental open-addressed hash table without reordering that overcomes this coupon-collecting bottleneck. The hash table achieves $O(1)$ for both amortized expected positive query time and worst-case expected insertion time (while also matching the $\Omega(\log n)$ lower bound on worst-case expected positive query time without reordering~\cite{DBLP:conf/focs/Farach-ColtonKK24}).

\begin{restatable}{theorem}{thmintrostable}\label{thm:intro-stable}
There is an incremental open-addressed hash table without reordering that stores no metadata besides a counter of the number of elements present and, at every load factor up to and including $1$, supports insertions in $O(1)$ worst-case expected time and positive queries in $O(1)$ amortized expected time. Moreover, the worst-case expected positive query time is $O(\log n)$.
\end{restatable}

To prove \Cref{thm:intro-stable}, our construction (\Cref{sec:stable}) borrows the high-level structure of~\cite{DBLP:conf/focs/Farach-ColtonKK24}, but replaces each of its subarrays (formerly just a uniform-probing hash table) with a new tree-based structure. The basic idea is to organize each subarray as a binary tree that fills bottom up, maintaining the invariant that a node can only contain elements if \emph{all} of its descendant nodes are full. This means that an insertion can test whether a subtree contains \emph{any} empty slots by inspecting just a single node (the root of the subtree): it can, in effect, collect a superconstant number of coupons in constant time. We then show how to design the hash table in such a way that the majority of insertions first attempt to insert into a subtree that is \emph{already} full (and observe that it is full in $O(1)$ time), and then immediately fall back to a much cheaper option (inserting into a tree that is only a constant fraction full). Finally, we show that with the right query probe sequence, this construction continues to allow for $O(1)$ amortized queries, which allows us to obtain Theorem \ref{thm:intro-stable}.

\paragraph{From the warmup to the full theorem: Two challenges to overcome.}
Having established our warmup result without reordering (\Cref{thm:intro-stable}), we next turn to the problem of proving our full result (\Cref{thm:intro-full}), where reordering is permitted and we require $O(1)$ worst-case expected time for insertions, positive queries, and negative queries. To get from the warmup to our full construction, there are two main challenges that we must overcome.

\paragraph{Challenge 1: Amortized vs.\ worst-case expected positive queries.}
The first challenge is to upgrade our bound on positive queries from $O(1)$ \emph{amortized} expected time to $O(1)$ \emph{worst-case} expected time for every element present in the table. In the warmup construction, an element's query cost depends heavily on when it is inserted. Elements inserted when the hash table is very full have much more expensive queries. This is fine for \emph{amortized} query time because the vast majority of insertions are performed when the hash table is not very full. But to achieve an $O(1)$ \emph{worst-case} expected time bound, we must eliminate this inequity.

In \Cref{sec:reorder}, we give a simple solution to this problem. The idea is to swap each new element $x$ with a (close to) random \emph{victim} element $y$. We place $x$ in a position that is guaranteed to be cheap to query, we evict the element $y$ formerly in that position, and we complete the rest of the insertion as though $y$ is the element being inserted. This allows us to randomize which elements incur a large query cost, resulting in the following theorem:

\begin{restatable}{theorem}{thmintroreorder}\label{thm:intro-reorder}
There is an incremental open-addressed hash table with reordering that stores no metadata besides a counter of the number of elements present and, at every load factor up to and including $1$, supports insertions and positive queries in $O(1)$ worst-case expected time.
\end{restatable}

\paragraph{Challenge 2: Supporting negative queries.}
The second, and more difficult, challenge is to support $O(1)$-time \emph{negative} queries. Whereas positive queries can stop as soon as they find the element they are searching for, negative queries have no way \emph{a priori} of determining when to stop. Naively, when the hash table is 100\% full, a negative query would need to traverse the entire table before it can safely conclude that the queried element is not present.

To make negative queries fast, we combine two high-level ideas. The first is to design the hash table so that within each tree, almost all elements $y$ that reside in some high level $j$ of the tree satisfy $g(y)< 1/j^4$, where $g$ is a hash function. This means that, when querying an element $x$, in the common case where $g(x) > 1/j^4$, we can be relatively confident that $x$ is not in any of levels $j, j +1 , \ldots$ of the tree. (There will be some exceptions to this rule, but they will be rare enough that we can handle them with alternative techniques.) This simple idea is not enough to make negative queries fast, however, because a given element could still be in any of many trees.

The second high-level idea is to design insertions in such a way that they strategically leave ``breadcrumbs'' throughout the data structure (encoded implicitly within the arrangement of the stored elements themselves), so that negative queries can efficiently figure out when to stop.\footnote{Note that the query sequence is fixed ahead of time, so the algorithm is still oblivious. The role of the breadcrumbs is only to allow early termination.} Roughly speaking, the job of these breadcrumbs is to tell a (negative) query how many trees it needs to consider before it can give up.

What makes these ``breadcrumbs'' tricky is that we need insertions to be able to encode them very efficiently (in $O(1)$ time, even if $\omega(1)$ bits of information are being encoded!). Here, we make use of the following basic idea: to encode a value $j$ at a leaf $L$, we swap an element $u$ currently stored at the leaf with an element $v$ stored in the leaf's level-$j$ ancestor. We then make use of the fact that elements $v$ stored at level-$j$ nodes (usually) satisfy $g(v) < 1/j^4$ (where $g$ is the hash function from the previous paragraph). So, the simple fact that the leaf $L$ now contains an element $v$ with such an unusually small $g(v)$ value can serve as the ``breadcrumb'' to encode the number $j$ (albeit in a slightly imprecise way). This simple idea comes with many challenges to implement, but is what ends up allowing us to support negative queries efficiently.
Putting the pieces together, we are able to obtain the main result of the paper, \Cref{thm:intro-full}.

\paragraph{The cost of dynamic resizing.}
A natural extension of incremental open addressing is to allow the hash table to \emph{grow} over time, preserving the invariant that at any given moment a $1 - \epsilon$ fraction of slots are occupied. Given the results of this paper, one might hope that in this stronger setting $O(1)$-time operations continue to be possible.

In \Cref{app:resizing}, we show that this is not the case. Dynamic resizing, it turns out, is just as difficult as supporting deletions: any growing open-addressed hash table with load factor $1 - \epsilon$ (while maintaining $O(1)$-time queries) must incur $\Omega(\log \log \epsilon^{-1})$ amortized expected time per operation, matching the bounds that have already been achieved in~\cite{DBLP:conf/focs/BenderKZ24} (and that hold for a hash table that is both dynamically resized \emph{and} dynamic).

\subsection{Related Work}
\label{subsec:related}

Since the introduction of hash tables in the 1950s~\cite{DBLP:journals/ibmrd/Peterson57,Knu63,Knu98}, the study of open-addressed hash tables has been one of the most active and foundational areas in data structures. Below, we situate our results within the broader literature on open addressing and space-efficient dictionaries.

\paragraph{Greedy open addressing and open addressing without reordering.}
The study of open addressing began with Peterson's 1957 paper~\cite{DBLP:journals/ibmrd/Peterson57}, the first published study of open addressing (including linear probing and uniform probing), and Knuth's seminal 1963 memorandum~\cite{Knu63} analyzing the amortized expected query cost of linear probing. Over the subsequent decades, a large body of work analyzed specific greedy collision-resolution strategies (in which each insertion places its element in the first empty slot of its probe sequence), including linear probing~\cite{DBLP:journals/ibmrd/Peterson57,Knu63,KW66,DBLP:journals/cj/AmbleK74,DBLP:conf/stoc/SchmidtS90,DBLP:journals/siamcomp/PaghPR09,DBLP:journals/jacm/PatrascuT12,DBLP:journals/talg/PatrascuT16,DBLP:conf/focs/BenderKK21,DBLP:conf/focs/BravermanK24}, quadratic probing~\cite{DBLP:journals/cacm/Radke70,DBLP:journals/cj/HopgoodD72,DBLP:journals/cj/Ecker74,DBLP:journals/cacm/Batagelj75,DBLP:journals/cacm/Maurer83,DBLP:conf/icalp/KuszmaulX24,DBLP:journals/corr/abs-2607-13247,DBLP:journals/corr/abs-2608-28512,DBLP:journals/corr/abs-2608-08013,GPW26b}, double hashing~\cite{DBLP:journals/jcss/GuibasS78,DBLP:journals/combinatorica/LuekerM93} (see also~\cite{DBLP:conf/spaa/Mitzenmacher14,DBLP:conf/analco/Mitzenmacher16} for double hashing in balanced allocations), and uniform probing~\cite{DBLP:journals/ibmrd/Peterson57,DBLP:journals/jacm/Ullman72,DBLP:journals/jacm/Yao85,DBLP:conf/focs/Farach-ColtonKK24}.

In 1972, Ullman~\cite{DBLP:journals/jacm/Ullman72} formalized the notion of greedy open addressing and conjectured that no greedy probe distribution can beat the $\Theta(\log \epsilon^{-1})$ amortized expected insertion and positive query time of uniform probing at load factor $1 - \epsilon$. Knuth~\cite{Knu74} posed a special case of Ullman's conjecture for single-hashing strategies, which was resolved by Ajtai, Koml\'os, and Szemer\'edi~\cite{DBLP:journals/ipl/AjtaiKS78} in 1978; building on their potential-function technique, Yao~\cite{DBLP:journals/jacm/Yao85} proved Ullman's full conjecture in 1985.

To bypass Yao's $\Omega(\log \epsilon^{-1})$ amortized lower bound without reordering elements over time, an insertion algorithm must be \emph{non-greedy} (i.e., it cannot simply place each element into the first empty slot along its query probe sequence). Two such algorithms have been proposed in prior work, \emph{elastic hashing} \cite{DBLP:conf/focs/Farach-ColtonKK24} and \emph{pretzel probing} \cite{DBLP:conf/stoc/Farach-ColtonKK26}, both of which achieve $O(1)$ amortized expected query time, and $O(\log \epsilon^{-1})$ worst-case expected query time (for positive queries). Non-greedy insertions without reordering have also recently been studied for linear probing~\cite{DBLP:journals/corr/abs-2607-17494}, as has the locality of open addressing without reordering~\cite{DBLP:journals/corr/abs-2607-16390}.

Both elastic hashing~\cite{DBLP:conf/focs/Farach-ColtonKK24} and pretzel probing~\cite{DBLP:conf/stoc/Farach-ColtonKK26} require $\Omega(\log \epsilon^{-1})$ amortized expected insertion time due to the coupon collecting bottleneck discussed earlier. The first result of this paper, \Cref{thm:intro-stable}, resolves this bottleneck, showing that $O(1)$ worst-case expected insertion time and $O(1)$ amortized expected positive query time can be achieved simultaneously without reordering, even at load factor $1$.

\paragraph{Open addressing with reordering.}
Starting in the 1970s, there has also been a great deal of interest in whether open-addressed hash tables \emph{with} reordering can do better. It has been known since Brent's 1973 breakthrough~\cite{DBLP:journals/cacm/Brent73} (see~\cite{DBLP:journals/corr/abs-2608-00762} for a formal analysis) that one can achieve $O(1)$ amortized expected positive query time at any load factor. Brent's paper inspired a rich line of work in the 1970s and 1980s on reordering-based incremental hash tables, including ordered hash tables~\cite{DBLP:journals/cj/AmbleK74}, Mallach's method~\cite{DBLP:journals/cj/Mallach77}, binary tree hashing~\cite{DBLP:journals/siamcomp/GonnetM79}, Robin Hood hashing~\cite{DBLP:conf/focs/CelisLM85}, and last-come-first-served hashing~\cite{DBLP:journals/jal/PobleteM89} (as well as optimal arrangements of elements computed via the assignment problem~\cite{DBLP:journals/jacm/Rivest78}). In all of these classical reordering schemes, fast positive queries came at the expense of slow insertions (requiring $\Omega(\epsilon^{-1})$ or $\Omega(\epsilon^{-2})$ expected insertion time at load factor $1 - \epsilon$). The main contribution of the current paper is to show that such a tradeoff is not necessary.

Over the past two decades, much of the work on reordering has centered on \emph{cuckoo hashing}~\cite{DBLP:journals/jal/PaghR04} and its generalizations to high load factors, such as $d$-ary cuckoo hashing~\cite{DBLP:journals/mst/FotakisPSS05} and bucketized cuckoo hashing~\cite{DBLP:journals/tcs/DietzfelbingerW07} (which have also been the subject of extensive probabilistic analysis establishing tight load thresholds and insertion times~\cite{DBLP:conf/soda/CainSW07,DBLP:conf/soda/FernholzR07,DBLP:conf/esa/LehmanP09,DBLP:conf/icalp/FountoulakisP10,DBLP:conf/stoc/GaoW10,DBLP:journals/siamcomp/KirschMW09,DBLP:conf/icalp/DietzfelbingerGMMPR10,DBLP:journals/siamcomp/FriezeMM11,DBLP:journals/rsa/FriezeM12,DBLP:conf/soda/Lelarge12,DBLP:journals/siamcomp/FountoulakisPS13,DBLP:journals/cpc/FountoulakisKP16,DBLP:journals/ipl/FriezeP18,DBLP:journals/rsa/FriezeJ19,DBLP:journals/algorithmica/KhoslaA19,DBLP:journals/talg/Walzer23,DBLP:conf/esa/Walzer22,DBLP:conf/focs/BellF24,DBLP:conf/soda/KuszmaulM25,DBLP:journals/corr/abs-2607-24545}). When configured with $d = \Theta(\log \epsilon^{-1})$ hash functions~\cite{DBLP:journals/mst/FotakisPSS05,DBLP:conf/focs/BellF24,DBLP:conf/soda/KuszmaulM25}, cuckoo hashing supports \emph{worst-case} $O(\log \epsilon^{-1})$-time positive and negative queries at load factor $1 - \epsilon$. Nevertheless, prior to~\cite{DBLP:conf/focs/BenderKZ24} (which we discuss next), every known dynamic open-addressed hash table with $O(t)$-time queries required at least $\Omega(\epsilon^{-1}/t)$ expected time per insertion just to find empty slots.

Recently, Bender, Kuszmaul, and Zhou~\cite{DBLP:conf/focs/BenderKZ24} broke the $\Omega(\epsilon^{-1})$ insertion barrier by introducing \emph{rainbow hashing}, which supports $O(1)$ expected-time positive and negative queries and $O(\log \log \epsilon^{-1})$ expected-time insertions and deletions (even with dynamic resizing), and proved a matching $\Omega(\log \log \epsilon^{-1})$ amortized update lower bound for any dynamic open-addressed hash table with $O(2^{\sqrt{\log \epsilon^{-1}}})$ expected query time. Our \Cref{thm:intro-full,thm:intro-reorder} show that in the incremental setting with fixed capacity, even this $\Theta(\log \log \epsilon^{-1})$ barrier disappears.

\paragraph{Implicit data structures, non-oblivious open addressing, and partner hashing.}
Our construction in \Cref{sec:full} supports $O(1)$-time negative queries by encoding structural information directly into the relative arrangement of the elements in the array. The technique of encoding metadata in the permutation of stored elements dates back to the classic literature on \emph{implicit data structures} by Munro and Suwanda~\cite{DBLP:journals/jcss/MunroS80} and Yao~\cite{DBLP:journals/jacm/Yao81a}. In the context of open addressing, this paradigm has traditionally been studied in the \emph{non-oblivious} model, where queries are allowed to inspect elements, decode information from their arrangement, and use that information to \emph{adaptively} decide which array slot to probe next. In the static setting, Fiat, Naor, Schmidt, and Siegel~\cite{DBLP:journals/jacm/FiatNSS92} and Fiat and Naor~\cite{DBLP:journals/siamcomp/FiatN93} used non-oblivious probing to achieve $O(1)$ worst-case lookup time in a completely full table with little or no auxiliary memory.

Recently, Bender, Kuszmaul, and Zhou~\cite{DBLP:conf/stoc/BenderKZ25} resolved the dynamic non-oblivious setting by introducing \emph{partner hashing}, a dynamic non-oblivious open-addressed hash table that supports $O(1)$-time insertions, deletions, and queries with high probability even at load factor $1$ and under dynamic resizing. At the heart of partner hashing is a swapping trick in which elements are paired up as ``partners'' and swapped out of their home locations so that the displacement of one element encodes routing information for the other.

The swapping trick that we use to support negative queries in \Cref{sec:full} can be viewed as a cousin of the swapping trick in partner hashing~\cite{DBLP:conf/stoc/BenderKZ25}. However, there is a fundamental difference between the two settings. In partner hashing~\cite{DBLP:conf/stoc/BenderKZ25}, when an element is swapped out of its home location in order to encode information, a query for that element can no longer find it along a fixed, oblivious probe sequence (instead, the query must read an encoded pointer and adaptively jump to the element's new slot). In our classical (oblivious) setting, adaptive jumps are forbidden, so moving an element off its fixed probe sequence would destroy its query time. Somewhat remarkably, we are able to design our tree-swapping mechanism so that every swap moves elements only between nearby locations on their fixed probe sequences, and in such a way that it does not hurt the overall worst-case $O(1)$ expected query time. This allows the encoding trick to be used as part of an oblivious solution, where the decoded information is used solely to determine \emph{when to stop} probing rather than \emph{where to probe next}.

\paragraph{Succinct dictionaries, retrieval, and other hashing models.}
Beyond open addressing, there has been a great deal of work on understanding the optimal time-space tradeoffs for \emph{succinct dictionaries}, which aim to store $n$ elements from a universe $U$ using $\mathrm{OPT} + R$ bits of space, where $\mathrm{OPT} = \lceil \log_2 \binom{|U|}{n} \rceil$ is the information-theoretic minimum and $R$ is the \emph{redundancy}~\cite{DBLP:journals/siamcomp/Pagh01,DBLP:conf/icalp/RamanR03,DBLP:conf/focs/Patrascu08,DBLP:conf/focs/ArbitmanNS10,DBLP:journals/siamcomp/Yu22,DBLP:conf/stoc/HuLYZZ25,DBLP:conf/stoc/BenderFKKL22,DBLP:journals/jacm/BenderCFKT23,DBLP:journals/algorithmica/BerceaE23,DBLP:conf/focs/LiLYZ23a,DBLP:conf/focs/LiLYZ23,DBLP:conf/soda/LiLYZ24,DBLP:journals/talg/BenderCFKT25,DBLP:conf/icalp/LarsenPPPYZ24}. For dynamic dictionaries, Bender, Farach-Colton, Kuszmaul, Kuszmaul, and Liu~\cite{DBLP:conf/stoc/BenderFKKL22} and Li, Liang, Yu, and Zhou~\cite{DBLP:conf/focs/LiLYZ23a,DBLP:conf/soda/LiLYZ24} established a tight cell-probe tradeoff showing that achieving $O(1)$ wasted bits per element ($R = O(n)$) requires $\Theta(\log^* n)$ time per operation. Just as our work establishes a separation between incremental and dynamic open addressing, Kuszmaul, Putterman, Xu, Zhou, and Zhou~\cite{DBLP:conf/soda/KuszmaulPXZZ25} recently established an analogous separation between incremental and dynamic \emph{retrieval} data structures (see also~\cite{DBLP:journals/corr/abs-2606-15944} for resizable retrieval).
Other related directions in hash-table design include history-independent hashing~\cite{DBLP:journals/iacr/NaorT01,DBLP:conf/focs/BlellochG07,DBLP:conf/icalp/NaorSW08,DBLP:conf/focs/Kuszmaul23}, external-memory hashing~\cite{DBLP:journals/algorithmica/JensenP08,DBLP:conf/soda/IaconoP12,DBLP:journals/siamcomp/VerbinZ13,DBLP:conf/icalp/ConwayFS18}, and hash tables with high-probability or deterministic guarantees~\cite{DBLP:conf/focs/Sundar91,DBLP:journals/jal/HagerupMP01,DBLP:conf/latin/DemaineHPP06,DBLP:journals/talg/Ruzic08,DBLP:conf/icalp/ArbitmanNS09,DBLP:conf/medalg/GoodrichHMT12,DBLP:conf/focs/Kuszmaul22,DBLP:journals/jacm/BenderCFKT23}. See~\cite{DBLP:conf/sosa/BenderFKK24} for a simple presentation of some of the basic techniques that have been developed in these settings.

\paragraph{Random hash functions.}
As is standard for this problem, we assume random hash functions throughout the paper, but there is also a large body of work on how to simulate them efficiently. This line of work began with the universal hash functions of Carter and Wegman~\cite{DBLP:journals/jcss/CarterW79,DBLP:journals/jcss/WegmanC81}, and has since produced constant-time hash functions with high independence~\cite{DBLP:journals/siamcomp/Siegel04,DBLP:conf/focs/Thorup13,DBLP:conf/stoc/ChristianiPT15}, hash functions that are fully random on any fixed set of elements with high probability~\cite{DBLP:conf/stoc/OstlinP03,DBLP:journals/siamcomp/PaghP08,DBLP:conf/stoc/DietzfelbingerW03}, and tabulation-based hash functions~\cite{Zob70,DBLP:journals/jacm/PatrascuT12,DBLP:conf/soda/PatrascuT13,DBLP:conf/focs/Thorup13,DBLP:conf/focs/DahlgaardKRT15,DBLP:journals/cacm/Thorup17,DBLP:conf/stoc/AamandKKRT20}, including recent work on \emph{locally uniform} hashing~\cite{DBLP:conf/focs/DahlgaardKRT15,DBLP:conf/focs/Bercea0KHT23,DBLP:journals/corr/abs-2411-19394}. Techniques such as the splitting trick~\cite{DBLP:conf/icalp/DietzfelbingerR09} allow these hash functions to simulate full randomness in essentially any hash table. There has also been work specifically on open addressing under limited independence~\cite{DBLP:conf/stoc/SchmidtS90,DBLP:journals/siamcomp/PaghPR09,DBLP:journals/talg/PatrascuT16}.

\section{An Optimal Construction Without Reordering}
\label{sec:stable}

In this section, we establish that there is no tension between insertion time and amortized positive query time in incremental open addressing without reordering.  This is one of our two main results and sets the stage for the other.

In particular, we establish that there is
a table that never moves an element after placing it, that fills all $n$
slots, that spends only constant expected time per insertion, and that achieves constant amortized expected positive
query time.

\thmintrostable*

\subsection{The Algorithm}
\label{subsec:layout}

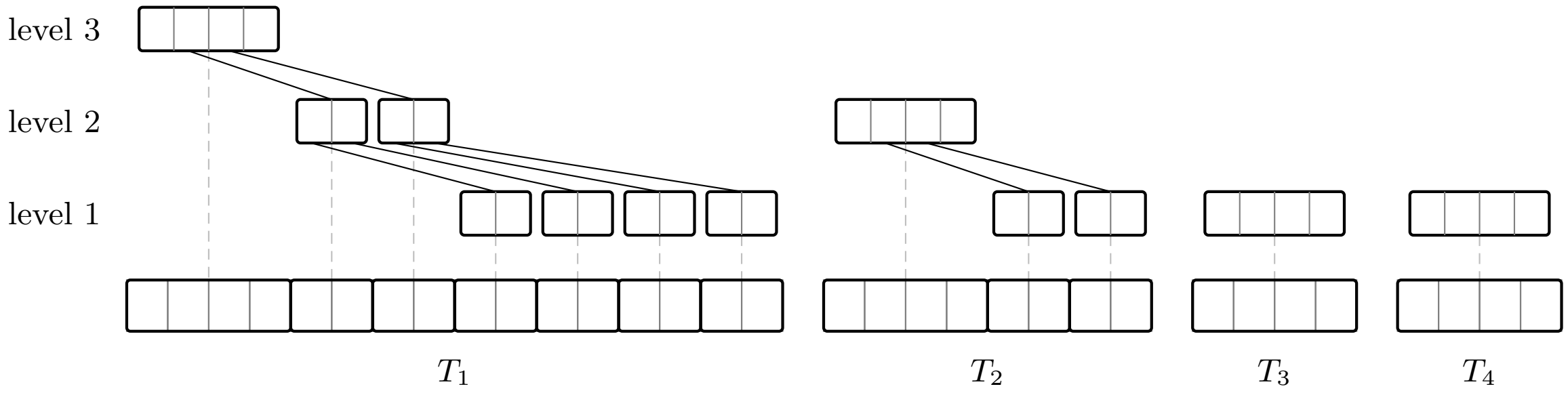
\begin{figure}[t]
\centering
\begin{tikzpicture}[x=0.4cm,y=0.5cm]
  \def\nodes{2/5.9/1.7, 5/4.1/0.85, 7/4.1/0.85,
             9/2.3/0.85, 11/2.3/0.85, 13/2.3/0.85, 15/2.3/0.85,
             19/4.1/1.7, 22/2.3/0.85, 24/2.3/0.85,
             28/2.3/1.7, 33/2.3/1.7}
  \foreach \x/\y/\w in \nodes
    {\draw[densely dashed,gray!50] (\x,\y-0.425) -- (\x,1);}
  \foreach \px/\py/\o/\cx/\cy in {2/5.9/-1/5/4.1, 2/5.9/1/7/4.1,
      5/4.1/-1/9/2.3, 5/4.1/1/11/2.3, 7/4.1/-1/13/2.3, 7/4.1/1/15/2.3,
      19/4.1/-1/22/2.3, 19/4.1/1/24/2.3}
    {\draw (\px+0.5*\o,\py-0.425) -- (\cx,\cy+0.425);}
  \foreach \x/\y/\w in \nodes
    {\draw[thick,rounded corners=1pt,fill=white]
       (\x-\w,\y-0.425) rectangle (\x+\w,\y+0.425);
     \draw[gray] (\x,\y-0.425) -- (\x,\y+0.425);
     \ifdim\w pt>1pt
       \foreach \k in {-1,1}
         {\draw[gray] (\x+0.85*\k,\y-0.425) -- (\x+0.85*\k,\y+0.425);}
     \fi}
  \foreach \s in {0,...,15,17,18,...,24,26,27,28,29,31,32,33,34}
    {\draw[gray] (\s,0) rectangle +(1,1);}
  \foreach \s/\w in {0/4,4/2,6/2,8/2,10/2,12/2,14/2,
                     17/4,21/2,23/2,26/4,31/4}
    {\draw[thick,rounded corners=1pt] (\s,0) rectangle +(\w,1);}
  \foreach \l/\y in {1/2.3,2/4.1,3/5.9}
    {\node[anchor=east,font=\footnotesize] at (-0.3,\y) {level \l};}
  \foreach \x/\t in {8/1,21/2,28/3,33/4}
    {\node[font=\footnotesize] at (\x,-0.8) {$T_\t$};}
\end{tikzpicture}%
\caption{The layout for $n=32$: trees $T_1,\ldots,T_4$ with $16$, $8$,
$4$, and $4$ slots and heights $3$, $2$, $1$, and $1$. The thick boxes are
nodes, of two slots each except that each root has four. Each tree is
stored as an implicit binary heap: its root first, then the nodes of the
next level down, and so on, with the leaves last.}
\label{fig:forest}
\end{figure}

We assume for now that $n$ is a power of two; \Cref{app:addresses}
removes this assumption. We logically organize the array as a forest of
perfect binary trees $T_1,T_2,\ldots,T_\alpha$, where $\alpha=\log_2n-1$.
For $i<\alpha$, the tree $T_i$ has $n/2^i$ slots, so $T_{\alpha -1}$ has four slots, and $T_\alpha$ has the remaining four slots. The slots are mapped to the nodes of the tree so that each node covers two
slots, except for the root of each tree, which covers four.\footnote{The root node has extra slots just to make sure $|T_i|$ is a power of 2.} See
\Cref{fig:forest} for an illustration.
We count the levels of each tree from the bottom: leaves are at level
$1$, their parents are at level $2$, and so on.
Each tree covers a contiguous part of the array in breadth-first order; \Cref{app:addresses}
gives the exact addresses, which are slightly involved when $n$ is not a power of 2.

\paragraph{Hashing.}
Each element $x$ selects a candidate set of leaves, one per tree.  Specifically, it selects leaf $L_i(x)$ in tree $T_i$ via a set of uniformly and independently random hash functions $L_i$.
The level-$j$ ancestor of
$L_i(x)$ is $x$'s \defn{level-$j$ ancestor in $T_i$}, and its subtree is
$x$'s \defn{level-$j$ home subtree}. In addition, each $x$ is given a uniformly random \defn{priority} $g_i(x) \in [0,1]$ for each tree $T_i$.

\subsubsection{Insertions.}
The primary building block used in insertions is $\Fill(x,i,j)$ (detailed in \Cref{alg:fill}), which
inserts $x$ into its level-$j$ home subtree in $T_i$ and returns true if that subtree has
room, in time proportional to the height of the subtree. If the subtree is full, $\Fill(x,i,j)$ returns false in $O(1)$ time. We call $x$'s level-$j$ ancestor in $T_i$, the root of this subtree, the target node of the call.

Na\"{\i}vely checking every slot would take time $O(2^j)$. To avoid this issue, we will enforce the invariant that \emph{a node
hold an element only if both of its child subtrees are full.}
Thus, a subtree is full if and only if its root node is full, so we can return false after checking the $O(1)$ slots in the root node of the subtree. To maintain the invariant, we make sure to insert into a node only if there is no room lower down in its subtree.

\begin{algorithm}[t]
\caption{$\Fill(x,i,j)$}
\label{alg:fill}
\begin{algorithmic}[1]
\State $V\gets x$'s level-$j$ ancestor in $T_i$
\If{$V$ is full}
  \State \Return false \Comment{the whole subtree of $V$ is full}
\EndIf
\State $U\gets V$
\While{$U$ is not a leaf and some child of $U$ is nonfull}
  \State $U\gets$ the child of $U$ above $L_i(x)$ if there is one and it is nonfull, else a nonfull child
\EndWhile
\State place $x$ in an empty slot of $U$ \Comment{$U$ is nonfull}
\State \Return true
\end{algorithmic}
\end{algorithm}

\paragraph{The schedule.}
Insertions are grouped into $\alpha$ \defn{epochs}. Epoch
one just fills $T_1$ until it is exactly a $1-\eta$ fraction full (up to roundoff error), for a small
constant $\eta>0$ to be specified in the analysis. Every later epoch $i$ has two jobs: it \emph{cleans up} $T_{i-1}$, getting it from exactly a $1-\eta$ fraction full to
completely full while filling $T_i$ from empty to exactly a $1-\eta$ fraction full (the last epoch fills $T_\alpha$ completely).

Each epoch $i\geq2$ is further subdivided into \defn{phases} $1, 2, \ldots, h+1$,
where $h$ is the height of $T_{i-1}$; the last of these, phase $h+1$, is the \defn{final phase}.
Each nonfinal phase $j \leq h$ consists of a $\Theta(1/j^2)$ fraction of the insertions in the epoch.

Conceptually, at the start of epoch \(i\), \(T_{i-1}\) is already nearly full,
but its remaining empty slots may be scattered throughout the tree.
The phases gradually fill these empty slots, first cleaning up small subtrees
and then progressively larger ones. More precisely,
for each fixed subtree rooted at level $j \leq h$ of $T_{i-1}$,
the probability that it is full by the end of phase $j$ of epoch $i$ is at least $1-e^{-\Omega(j)}$.
Thus, by the end of phase $h$, $T_{i-1}$ is completely full with all but exponentially small probability.
The final phase then deterministically ensures that $T_{i-1}$ becomes completely full.

Concretely, the epoch boundaries are
\begin{equation}\label{eq:stages}
 t_0=0,\qquad
 t_i=\sum_{r<i}|T_r|+\floor{(1-\eta)|T_i|}\quad(i<\alpha),\qquad
 t_\alpha=n,
\end{equation}
so that epoch $i$ consists of insertions $t_{i-1}+1,\ldots,t_i$.
If $r$ insertions have already been performed in epoch $i\geq2$, the next
insertion belongs to the final phase if $5r\geq|T_{i-1}|$, and to
phase $j=\min\{h+1,\floor{|T_{i-1}|/(|T_{i-1}|-5r)}\}$ otherwise.

Note that, for $i\geq2$, the epoch boundaries give at most $|T_{i-1}|+|T_i|$
insertions in epoch $i$, and epoch one has at most $|T_1|$. Thus the
epoch lengths decrease geometrically:
\begin{equation}\label{eq:epoch-length}
 t_i-t_{i-1}=O(n(3/4)^i).
\end{equation}
For $n$ a power of two, the stronger bound $O(n/2^i)$ trivially holds. We use the
weaker bound in \eqref{eq:epoch-length} because it also holds for arbitrary
$n$, as shown in \Cref{app:addresses}.

The following simple lemma establishes that each nonfinal phase $j$
indeed consists of $\Theta(|T_{i-1}|/j^2)$ insertions (up to rounding),
and the final phase consists of at least as many insertions as $T_{i-1}$
has empty slots when the epoch begins.

\begin{lemma}[Phase lengths]\label{lem:phases}
In every epoch $i\geq2$:
\begin{enumerate}
\item\label{prop:phase-length} each nonfinal phase $j$ consists of at
least $\floor{|T_{i-1}|/(10j^2)}$ and at most $|T_{i-1}|/(5j^2)+1$ insertions;
\item\label{prop:phase-final} the final phase consists of at least
$|T_{i-1}|-\floor{(1-\eta)|T_{i-1}|}$ insertions.
\end{enumerate}
\end{lemma}
\begin{proof}
Since $j(|T_{i-1}|-5r)\leq|T_{i-1}|$ if and only if $r\geq\ceil{|T_{i-1}|(j-1)/(5j)}$,
phase $j\leq h$ consists of the insertions with
$\ceil{|T_{i-1}|(j-1)/(5j)}\leq r<\ceil{|T_{i-1}|j/(5(j+1))}$. Since
$\floor{a-b}\leq\ceil{a}-\ceil{b}\leq\ceil{a-b}$, its length is at least
$\floor{|T_{i-1}|/(5j(j+1))}\geq\floor{|T_{i-1}|/(10j^2)}$ and at most
$|T_{i-1}|/(5j(j+1))+1$.

For the final phase, write $H=|T_{i-1}|-\floor{(1-\eta)|T_{i-1}|}$. By
\eqref{eq:stages}, the epoch consists of $H+\floor{(1-\eta)|T_i|}$
insertions ($H+|T_\alpha|$ in the last epoch), and the nonfinal phases
consist of the first $M=\ceil{|T_{i-1}|/5-|T_{i-1}|/(5(h+1))}$ of them, where
$|T_{i-1}|=2^{h+1}$ because of the larger root. So it suffices to show that
$M\leq\floor{(1-\eta)|T_i|}$ (or $M\leq|T_\alpha|$). If $h\geq4$, then
$|T_{i-1}|/(5(h+1))\geq1$, so $M\leq\floor{|T_{i-1}|/5}\leq\floor{(1-\eta)|T_i|}$ by
\eqref{eq:sizes}. For $h\leq3$, direct calculation yields the desired
bound, including when $T_i$ is the final single-node tree of
\Cref{app:addresses}.
\end{proof}

\paragraph{The insertion algorithm.}
Suppose $x$ arrives during epoch $i$, phase $j$. We try two things in
order.
\begin{enumerate}
\item\label{step:cleanup} \emph{Cleanup attempt.} Which call we make depends on the epoch, the phase, and $x$'s priority:
\begin{itemize}
\item {\boldmath\textbf{Final phase of epoch $i\geq2$:}} run \mbox{$\Fill(x,i-1,k)$}, where $k$ is the
height of $T_{i-1}$; that is, try the whole of $T_{i-1}$.
\item {\boldmath\textbf{Nonfinal phase $j$ of epoch $i\geq2$, and $g_{i-1}(x)<1/j^4$:}} try $x$'s level-$j$ home subtree in $T_{i-1}$ by running \mbox{$\Fill(x,i-1,j)$}.
\end{itemize}
A call made in this step is a \defn{cleanup attempt}: it is $x$'s
contribution to cleaning up the previous tree. If it succeeds, we are
done; otherwise proceed to step~2.
\item\label{step:current} \emph{Current-array insertion.} %
Place $x$ in $T_i$ by trying
$\Fill(x,i,\ell)$ for $\ell=1,2,\ldots$ in turn, that is, up the home
path, stopping at the first success. If we never succeed, the algorithm
is said to have \emph{failed} (but this never happens, as shown below). We call this a \defn{current-array
insertion}.
\end{enumerate}

We remark that the condition $g_{i - 1}(x) < 1/j^4$ used in the cleanup attempt
ends up playing a somewhat subtle role in the paper. For the main purpose of this section, obtaining
an amortized expected positive query bound of $O(1)$, the condition is actually not necessary at all (although it is used to obtain an $O(\log n)$ worst-case expected positive-query bound). The condition ends up being crucial, however, for supporting \emph{negative} queries in Section~\ref{sec:full}; there, we will make extensive use of the fact that most elements $x$ stored at a high level $j$ of a tree $T_i$ satisfy $g_i(x)<1/j^4$.

Before continuing, we verify the correctness of the insertion algorithm. To see that insertions never fail, we must show that each
epoch $i$ succeeds at its two jobs: it cleans up $T_{i-1}$, and it fills
$T_i$ to a $1-\eta$ fraction. The first is simple to see. $T_{i-1}$
begins the epoch only $\eta$-short of full, while the final phase of
epoch $i$ makes roughly $.6|T_i|$ cleanup attempts at the root of
$T_{i-1}$, that is, into the entire tree, and each of them fills an empty slot
as long as one remains. So even if all the earlier phases do absolutely
nothing, the final phase alone fills $T_{i-1}$ completely. The second job
then follows by counting: the remaining insertions of the epoch all go to
$T_i$, and \eqref{eq:stages} makes their number exactly
$\floor{(1-\eta)|T_i|}$.

\begin{lemma}[Insertions never fail]\label{lem:completion}
Deterministically, the algorithm never fails. Specifically, for each $i<\alpha$, the
tree $T_i$ holds exactly $\floor{(1-\eta)|T_i|}$ elements at time $t_i$,
all of them placed by current-array insertions made during epoch $i$, and
$T_i$ is full at time $t_{i+1}$.
\end{lemma}
\begin{proof}
We show by induction on $i<\alpha$ that no insertion of epochs
$1,\ldots,i$ fails and that, at time $t_i$, the trees $T_1,\ldots,T_{i-1}$
are full and $T_i$ holds exactly $\floor{(1-\eta)|T_i|}$ elements, all
placed by current-array insertions of epoch $i$. Note that each tree
$T_i$ is empty when epoch $i$ begins, since nothing before epoch $i$
targets it. Every insertion of epoch one is a current-array insertion into
$T_1$, and by \eqref{eq:stages} there are $\floor{(1-\eta)|T_1|}$ of them
(or $|T_1|$ if $\alpha=1$), so $T_1$ always has an empty slot, and no
insertion in epoch one fails.

Now suppose the claim holds for some $i<\alpha$, so that $T_i$ has
$H=|T_i|-\floor{(1-\eta)|T_i|}$ empty slots at time $t_i$. By
\eqref{eq:stages}, epoch $i+1$ consists of $H+L$ insertions, where
$L=\floor{(1-\eta)|T_{i+1}|}$, or $L=|T_\alpha|$ if $i+1=\alpha$.
First, $T_i$ gets filled. Every insertion of the final phase makes a cleanup attempt at the root of
$T_i$, whatever its priority, and this fills an empty slot unless $T_i$ is
already full. By \Cref{lem:phases} there are at least $H$ such
insertions, so exactly $H$ cleanup attempts succeed, and $T_i$ is full at
time $t_{i+1}$. All of the other $L$ insertions of the epoch are
current-array insertions into $T_{i+1}$. Before each of them, $T_{i+1}$
holds fewer than $L\leq|T_{i+1}|$ elements, so it has an empty slot and
the insertion succeeds. Thus no insertion of epoch $i+1$ fails, and
$T_{i+1}$ ends the epoch holding exactly $L$ elements, all placed by its
current-array insertions, which for $i+1<\alpha$ is the claim for $i+1$.
\end{proof}

\subsection{The Query Sequence and an Element's Search Bound}
\label{subsec:search}

A query for $u$ probes a fixed sequence computed from $u$'s hashes
alone; it does not know which tree holds $u$, let alone which node.
Generally, the query should spend its probes where elements are likely
to be, and two facts tell us where that is. First, within a tree $T_i$,
most elements are placed near their home leaves. Second, the trees
shrink geometrically, so most elements lie in the larger, earlier trees.
Accordingly, for each tree $T_i$ we fix a \defn{local list} of probes
that explores the places in $T_i$ where $\Fill$ could have put $u$,
roughly in order of how likely they are. Then, we merge the local lists
of all the trees, giving the larger trees priority.

The
local list for $T_i$ interleaves the following two probe sequences:
\begin{enumerate}
  \item the \defn{ancestor list}: $u$'s home leaf $L_i(u)$, followed by
  the nodes on the path from it up to the root;
  \item the \defn{expanding-subtree list}: $u$'s entire level-$1$ home
  subtree, then its entire level-$2$ home subtree, and so on, ending
  with the whole tree.
\end{enumerate}
The ancestor list finds $u$ quickly when $u$ was placed in
one of its own ancestor nodes, which is the typical outcome of a cleanup
attempt. The expanding-subtree list catches the case where $\Fill$
descended from $u$'s ancestor into some other part of a home
subtree. Within $u$'s local list for $T_i$,
$u$'s level-$j$ ancestor is read within the first $O(j)$ probes, and its
level-$h$ home subtree within the first $O(2^h)$.

These local lists are merged into one sequence, $u$'s \defn{query probe
sequence}, such that
\begin{equation}\label{eq:weights}
 \text{the }k\text{th probe of }T_i\text{'s local list occurs within the
 first }16ik^2\text{ probes of }u\text{'s query probe sequence}.
\end{equation}
An injection achieving this bound is given
in~\cite{DBLP:conf/focs/Farach-ColtonKK24}. To maintain constant-time
operations using $O(1)$ words of working memory, we use a different
interleaving with the same bound, described in \Cref{app:probes}.

It will be convenient to have a name for how far down its sequence an element
can possibly sit. We call this a \emph{search bound}. The definition is
in terms of the destination \emph{node}, not the particular slot inside
that node, so that we can later shuffle elements inside a node without
invalidating the bound.

\begin{definition}\label{def:search-bound}
The \defn{search bound} for an element $u$ (that is in the hash table)
is an integer $q(u)$ defined as follows. Let $D$ be a sufficiently
large constant, and let $T_i$ be the tree
containing $u$. If the node containing $u$ is the level-$j$
ancestor of $u$, then $q(u) = Dij^2$. Otherwise, if $h$ is the
smallest integer such that the node containing $u$ is
 in the level-$h$ home subtree
of $u$, then $q(u) = Di4^h$. Either way, $q(u)$ serves as an upper
bound for the number of probes to query $u$.
\end{definition}

\subsection{Probabilistic Preliminaries}
\label{subsec:probability}

Our analyses will make use of two standard probabilistic facts.
 The first is a multiplicative
Chernoff bound:
\begin{lemma}[Chernoff bounds~{\cite[Ch.~4]{MU17}}]\label{lem:chernoff}
Let $X$ be a sum of independent zero-one random variables, and let
$\mu=\E[X]$. Then for every $\delta>0$,
\begin{equation}\label{eq:chernoff}
 \Prb(X\leq(1-\delta)\mu)\leq e^{-\delta^2\mu/2},\qquad
 \Prb(X\geq(1+\delta)\mu)\leq e^{-\delta^2\mu/(2+\delta)}.
\end{equation}
Moreover, if $\mu\leq\bar\mu<a$, then
\begin{equation}\label{eq:count-tail}
             \Prb(X\geq a)\leq(e\bar\mu/a)^a.
\end{equation}
\end{lemma}

The second fact allows us to apply Chernoff bounds in settings where
the randomness comes from balls being thrown into bins (meaning that
the bin loads are technically not quite independent).
The lemma follows from standard negative association arguments~\cite{Waj17}.

\begin{lemma}[Monotone functions of bin loads~\cite{DBLP:journals/rsa/DubhashiR98,Waj17}]\label{lem:bins}
Throw a fixed number of balls into bins independently, each ball
according to its own distribution (a ball may land in no bin). Let
$G_1,\ldots,G_k$ be disjoint sets of bins, and let each
$Y_\ell\in\{0,1\}$ be a function of the loads of the bins in $G_\ell$,
where either every $Y_\ell$ is nondecreasing in these loads or every
$Y_\ell$ is nonincreasing. Then $Y=\sum_\ell Y_\ell$ satisfies the bounds
of \Cref{lem:chernoff}, with $\mu=\E[Y]$, as if the $Y_\ell$ were
independent; and
$\Prb(Y_\ell=1\text{ for all }\ell\in S)\leq\prod_{\ell\in S}\Prb(Y_\ell=1)$
for every $S\subseteq[k]$.
\end{lemma}

\subsection{\texorpdfstring{Phase $j$ Fills Almost Every Height-$j$ Subtree}{Phase j Fills Almost Every Height-j Subtree}}
\label{subsec:fullness}

We now show that the cleanup attempts do their job, meaning that, by the end of phase
$j$ of epoch $i+1$, essentially every level-$j$ subtree of $T_i$
is completely full. This is the technical heart of the construction, and
it is what an insertion relies on when it assumes that a cheap fullness
test will usually come back ``full.''

The whole argument is a counting argument. During phase $j$, a
predictable number of insertions each, with probability $1/j^4$, pick a
uniformly random level-$j$ node of $T_i$ and
fill an empty slot somewhere beneath it, unless its subtree is already full. If
a node receives at least as many attempts as its subtree has empty slots, the
subtree ends up full. So we need two things: a lower bound on the
number of attempts a node receives, which is a simple balls-and-bins
calculation,
and an upper bound on the number of empty slots still present when the phase
begins, which we will obtain by an induction argument. The counts
concentrate only once a subtree is large, so the induction starts at a
level $\lambda$, a large constant that appears only in the analysis.
 The
phases below level $\lambda$ are not needed at all: there the
current-array attempts of epoch $i$ have already left so few empty slots,
because $\eta$ is small, that phase $\lambda+1$ can happily fill them.

\paragraph{Some basic notation and calculations.}
A tree of height $h$ has exactly
$2(2^h-1)+2=2^{h+1}$ slots (recall the root has 4 slots).
So $T_i$ has height $h_i=\log_2n-i-1$ for
$i<\alpha$, and the final tree is a single four-slot root with
$h_\alpha=1$. We write
$T_i[j,k]$ for the $k$th node at level $j$ of $T_i$, counting from $k=0$
on the left, so that the children of $T_i[j,k]$ are $T_i[j-1,2k]$ and
$T_i[j-1,2k+1]$.
The subtree of a level-$j$ node has $2^{j-1}$ leaves,
$2^j-1$ nodes, and $2(2^j-1)$ slots, or two more if the node is a root.

From now on, $\lambda$ is a sufficiently large constant, and we fix
the constant $\eta$ in the schedule as a function of $\lambda$:
\begin{equation}\label{eq:eps}
                     \eta\leq\frac1{400(\lambda+1)^6}.
\end{equation}
As we will see in \Cref{lem:Y}, this makes the current-array insertions
of epoch $i$ leave very few empty slots in each level-$\lambda$ subtree, except
with doubly exponentially small probability as a function of $\lambda$.

For each node in $T_i$, we will be interested in the number of
\defn{direct attempts} it receives, defined as follows.
During epoch $i+1$ and
phase $j>\lambda$, if the insertion of an element $x$
makes a cleanup attempt (i.e., calls $\Fill(x,i,j)$),
the level-$j$ ancestor of $x$ is said to have received a direct attempt.
Additionally, during epoch $i$, whenever there is a current-array insertion
of an element $x$,
the level-$\lambda$ ancestor of $x$ is said to have received a direct attempt.
In both cases, whenever a direct attempt is made on some node $V$,
if the subtree rooted at $V$ contains any empty slots, then one of those
empty slots is guaranteed to be filled.

\paragraph{Analyzing the probability that subtrees are filled.}
We want to bound the probability that a given subtree
$T_i[j,k]$ has any empty slots at the end of phase $j$.
Instead of directly computing the probability of this event, we consider a slightly weaker notion of
failure (a subtree being \emph{bad}, as we define below) that can be analyzed
in a recursive fashion. It is built from a simpler per-node event,
that the node receives too few direct attempts.

Let $X_{i,j,k}$ be the number of direct attempts on node
$T_i[j,k]$. Let $Y_{i,j,k}$ be the indicator of the
event that these attempts are too few to be useful: for any $j$, set
\[
                        \theta_j=\frac{2^j}{50j^6},
\]
a constant fraction of the number of direct attempts we expect;
for $j > \lambda$, let
$Y_{i,j,k}=1$ if $X_{i,j,k}<\theta_j$; and
let $Y_{i,\lambda,k}=1$ if $X_{i,\lambda,k}<2^{\lambda+1}-\theta_{\lambda+1}/4$ (just below the number of slots in a level-$\lambda$
subtree).

The next lemma bounds the probability $Y_{i,j,k} = 1$ for
any $i,j,k$.
\begin{lemma}[Each node is rarely bad]\label{lem:Y}
For every $j\geq\lambda$
and every $k$,
\begin{equation}\label{eq:Y}
                  \Prb(Y_{i,j,k}=1)\leq e^{-\Omega(2^{j/2})},
\end{equation}
where the hidden constant is absolute (in particular, independent of
$\lambda$).
\end{lemma}
\begin{proof}
We apply the Chernoff bound \eqref{eq:chernoff}.
By symmetry, the number of direct attempts each subtree
at level $\lambda$ receives is in expectation
at least $(1-\eta)2^{\lambda+1} - 1$ (the minus
1 handles rounding issues from the number of insertions
into $T_i$ during epoch $i$).

Since \eqref{eq:eps} gives $\eta2^{\lambda+1}\leq\theta_{\lambda+1}/8$,
this mean exceeds the threshold $2^{\lambda+1}-\theta_{\lambda+1}/4$ by
at least $\theta_{\lambda+1}/8-1$. That is a fraction
$\Omega(\lambda^{-6})$ of the mean, so the probability that
$X_{i,\lambda,k}$ is below the threshold is
$e^{-\Omega(2^\lambda/\lambda^{12})}$.

For $j>\lambda$, phase $j$ of epoch $i+1$ is a nonfinal phase, so it
consists of at least $\floor{|T_i|/(10j^2)}\geq|T_i|/(20j^2)$ insertions. Each of
them makes a direct attempt on $T_i[j,k]$ exactly when its priority
$g_i(x)$ is below $1/j^4$ and its level-$j$ home subtree is the one
rooted at $T_i[j,k]$, which happens with probability
$2^{j+1}/(j^4|T_i|)$, since that subtree contains a $2^{j+1}/|T_i|$
fraction of the leaves. So the mean of $X_{i,j,k}$ is at least
$2^j/(10j^6)=5\theta_j$, and the probability that
$X_{i,j,k}<\theta_j$ is
$e^{-\Omega(2^j/j^6)} = e^{-\Omega(2^{j/2})}$.
\end{proof}

We now define bad subtrees. For
$\lambda<j\leq2\lambda+1$, let $Z_{i,j,k}=1$ if some $Y$ in the subtree of
$T_i[j,k]$, at levels $\lambda$ through $j$, equals one. For
$j>2\lambda+1$, set $h=\floor{j/2}$, so that $h>\lambda$, and let
$Z_{i,j,k}=1$ if
\begin{equation}\label{eq:Z}
 Y_{i,j,k}=1\quad\text{or}\quad
 \sum_{k'=2^{j-h}k}^{2^{j-h}(k+1)-1} Z_{i,h,k'}\geq2^{j/4}.
\end{equation}
We call a subtree \defn{bad} if $Z_{i,j,k} = 1$. In words, we call a subtree bad if it receives too few attempts of
its own, or if too many of its half-height subtrees were already bad.

The next lemma says that if a subtree at level $j$ is not bad, then it
is full at the end of phase $j$.

\begin{lemma}\label{lem:full}
If $j>\lambda$ and $Z_{i,j,k}=0$, the subtree of $T_i[j,k]$ is full by the
end of phase $j$ of epoch $i+1$.
\end{lemma}
\begin{proof}
First, a fact about a single phase $\ell>\lambda$. Let $V$ be a
level-$\ell$ node with $Y=0$ whose subtree has at most $\theta_\ell$
empty slots when phase $\ell$ begins. Each attempt at $V$ fills an empty slot of its
subtree unless that subtree is already full, and $V$ receives at least
$\theta_\ell$ attempts, so its subtree is full when the phase ends.

We now prove the lemma by induction on $j$. The base case is
$\lambda<j\leq2\lambda+1$, where $Z_{i,j,k}=0$ means that every $Y$ in the
subtree, at levels $\lambda$ through $j$, is zero. The base case is
itself an induction, on the level $\ell=\lambda+1,\ldots,j$: we show that
every level-$\ell$ subtree inside ours is full after phase $\ell$. Each
level-$\lambda$ subtree has fewer than $\theta_{\lambda+1}/4$ empty slots at
time $t_i$, and so also when phase $\lambda+1$ begins. Each
level-$(\lambda+1)$ subtree then has fewer than
$\theta_{\lambda+1}/2+2\leq\theta_{\lambda+1}$ empty slots, counting the node's
own two slots, so by the fact above it is full after phase $\lambda+1$.
For $\ell>\lambda+1$, both children of each level-$\ell$ node are full
after phase $\ell-1$, so only the node's own slots can be empty when phase
$\ell$ begins, and the fact applies again.

For the inductive step, take $j>2\lambda+1$ and let $h=\floor{j/2}$, so
that $h>\lambda$. By the induction hypothesis applied to each level-$h$
subtree, after phase $h$ the only level-$h$ subtrees that can still
contain empty slots are the bad ones, and since $Z_{i,j,k}=0$, the second
clause of \eqref{eq:Z} tells us there are fewer than $2^{j/4}$ of them
inside our subtree. Each such subtree has fewer than $2^{h+1}$ slots.
Above the level-$h$ subtrees, our subtree contains fewer than
$2^{j-h}\leq2\cdot2^{j/2}$ nodes, holding at most $2\cdot2^{j-h}+2\leq
4\cdot2^{j/2}$ slots. So when phase $j$
begins, the number of empty slots is at most
\[
 2^{j/4}2^{h+1}+4\cdot2^{j/2}
 \leq 2\cdot2^{3j/4}+4\cdot2^{j/2}
 \leq \theta_j,
\]
where the last inequality, which is equivalent to
$100j^62^{-j/4}+200j^62^{-j/2}\leq1$, holds because $j>2\lambda+1$ and
$\lambda$ is large. Since
$Y_{i,j,k}=0$, the fact above applies, and filled slots never empty again, so the
subtree is full when the phase ends.
\end{proof}

In the following lemma, we bound the probability that a given
subtree is bad. This is where it matters that \eqref{eq:Z} compares a
subtree against its half-height subtrees rather than against its two
children. A bad child of $T_i[j,k]$ may leave as many as $2^{j}$
empty slots, far more than the $\theta_j$ empty slots that phase $j$ can be relied
on to fill, so a recursion through the children would have to treat a
single bad child as fatal. The probability that a level-$j$ subtree is
bad would then be at least the probability that its child is, and it
would not decay with $j$ at all. A bad level-$h$ subtree, by contrast,
has fewer than $2^{h+1}\leq2\cdot2^{j/2}$ slots, and the proof of
\Cref{lem:full} tolerates up to $2^{j/4}$ of them. So for $T_i[j,k]$ to
be bad, many rare events must occur at once.

These events are not independent, because all of the half-height
subtrees draw their attempts from the same insertions. But they concern
disjoint sets of nodes and are monotone in the numbers of attempts those
nodes receive, so \Cref{lem:bins} lets us treat them as negatively
associated, and apply the tail bound \eqref{eq:count-tail} to their
number as if they were independent. Because there are $2^{j-h}\geq2^{j/2}$
of them, each bad with probability at most $e^{-2^{h/5}}$, the
probability that $2^{j/4}$ of them are bad is essentially the product of
$2^{j/4}$ such probabilities, about $e^{-2^{j/4}2^{h/5}}\approx
e^{-2^{7j/20}}$. This comfortably beats the target $e^{-2^{j/5}}$, which
is what lets us complete the induction.

\begin{lemma}[Probability of a bad subtree]\label{lem:full-tail}
If $\lambda$ is a sufficiently large constant and $\eta$ satisfies
\eqref{eq:eps}, then for every $j>\lambda$,
\begin{equation}\label{eq:full-tail}
                      \Prb(Z_{i,j,k}=1)\leq e^{-2^{j/5}}.
\end{equation}
\end{lemma}
\begin{proof}
We first show that, at any fixed level, \Cref{lem:bins} applies to the
$Z$ variables. Each $Y_{i,\ell,k'}$ is a
nonincreasing function of its own node's count $X_{i,\ell,k'}$. By
induction on $j$, each $Z_{i,j,k}$ is a nondecreasing function of the $Y$
variables of the nodes in its subtree at levels $\lambda$ through $j$.
In the base case it is their maximum, and in \eqref{eq:Z} it is
nondecreasing in $Y_{i,j,k}$ and in the $Z_{i,h,k'}$, which by induction
are nondecreasing functions of the $Y$ variables lower down in the
subtree. So $Z_{i,j,k}$ is a nonincreasing function of the counts of the
nodes in its subtree. The direct attempts on $T_i$ form a single
balls-and-bins experiment with the nodes of $T_i$ as bins: each of the
$\floor{(1-\eta)|T_i|}$ current-array insertions into $T_i$
(\Cref{lem:completion}) lands in a uniformly random level-$\lambda$ node,
and each insertion of phase $\ell>\lambda$ of epoch $i+1$ lands in a
uniformly random level-$\ell$ node with probability $1/\ell^4$, and in
no bin otherwise. Each ball's bin is determined by its own hashes
$L_i(x)$ and $g_i(x)$, so the balls are independent. So at any fixed
level, the $Z$ variables are monotone functions of the loads of disjoint
sets of bins, and \Cref{lem:bins} applies to them.

We prove \eqref{eq:full-tail} by strong induction on $j$. For the base
case, start with $\lambda<j\leq2\lambda+1$. If $Z_{i,j,k}=1$, then some
node of the subtree at levels $\lambda$ through $j$ has $Y=1$. There are
$2^{j-\lambda+1}-1<2^{\lambda+2}$ such nodes, and each of them is at
level at least $\lambda$, so by \eqref{eq:Y} each has $Y=1$ with
probability $e^{-\Omega(2^{\lambda/2})}$. A union bound gives
$\Prb(Z_{i,j,k}=1)\leq2^{\lambda+2}e^{-\Omega(2^{\lambda/2})}
< e^{-2^{(2\lambda+1)/5}}$ as $\lambda$ was chosen to be
sufficiently large
(recall the constant in the exponent does not depend on $\lambda$).

For the induction step, let $j>2\lambda+1$ and assume
\eqref{eq:full-tail} at every level strictly between $\lambda$ and $j$.
Let $h=\floor{j/2}$. Since
$\lambda<h<j$, the inductive hypothesis applies at level $h$: each of the
$2^{j-h}$ half-height subtrees inside $T_i[j,k]$ is bad with probability at
most $e^{-2^{h/5}}$. Let $S$ be the number of them that are bad, which is
the sum in \eqref{eq:Z}. By the first paragraph of this proof, applied at
level $h$, $S$ obeys the bounds of \Cref{lem:chernoff}, and its mean is
at most $\bar\mu=2^{j-h}e^{-2^{h/5}}$. Since $h\geq(j-1)/2$,
\[
 \frac{e\bar\mu}{2^{j/4}}
 \leq e2^je^{-2^{(j-1)/10}}
 \leq\frac12,
\]
where the last inequality holds because $j>2\lambda+1$ and $\lambda$ is
large. Thus, a Chernoff
bound~\eqref{eq:count-tail} bounds the probability of the second event
in \eqref{eq:Z} by
\[
 \Prb(S\geq2^{j/4})
 \leq\left(\frac{e\bar\mu}{2^{j/4}}\right)^{2^{j/4}}
 \leq 2^{-2^{j/4}}
 \leq e^{-2\cdot2^{j/5}},
\]
where the last inequality, $2^{j/20}\ln2\geq2$, again holds because $j$
is large.

For the first event in \eqref{eq:Z}, \eqref{eq:Y} gives
$\Prb(Y_{i,j,k}=1)\leq e^{-\Omega(2^{j/2})}\leq e^{-2\cdot2^{j/5}}$,
where the second inequality again holds because $j$ is large. A union bound over the two events gives
$\Prb(Z_{i,j,k}=1)\leq2e^{-2\cdot2^{j/5}}\leq e^{-2^{j/5}}$.
\end{proof}

\subsection{Analyzing Current-Array Insertions}
\label{subsec:overflow}
Next, we analyze the state of $T_i$ at the start of epoch
$i+1$, meaning that it has already received all of its
current-array insertions but none of its cleanup insertions.
We wish to show that, at this point in time, almost all
insertions have interacted with only small subtrees. This
ensures that those insertions (and corresponding queries)
are cheap.

Call a subtree
$P$ with at least two leaves \defn{overcrowded} if one of its two children
receives at least its own capacity in \defn{home arrivals}, that is,
current-array insertions into $T_i$ whose home leaf lies in that child.

\begin{lemma}[Where overflow can go]\label{lem:cover}
\begin{enumerate}[label=(\roman*)]
\item A specified subtree with $s$ leaves is overcrowded with
probability $e^{-\Omega(s)}$, where the constant depends only on
$\eta$. For disjoint subtrees, these upper bounds multiply.
\item In a process consisting only of current-array attempts, the target
subtree of every call $\Fill(x,i,\ell)$ with $\ell\geq2$ lies inside an
overcrowded subtree at level $\ell$ or above. In particular, an element whose
current-array attempt reaches level $J$ is placed inside an overcrowded
subtree at level $J$ or above.
\end{enumerate}
\end{lemma}
\begin{proof}
For (i), a child with $s/2$ leaves has capacity $2(s-1)$. Each of the
$\floor{(1-\eta)|T_i|}$ current-array insertions into $T_i$ has its home
leaf in that child with probability $(s/2)/(|T_i|/4)=2s/|T_i|$, so the
child's expected number of home arrivals is at most $(1-\eta)2s$, which
is below
$(1-\eta/2)2(s-1)$ once $s\geq4/\eta$. Apply
\eqref{eq:chernoff} and sum over the two children. Distinct parents that
are disjoint have disjoint sets of possible witness children, so
the product bound of \Cref{lem:bins}, followed by a sum over the two
choices of witness for each parent, gives the product bound.

Part (ii) is deterministic, and is the statement that overflow cannot
travel far without leaving evidence behind. A call $\Fill(x,i,\ell)$ with
$\ell\geq2$ is made only because $x$'s level-$(\ell-1)$ home subtree was
full. Consider the largest subtrees that such calls ever target; these are
disjoint. Look at the first call that targets such a maximal subtree $P$.
At that moment one child of $P$ was full. Since neither $P$ nor any
ancestor of $P$ had ever been targeted before, that child cannot have
received any element from outside itself, so all of the elements filling it
arrived at their own homes inside it: its home arrivals number at least
its capacity, and $P$ is overcrowded. Every other call, including where
it inserts its element below its target, stays inside some maximal $P$.
\end{proof}

\begin{lemma}[Early overflow]\label{lem:early}
Before epoch $i+1$, the probability that a call of $\Fill$ targets a
specified level-$j$ subtree of the tree $T_i$, for $j\geq2$, is
$e^{-\Omega(2^j)}$, where the constant depends only on $\eta$. The same bound holds for a fixed current-array
insertion to reach level $j$ on its home path.
\end{lemma}
\begin{proof}
Before epoch $i+1$ the only calls of $\Fill$ into $T_i$ are those of
the current-array attempts of epoch $i$, so by \Cref{lem:cover}(ii) a
specified subtree is targeted only if it or one of its ancestors is
overcrowded. There is at most one ancestor at each level from $j$ upward,
so summing $e^{-\Omega(2^\ell)}$ over $\ell\geq j$ proves the
first claim.

The second claim follows in the same way: a current-array insertion of
$x$ reaches level $j$ only through a call $\Fill(x,i,j)$, which targets
$x$'s level-$j$ home subtree.
\end{proof}

\subsection{Bounding Insertion and Query Time}
\label{subsec:accounting}
First, we show insertions are cheap.

\begin{lemma}[Insertion cost]\label{lem:insert}
Every fixed insertion takes $O(1)$ expected time.
\end{lemma}
\begin{proof}
Consider the two steps in turn. Checking the
priority takes constant time. A cleanup attempt in phase $j$ first
reads its target node $V$ and, if $V$ is nonfull, the two children of $V$.
If $V$ is full, or both children are full, the attempt ends after these
constant many probes. Only if a child of $V$ is
nonfull does it insert below $V$, at cost $O(j)$. For $j\leq\lambda+1$ this cost
is a constant outright. For $j\geq\lambda+2$, the children of $V$ are
level-$(j-1)$ subtrees with $j-1>\lambda$, and a nonfull one at this point was
already nonfull at the end of phase $j-1$. The incoming element's home is
fresh, so \Cref{lem:full} and \Cref{lem:full-tail} bound the probability
of this case, and the expected cost of the cleanup attempt is at most
\[
                       O(1)+O(j)\cdot2e^{-2^{(j-1)/5}}=O(1).
\]
In the final phase, the target is the root, and the same estimate holds
with $j$ replaced by the height of $T_{i-1}$.

For the current-array attempt, each call that fails costs one node read, and
the one that succeeds costs $O(\ell)$ at level $\ell$. \Cref{lem:early}
bounds the probability of reaching level $\ell$ by $e^{-\Omega(2^\ell)}$, so
summing $O(\ell)$ over the levels, up to and including the root, gives a
constant. In the last epoch the current tree has constant size.
\end{proof}

We next bound query time. The search bound of an element in
$T_i$ carries a factor $i$, and the next lemma shows that this factor is
the only thing that matters: up to constants, an element's expected search
bound is its epoch number. This is where the priority condition in the
cleanup attempt earns its keep.

\begin{lemma}[A fixed element's search bound]\label{lem:worst-query}
For a fixed element $u$ inserted during epoch $s$,
\begin{equation}\label{eq:worst-query}
                    \E[q(u)]=O(s)=O(\log n).
\end{equation}
\end{lemma}
\begin{proof}
We bound the contribution of each way in which $u$ can be placed. If
$s=1$, only the last case arises; otherwise let $j$ be the phase.

\begin{enumerate}
\item \emph{Step~\ref{step:cleanup}, nonfinal phase $j$, placement at or
within the target.} The cleanup attempt places $u$ in its
level-$j$ home subtree (phase $j\leq\lambda+1$) or in its target node
(later phases). The search bound is respectively $O(s4^{\lambda+1})$ and
$O(sj^2)$. In the second case the cleanup attempt must have been made at
all, which requires $g_{s-1}(u)<1/j^4$, an event of probability $1/j^4$
independent of the table and of $u$'s home, so the expected contribution
is $O(sj^2/j^4)=O(s)$.
\item \emph{Step~\ref{step:cleanup}, nonfinal phase $j\geq\lambda+2$,
placement below the target.} The cleanup attempt inserts
its element below its target node. As in \Cref{lem:insert}, this has
probability at most $2e^{-2^{(j-1)/5}}$, and the search bound is
$O(s4^j)$.
\item \emph{Step~\ref{step:cleanup}, final phase.} The cleanup attempt
places $u$ anywhere in $T_{s-1}$. Let $h$ be the height of $T_{s-1}$. Phase $h$ is a nonfinal
phase at the root, so this requires $T_{s-1}$ to be nonfull at the end
of phase $h$, which by \Cref{lem:full,lem:full-tail} has probability at
most $e^{-2^{h/5}}$. The search bound is $O(s4^h)$.
\item \emph{Step~\ref{step:current}, current-array insertion.} This
covers every element whose cleanup attempt was not made or failed. The
current-array insertion reaches level $\ell$, with probability at
most $e^{-\Omega(2^\ell)}$ by \Cref{lem:early}, and search bound $O(s4^\ell)$.
\end{enumerate}
Each of these contributions is $O(s)$, and the sum over $\ell$ converges because
doubly exponential decay beats the polynomial factors.
(A placement into one of the constant-size trees has search bound $O(s)$
outright.) Finally, $s\leq\alpha=O(\log n)$.
\end{proof}

The factor $s$ costs nothing on average, because late epochs are short.
So the total of all search bounds is linear.

\begin{lemma}[Total search bounds]\label{lem:moments}
Let $q_t$ be the search bound of the element placed by insertion $t$. Then
\begin{equation}\label{eq:moments}
                        \E\Bigl[\sum_{t=1}^n q_t\Bigr]=O(n).
\end{equation}
\end{lemma}
\begin{proof}
By \eqref{eq:epoch-length}, epoch $s$ has $O(n(3/4)^s)$ insertions,
and by \Cref{lem:worst-query} each has
expected search bound $O(s)$. Summing, $\E[\sum_tq_t]=O(\sum_ssn(3/4)^s)=O(n)$.
\end{proof}

Putting the pieces together, we obtain our first theorem.

\thmintrostable*
\begin{proof}
Correctness of insertion holds on every outcome, by
\Cref{lem:completion}, and \Cref{lem:insert} gives the expected
insertion cost. A query for a present element stops by the time it reaches
the element's search bound. For each fixed element, \Cref{lem:worst-query} bounds
the expected search bound by $O(s)=O(\log n)$, where $s$ is the epoch in
which the element was inserted. For the amortized bound:
during epoch one every element has $s=1$, and after epoch one we have
$t\geq t_1=\floor{(1-\eta)|T_1|}=\Omega(n)$, so by
\Cref{lem:moments} the expected search bounds of the first $t$ elements sum
to $O(n)=O(t)$. The only metadata stored outside the array is a counter
of the elements present, from which the epoch and phase of an insertion
are computed (\Cref{app:addresses}). (Queries do not read
it; the probe schedule of \Cref{app:probes} needs only $O(1)$ words of
scratch space during a query.) No inserted element is ever moved, so the table is stable.
\end{proof}

\section{\texorpdfstring{$O(1)$}{O(1)} Worst-Case Expected Positive Queries}
\label{sec:reorder}

In the previous section, we showed that we can get
an amortized expected query time of $O(1)$
without reordering---on average, each element is
cheap to query.
In this section, we show that the following simple modification to the
insertion algorithm
turns this amortized bound into a \emph{worst-case} expected bound:
on an insertion of an element $x$ during an epoch $s > 2$, we swap $x$ with
the element $y$ in the first slot of $L_1(x)$ (recall that $T_1$, and thus $L_1(x)$,
 is full by the end of epoch 2)
 and then run the algorithm of
\Cref{sec:stable} on $y$ instead of $x$. We say the element $y$ is
\defn{evicted during epoch $\boldsymbol{s}$}.

Intuitively, this modification makes it so that
some random previously inserted
element
($y$)
pays the $\Theta(s)$ expected query cost of an element inserted
during epoch $s$ instead of $x$.
There are $\Theta(n)$
 leaves of $T_1$, and there are only $O(n/2^s)$ elements
inserted in epoch $s$, so any given element has only a small chance of
having a high query cost. This intuition is made formal in the following lemma.
\begin{lemma}\label{lem:evict-prob}
      For any element $x$ and epoch $s\geq 3$, the probability that $x$ is evicted
      in epoch $s$ is $O((3/4)^s)$.
\end{lemma}
\begin{proof}
      At any insertion during epoch $s$, $x$ may only be evicted if it is in
      the first slot of a leaf of $T_1$, and, as $T_1$ has $\Theta(n)$ leaves,
      the chance $x$ is evicted is $O(1/n)$. By \eqref{eq:epoch-length}, epoch $s$
      consists of $O(n(3/4)^s)$ insertions, so $x$ is evicted in epoch $s$
      with probability $O((3/4)^s)$.
\end{proof}

\thmintroreorder*

\begin{proof}
      An element in $T_1$ has been placed using only $L_1$ and possibly $g_1$. The hashes
      $g_{s-1}(y)$, $L_{s-1}(y)$, and $L_s(y)$ used to reinsert an element
      evicted in epoch $s>2$ are still fresh and independent. Its reinsertion
      therefore has the same distribution as an insertion in \Cref{sec:stable}.
      Our modification only adds $O(1)$ work to insertions, so the expected insertion
      time remains $O(1)$. It remains to show the worst-case expected query time is $O(1)$.

      An element $x$ can be evicted at most once: after it is evicted from
      $T_1$ in epoch $s$, it is inserted either into $T_{s-1}$ or $T_s$ and thus can
      never be evicted again. The moment after $x$ is inserted, its expected query time
      is $O(1)$: if it is inserted in epoch $s \leq 2$, this follows from \Cref{lem:worst-query},
      and, otherwise, this follows from the fact that $x$ is inserted into $L_1(x)$.
      Further, conditioned on being evicted in epoch $s$, its expected query time is $O(s)$ by
      \Cref{lem:worst-query}. Therefore, at any given point in time, the
      expected query time for
      $x$ is at most
      \[O(1) + \sum_{s\geq3} \Prb\left[x \text{ was evicted in epoch } s\right] \cdot O(s).\]
      By \Cref{lem:evict-prob}, it follows that
      the expected query time for $x$ is
      \[O(1) + \sum_{s\geq3}  O(s(3/4)^s) = O(1).\]
\end{proof}

\section{\texorpdfstring{$O(1)$}{O(1)} Worst-Case Expected Negative Queries}
\label{sec:full}
Whereas a positive query for $x$ can simply stop
when $x$ is found, a negative query needs an explicit stopping rule
to avoid searching through the entire table. To do this, we leave
implicitly encoded ``breadcrumbs'' throughout the data structure
that let a negative query know when to stop (or, perhaps more
accurately, when to continue).

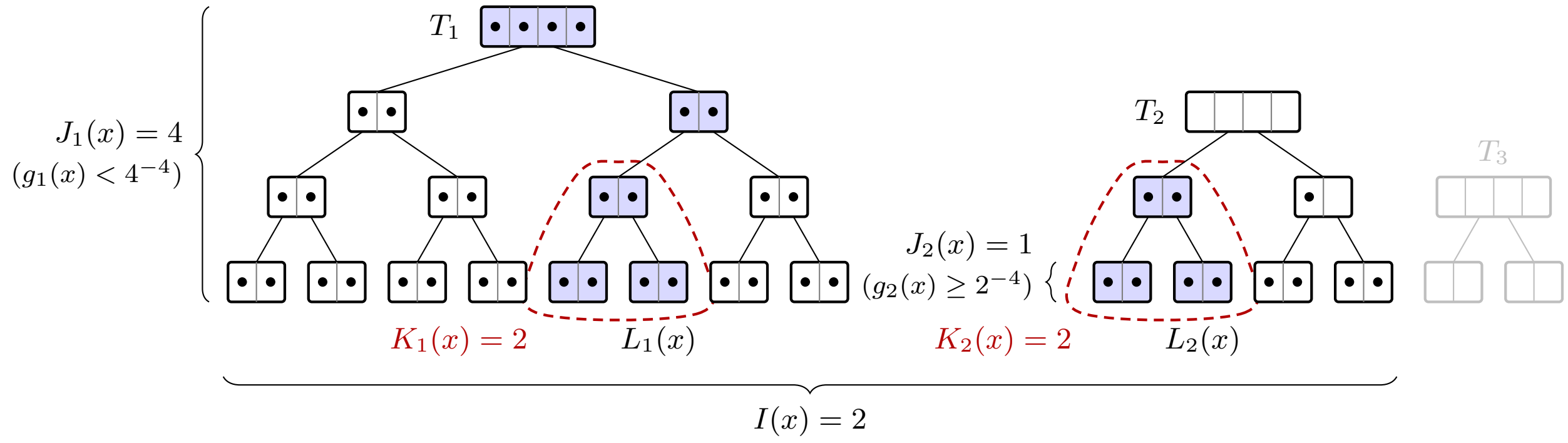
\begin{figure}[!b]
\centering
\begin{tikzpicture}[>=stealth]
  \def\nodes{0/1/2/2/0, 0.85/1/2/2/0, 1.7/1/2/2/0, 2.55/1/2/2/0,
             3.4/1/2/2/1, 4.25/1/2/2/1, 5.1/1/2/2/0, 5.95/1/2/2/0,
             0.425/2/2/2/0, 2.125/2/2/2/0, 3.825/2/2/2/1, 5.525/2/2/2/0,
             1.275/3/2/2/0, 4.675/3/2/2/1, 2.975/4/4/4/1,
             9.15/1/2/2/1, 10/1/2/2/1, 10.85/1/2/2/0, 11.7/1/2/2/0,
             9.575/2/2/2/1, 11.275/2/2/1/0, 10.425/3/4/0/0}
  \foreach \l/\r in {2.975/4.675, 8.725/10.425}
    {\pgfmathsetmacro\m{(\l+\r)/2}
     \draw[red!70!black,thick,densely dashed] plot[smooth cycle,tension=0.6]
       coordinates {(\l,-0.33) (\r,-0.33) (\r+0.05,0.25) (\m+0.425,1.15)
                    (\m,1.27) (\m-0.425,1.15) (\l-0.05,0.25)};}
  \foreach \px/\pl/\cx in {0.425/2/0, 0.425/2/0.85, 2.125/2/1.7,
      2.125/2/2.55, 3.825/2/3.4, 3.825/2/4.25, 5.525/2/5.1, 5.525/2/5.95,
      1.275/3/0.425, 1.275/3/2.125, 4.675/3/3.825, 4.675/3/5.525,
      2.975/4/1.275, 2.975/4/4.675,
      9.575/2/9.15, 9.575/2/10, 11.275/2/10.85, 11.275/2/11.7,
      10.425/3/9.575, 10.425/3/11.275}
    {\pgfmathsetmacro\o{\cx<\px ? -0.15 : 0.15}
     \draw (\px+\o,0.9*\pl-1.11) -- (\cx,0.9*\pl-1.59);}
  \foreach \x/\l/\k/\c/\h in \nodes
    {\pgfmathsetmacro\y{0.9*\l-0.9}
     \ifnum\h=1 \def\fillc{blue!15}\else\def\fillc{white}\fi
     \draw[thick,rounded corners=1pt,fill=\fillc]
       (\x-0.15*\k,\y-0.21) rectangle (\x+0.15*\k,\y+0.21);
     \foreach \i in {1,...,\k}
       {\ifnum\i<\k
          \draw[gray] (\x-0.15*\k+0.3*\i,\y-0.21) -- +(0,0.42);
        \fi
        \ifnum\i>\c\else
          \fill (\x-0.15*\k+0.3*\i-0.15,\y) circle (0.05);
        \fi}}
  \node[font=\footnotesize,anchor=east] at (2.3,2.7) {$T_1$};
  \node[font=\footnotesize,anchor=east] at (9.75,1.8) {$T_2$};
  \begin{scope}[gray!50]
    \draw (12.925,0.69) -- (12.65,0.21);
    \draw (13.225,0.69) -- (13.5,0.21);
    \foreach \x/\y/\k in {12.65/0/2, 13.5/0/2, 13.075/0.9/4}
      {\draw[thick,rounded corners=1pt]
         (\x-0.15*\k,\y-0.21) rectangle (\x+0.15*\k,\y+0.21);
       \foreach \i in {1,...,\k}
         {\ifnum\i<\k
            \draw (\x-0.15*\k+0.3*\i,\y-0.21) -- +(0,0.42);
          \fi}}
    \node[font=\footnotesize,anchor=south] at (13.075,1.11) {$T_3$};
  \end{scope}
  \foreach \x/\i in {4.25/1, 10/2}
    {\node[font=\footnotesize,anchor=north] at (\x,-0.33) {$L_\i(x)$};}
  \foreach \x/\i in {3.0/1, 8.75/2}
    {\node[font=\footnotesize,anchor=north east,red!70!black]
       at (\x,-0.33) {$K_\i(x)=2$};}
  \draw[decorate,decoration={brace,amplitude=4pt}]
    (-0.5,-0.21) -- (-0.5,2.91)
    node[midway,anchor=east,xshift=-4pt,font=\footnotesize,align=right]
      {$J_1(x)=4$\\[1pt] \scriptsize ($g_1(x)<4^{-4}$)};
  \draw[decorate,decoration={brace,amplitude=3pt}]
    (8.45,-0.21) -- (8.45,0.21)
    node[midway,anchor=east,xshift=-3pt,yshift=5pt,font=\footnotesize,
         align=right]
      {$J_2(x)=1$\\[1pt] \scriptsize ($g_2(x)\geq2^{-4}$)};
  \draw[decorate,decoration={brace,mirror,amplitude=5pt}]
    (-0.35,-1.0) -- (12.05,-1.0)
    node[midway,anchor=north,yshift=-5pt,font=\footnotesize] {$I(x)=2$};
\end{tikzpicture}%
\caption{The stopping condition for a query for $x$ with $I(x)=2$. (In
the construction, $I(x)\geq100$; the figure uses a smaller value for
illustration.) Each dot is an element. The shaded nodes are the only
ones that may hold $x$: in each tree $T_i$ with $i\leq I(x)$, the
ancestors of $L_i(x)$ at levels $1$ through $J_i(x)$ (braces), or
through $\max\{J_1(x),I(x)\}$ in $T_1$, together with all of $x$'s
level-$K_i(x)$ home subtree (dashed outline). Each $J_i(x)$ is the
largest $j$ with $g_i(x)<1/j^4$, so it is determined by $x$'s priority
alone. The later tree $T_3$ (grey) is not searched at all.}
\label{fig:stopping}
\end{figure}

We will use these breadcrumbs to keep track of three types of bounds on the query time
for an element $x$. First, we track an index $I(x)$ such that $x$ can
only reside in one of the trees $T_1,\ldots,T_{I(x)}$. Second, for each
$i\leq I(x)$, we track two levels $J_i(x)$ and $K_i(x)$
with the following guarantee:
\begin{itemize}
\item if $x$ resides in $T_1$, then it lies either in an ancestor of
$L_1(x)$ at level at most $\max\{J_1(x),I(x)\}$, or somewhere in its
level-$K_1(x)$ home subtree;
\item if $x$ resides in $T_i$ for some $i\geq2$, then it lies either
in an ancestor of $L_i(x)$ at level at most $J_i(x)$, or somewhere in its
level-$K_i(x)$ home subtree.
\end{itemize}
The level $J_i(x)$ covers where $x$ would
typically be placed, and it will depend only on $g_i(x)$; the
level $K_i(x)$, together with $I(x)$ in $T_1$, covers every other possibility.
A query can conclude $x$ is not in $T_i$ once it finishes probing all
of these slots: the ancestors of $L_i(x)$ up to $J_i(x)$ if $i \geq 2$ or $\max\{J_i(x), I(x)\}$ if $i = 1$, and
all slots in the level-$K_i(x)$ home subtree of $x$. A query can then
stop looking for $x$ once it has concluded $x$ is not in $T_i$ for all $i \leq I(x).$
See \Cref{fig:stopping} for an illustration.

It remains to explain how $I(x), J_i(x)$, and $K_i(x)$ are encoded. Their precise definitions
are given in \Cref{tab:bounds}; this paragraph covers some of the
intuition behind them. $J_i(x)$ is simply
the largest $j$ such that $g_i(x) < 1/j^4$ and represents the node a
successful nonfinal-phase cleanup attempt may place $x$ in.
$I(x)$ and $K_i(x)$, on the other hand, are calculated from
the breadcrumbs, which are stored in one of two ways.
First, placing an element of unusually
low priority in a leaf of $T_1$ (such elements usually
reside higher up in $T_1$)
allows us to encode (an upper bound for) $I(x)$.
Insertions can find these low-priority elements efficiently
by looking at higher levels of $T_1$ and can then swap such
an element into a leaf in order to update $I(x)$ efficiently.
This search can fail, however, with small
probability, which is why $I(x)$ ends up partially relying on
$K_1(x)$ in \Cref{tab:bounds}.
Second, the relative order of the two elements in a node encodes a
single bit: a node encodes $0$ if it holds zero elements,
one element,
or two elements in increasing order, and it encodes $1$
otherwise.\footnote{For a root node, we only consider the elements in
the first two slots of the node to calculate what bit it encodes.}
These bits are used to encode $K_i(x)$: specifically,
$K_i(x)=K+1$, where $K$ is the largest level
such that the level-$K$ ancestor of $x$ is set to $1$
and such that, for all $j \in [J_i(x), K]$,
the level-$j$ ancestor of $x$ is also set to $1$;
if no such $K$ exists then $K_i(x) = 1$.
The full details are given in \Cref{tab:bounds}.

\begin{table}[t]
\centering
\small
\begin{tabular}{@{}l p{0.8\textwidth}@{}}
\toprule
& Definition \\
\midrule
$J_i(x)$ & The largest $j$ with $g_i(x)<1/j^4$. Depends only on $x$'s
priority. \\
\addlinespace
$K_i(x)$ & $K+1$, where $K$ is the largest level such that the
level-$j$ ancestor of $x$ has bit $1$ for $j=K$ and for every
$j\in[J_i(x),K]$; or $1$ if there is no such $K$. \\
\addlinespace
$I(x)$ & $100\cdot\max\{K_1(x),C(L)\}$, where $L$ ranges over the leaves
of $T_1$ that the query waits for ($L_1(x)$ and the leaves of $x$'s
level-$K_1(x)$ home subtree), and $C(L)$ is the largest $J_1$ of an
element of $L$, or $1$ if $L$ is empty. \\
\bottomrule
\end{tabular}
\caption{The three bounds that a query for $x$ tracks. The query computes
$K_1(x)$ before $I(x)$, since $I(x)$ depends on it.}
\label{tab:bounds}
\end{table}

\paragraph{Insertions.}
An insertion of $x$ during epoch $s$ performs the following three steps.
As a rule, apart from the bits that steps~2 and~3 deliberately set,
we implicitly assume that the insertion is implemented so that every
node that the insertion touches keeps its bit as it was prior
to the insertion. Additionally, it is worth noting that the algorithm
does not explicitly reason about $I,J,K$; instead, it implicitly
updates the breadcrumbs so that queries continue to be correct.
The three steps of the insertion are as follows:
\begin{enumerate}
\item\label{step:neg-insert} \emph{Insert as in \Cref{sec:reorder}.} If $s\leq2$, insert $x$ with
the algorithm of \Cref{sec:stable}. Otherwise, put $x$ in place of the
element $y$ in the first slot of $L_1(x)$, and insert $y$ with the
algorithm of \Cref{sec:stable}, which places $y$ in $T_{s-1}$ or $T_s$.
\item\label{step:neg-trail} \emph{Leave a trail.} Let $u$ be the element that step~1 placed with
the algorithm of \Cref{sec:stable} (namely $x$ if $s\leq2$, and $y$
otherwise), let $T_i$ be the tree in which it was placed, and let $j$ be
the smallest level such that $u$ lies in its level-$j$ home subtree in
$T_i$. Unless $u$ lies in its own level-$j$ ancestor and $j\leq J_i(u)$
(i.e., $g_i(u) < 1/j^4$),
set the bits of $u$'s ancestors at levels $1$ through $j-1$ to $1$.
\item\label{step:neg-record} \emph{Record the eviction.}
If $s\leq100$, do nothing. Otherwise, let
$\ell=\ceil{s/100}\geq2$, let $V$ be
the level-$\ell$ ancestor of $L_1(x)$ in $T_1$, and let $w$ be the element
of $L_1(x)$ that is not $x$. If $V$ holds an element $v$ with
$g_1(v)<1/\ell^4$, that is, with $J_1(v)\geq\ell$, exchange $v$ and $w$,
and, if $V$ is an ancestor of $L_1(v)$, set to $1$ the bit of the child
of $V$ that is an ancestor of $L_1(v)$. Otherwise, set
the bits of $L_1(x)$ and of all of its ancestors below $V$ to
$1$.
\end{enumerate}

It is worth briefly justifying that this algorithm always succeeds. Setting
the bit of a node to $1$ requires that node to be full, and the
following lemma establishes that this is always the case.

\begin{lemma}\label{lem:trail}
The insertion algorithm above never tries to set the bit of a nonfull
node to $1$.
\end{lemma}
\begin{proof}
For step~2: if $j=1$, step~2 sets no bits, so assume $j\geq2$. $\Fill$ places $u$ in its level-$j$ ancestor, or
leaves $u$'s path at that ancestor, only if the child on $u$'s path, at
level $j-1$, is full (this is where we use the preference in
\Cref{alg:fill} for the child above $L_i(u)$). A node holds an
element only if both of its child subtrees are full, so the whole
level-$(j-1)$ home subtree is full, and it contains every node whose bit
step~2 sets. For step~3, which acts only on $T_1$: $T_1$ is full by the end of
epoch two (\Cref{lem:completion}) and stays full, since step~1 only swaps
one element of $T_1$ for another and step~3 only exchanges elements.
\end{proof}

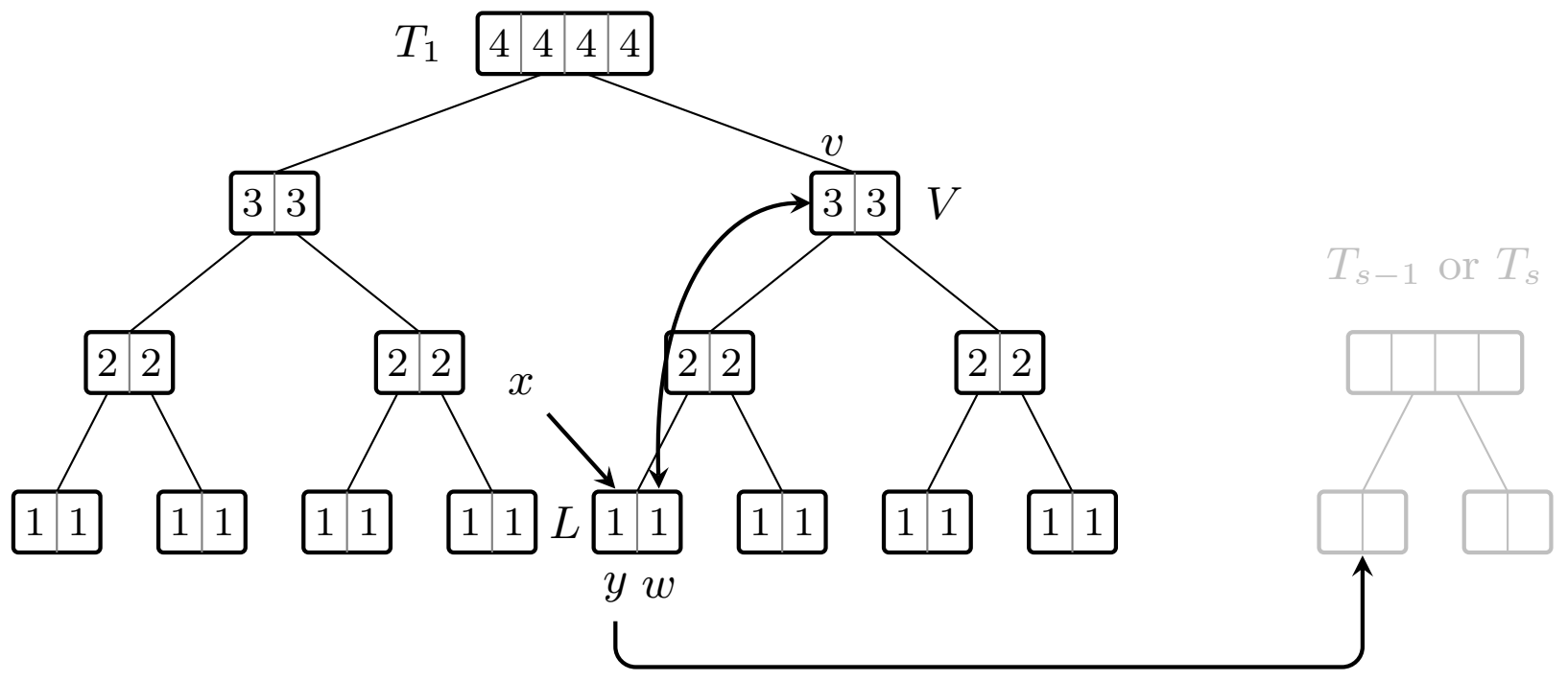
\begin{figure}[t]
\centering
\begin{tikzpicture}[>=stealth]
  \def\nodes{0/1/2, 1/1/2, 2/1/2, 3/1/2, 4/1/2, 5/1/2, 6/1/2, 7/1/2,
             0.5/2/2, 2.5/2/2, 4.5/2/2, 6.5/2/2,
             1.5/3/2, 5.5/3/2, 3.5/4/4}
  \foreach \px/\pl/\cx in {0.5/2/0, 0.5/2/1, 2.5/2/2, 2.5/2/3,
      4.5/2/4, 4.5/2/5, 6.5/2/6, 6.5/2/7,
      1.5/3/0.5, 1.5/3/2.5, 5.5/3/4.5, 5.5/3/6.5,
      3.5/4/1.5, 3.5/4/5.5}
    {\pgfmathsetmacro\o{\cx<\px ? -0.15 : 0.15}
     \draw (\px+\o,1.1*\pl-1.31) -- (\cx,1.1*\pl-1.99);}
  \foreach \x/\l/\k in \nodes
    {\pgfmathsetmacro\y{1.1*\l-1.1}
     \draw[thick,rounded corners=1pt,fill=white]
       (\x-0.15*\k,\y-0.21) rectangle (\x+0.15*\k,\y+0.21);
     \foreach \i in {1,...,\k}
       {\node[font=\scriptsize,inner sep=0pt]
          at (\x-0.15*\k+0.3*\i-0.15,\y) {$\l$};
        \ifnum\i<\k
          \draw[gray] (\x-0.15*\k+0.3*\i,\y-0.21) -- +(0,0.42);
        \fi}}
  \node[font=\footnotesize,anchor=east] at (2.8,3.3) {$T_1$};
  \begin{scope}[gray!50]
    \draw (9.35,0.89) -- (9.0,0.21);
    \draw (9.65,0.89) -- (10.0,0.21);
    \foreach \x/\y/\k in {9.0/0/2, 10.0/0/2, 9.5/1.1/4}
      {\draw[thick,rounded corners=1pt]
         (\x-0.15*\k,\y-0.21) rectangle (\x+0.15*\k,\y+0.21);
       \foreach \i in {1,...,\k}
         {\ifnum\i<\k
            \draw (\x-0.15*\k+0.3*\i,\y-0.21) -- +(0,0.42);
          \fi}}
    \node[font=\footnotesize] at (9.5,1.75) {$T_{s-1}$ or $T_s$};
  \end{scope}
  \node[font=\footnotesize,anchor=east] at (3.75,0) {$L$};
  \node[font=\footnotesize,anchor=west] at (5.85,2.2) {$V$};
  \node[font=\footnotesize] (x) at (3.2,0.95) {$x$};
  \draw[->,thick] (x) -- (3.85,0.23);
  \node[font=\footnotesize] (y) at (3.85,-0.45) {$y$};
  \node[font=\footnotesize] at (4.15,-0.45) {$w$};
  \node[font=\footnotesize] at (5.35,2.6) {$v$};
  \draw[->,thick,rounded corners=4pt] (y) |- (9.0,-1.0) -- (9.0,-0.23);
  \draw[<->,thick]
    (4.15,0.23) .. controls (4.1,1.6) and (4.6,2.2) .. (5.2,2.2);
\end{tikzpicture}%
\caption{Step~3 of a late insertion of $x$ during an epoch $s$ with
$\ceil{s/100}=3$. Each element $u$ of $T_1$ is labeled with $J_1(u)$;
for simplicity, every element at level $j$ has $J_1(u)=j$. The element
$x$ takes the first slot of the leaf $L=L_1(x)$, and the element $y$ that
it displaces is reinserted by the algorithm of \Cref{sec:stable} into a
later tree (grey). The level-$3$ ancestor $V$ of $L$ holds an element
$v$ with $J_1(v)\geq3$, which is exchanged with the other element $w$ of
$L$; if $V$ is an ancestor of $L_1(v)$, the bit of the child of
$V$ above $L_1(v)$ is set to $1$. Afterwards $L$ holds $v$, so
$C(L)\geq3$.}
\label{fig:late-insertion}
\end{figure}

\begin{figure}[t]
\centering
\begin{tikzpicture}[>=stealth]
  \def\sp{0.85}
  \draw[red!70!black,thick,densely dashed,rounded corners=6pt]
    (-0.45,-0.33) -- (6.4,-0.33) -- (6.4,2.05) -- (3.65,3.05)
    -- (2.3,3.05) -- (-0.45,2.05) -- cycle;
  \foreach \l in {2,...,5}
    {\pgfmathtruncatemacro\n{2^(5-\l)-1}
     \foreach \k in {0,...,\n}
       {\pgfmathsetmacro\px{(\k*2^(\l-1)+(2^(\l-1)-1)/2)*\sp}
        \pgfmathsetmacro\lx{((2*\k)*2^(\l-2)+(2^(\l-2)-1)/2)*\sp}
        \pgfmathsetmacro\rx{((2*\k+1)*2^(\l-2)+(2^(\l-2)-1)/2)*\sp}
        \draw (\px-0.15,0.9*\l-1.11) -- (\lx,0.9*\l-1.59);
        \draw (\px+0.15,0.9*\l-1.11) -- (\rx,0.9*\l-1.59);}}
  \foreach \px/\py/\cx/\cy in {3.975/0.69/4.25/0.21,
                                4.525/1.59/3.825/1.11,
                                3.125/2.49/4.675/2.01}
    {\draw[blue!60!black,very thick] (\px,\py) -- (\cx,\cy);}
  \foreach \l in {1,...,5}
    {\pgfmathtruncatemacro\n{2^(5-\l)-1}
     \pgfmathtruncatemacro\anc{floor(5/2^(\l-1))}
     \ifnum\l=5 \def\k{2}\else\def\k{1}\fi
     \pgfmathtruncatemacro\ns{2*\k}
     \pgfmathtruncatemacro\nsm{2*\k-1}
     \foreach \kk in {0,...,\n}
       {\pgfmathsetmacro\x{(\kk*2^(\l-1)+(2^(\l-1)-1)/2)*\sp}
        \pgfmathsetmacro\y{0.9*\l-0.9}
        \ifnum\kk=\anc \ifnum\l<5 \def\onpath{1}\else\def\onpath{0}\fi
        \else\def\onpath{0}\fi
        \def\fillc{white}
        \ifnum\l<5
          \pgfmathtruncatemacro\subtreenodes{2^(4-\l)}
          \ifnum\kk<\subtreenodes \def\fillc{blue!15}\fi
        \fi
        \ifnum\onpath=1 \def\drawc{blue!60!black}\else\def\drawc{black}\fi
        \draw[thick,draw=\drawc,rounded corners=1pt,fill=\fillc]
          (\x-0.3*\k,\y-0.21) rectangle (\x+0.3*\k,\y+0.21);
        \foreach \i in {1,...,\nsm}
          {\draw[gray] (\x-0.3*\k+0.3*\i,\y-0.21) -- +(0,0.42);}
        \ifnum\onpath=0
          \foreach \i in {1,...,\ns}
            {\fill (\x-0.3*\k+0.3*\i-0.15,\y) circle (0.05);}
        \fi}}
  \foreach \x/\l/\a/\b/\bit in {4.25/1/7/2/1, 3.825/2/3/6/0,
                                4.675/3/9/4/1, 2.975/4/1/8/0}
    {\pgfmathsetmacro\y{0.9*\l-0.9}
     \node[font=\scriptsize,inner sep=0pt] at (\x-0.15,\y) {$\a$};
     \node[font=\scriptsize,inner sep=0pt] at (\x+0.15,\y) {$\b$};
     \ifnum\bit=1
       \node[font=\scriptsize\bfseries,anchor=west,red!70!black,inner sep=1pt]
         at (\x+0.32,\y) {$1$};
     \else
       \node[font=\scriptsize,anchor=west,gray,inner sep=1pt]
         at (\x+0.32,\y) {$0$};
     \fi}
  \node[font=\footnotesize,anchor=north] at (4.25,-0.36) {$L_i(x)$};
  \node[font=\footnotesize,anchor=south east,red!70!black]
    at (2.3,3.0) {$K_i(x)=4$};
  \node[font=\footnotesize,anchor=south] at (6.375,4.05) {$T_i$};
  \draw[decorate,decoration={brace,amplitude=4pt}]
    (-0.65,-0.21) -- (-0.65,2.01)
    node[midway,anchor=east,xshift=-4pt,font=\footnotesize] {$J_i(x)=3$};
\end{tikzpicture}%
\caption{Computing $K_i(x)$ in a tree $T_i$ with $i\geq2$ and
$J_i(x)=3$. The blue ancestor path has bits $1,0,1,0$ from bottom to
top; the displayed elements illustrate their encoding. The query must wait
for levels $1$ through $3$ despite the $0$ at level $2$. The $1$ at
level $3$ makes it wait for level $4$, whose bit is $0$, and thus $K_i(x)=4$.
The query then waits for the entire level-$4$ home subtree.}
\label{fig:query}
\end{figure}
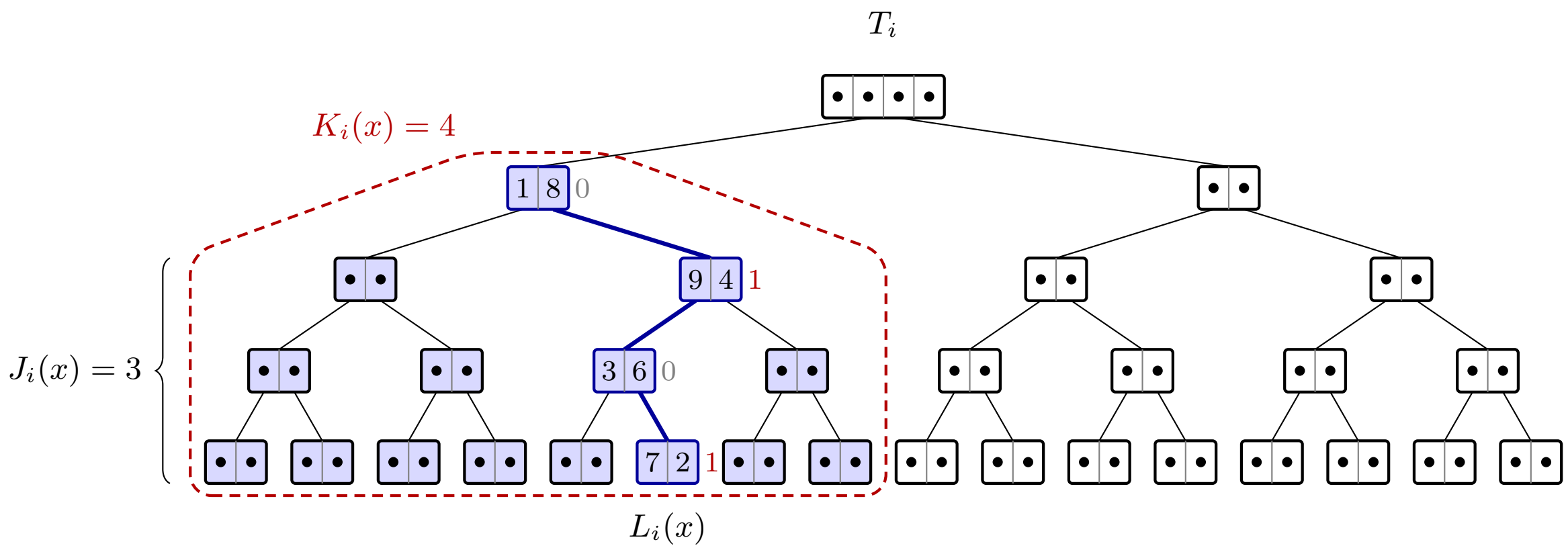

\paragraph{Queries.}
A query for $x$ probes its query probe sequence, as always, and reports
$x$ as soon as it sees it. We say that the query \defn{waits for} a
node if it does not report $x$ as absent until that node has been read.
To decide which nodes to wait for, it computes the bounds of
\Cref{tab:bounds} from the slots it has read so far. Fix a tree $T_i$,
and let $a_1,a_2,a_3,\ldots$ be the
ancestors of $L_i(x) = a_1$ in $T_i$, with $a_h$ at level $h$. To compute
$K_i(x)$, the query waits for $a_1,\ldots,a_{J_i(x)}$, and
then, as long as the last of these ancestors has bit $1$, it also waits
for the next one up; that is, it waits up to the first $a_h$ with
$h\geq J_i(x)$ whose bit is $0$, or up to the root. The level $K$ in the
definition of $K_i(x)$ is the highest level among these ancestors whose
bit is $1$, so once they have been read, the query knows $K_i(x)$, and
it waits for all of $x$'s level-$K_i(x)$
home subtree (the whole tree, if $K$ is the level of the root). In
$T_1$, once that subtree has been read, the query knows $I(x)$, since the
leaves in its definition are exactly those that it waits for in $T_1$,
and it waits for the ancestors of
$L_1(x)$ up to level $I(x)$. It reports that $x$ is absent once it has
read every node that it waits for in $T_i$, for every
$i\leq\min\{I(x),\alpha\}$ (recall $\alpha$ is the number of trees).
See \Cref{fig:query}.

So the query waits for exactly what the stopping condition at the start of
this section prescribes. Every quantity here depends only on
slots that the query has already read, and these quantities only decide
when the query stops, never where it probes next.

\subsection{Correctness}
\label{subsec:final-query}
We now show that queries are correct. That is, a query for a previously
inserted element $x$ does not stop prematurely and conclude $x$ is absent
from the table. Let $S_i(x)$ be $x$'s level-$K_i(x)$ home subtree in
$T_i$. Restating the stopping condition, a query for $x$ waits, in each
tree $T_i$ with $i\leq I(x)$, for all the nodes in
$S_i(x)$ and for the path from $L_i(x)$
to its level-$J_i(x)$ ancestor, or to its level-$\max\{J_1(x),I(x)\}$
ancestor if $i=1$. We say that $x$ is \defn{safe} if it lies in one of
these nodes.

Recall that the insertion algorithm never computes the search bounds
explicitly; it updates them implicitly by moving elements and setting
bits. The bounds $J_i(x)$ and $K_i(x)$ behave simply: $J_i(x)$ depends
only on $x$'s priority and is fixed, while $K_i(x)$ can only increase
as bits are set to $1$. The bound $I(x)$ requires more care: changing
the contents of a leaf $L$ can decrease $C(L)$ (the largest $J_1$-value
stored there) and thereby decrease $I(x)$ for other elements $x$ whose
queries include $L$, even if those elements do not move. Thus we cannot
rely on $I(x)$ being nondecreasing. However, during epoch $s$, all
elements reside in $T_1,\ldots,T_s$, so it suffices to track
$\min\{I(x),s\}$.
The next lemma shows that this quantity never decreases during an
insertion and provides the lower bounds on $I$ we need to prove correctness.

\begin{lemma}\label{lem:certificate-stability}
Consider an insertion of $x$ during epoch $s$, and let $L=L_1(x)$.
After the insertion,
\begin{equation}\label{eq:certify}
I(z)\geq s\quad\text{for every element $z$ with $L$ a leaf of $S_1(z)$.}
\end{equation}
Moreover, the insertion does not decrease $\min\{I(z),s\}$ for any
element $z$.
\end{lemma}
\begin{proof}
Once a node’s bit is set to \(1\), it remains \(1\), so each $K_i(z)$ never decreases, and each
$S_i(z)$ can only grow. If $s\leq100$, both claims follow from
$I(z)\geq100$. Otherwise, let $\ell=\ceil{s/100}$, and consider step~3.
If the exchange succeeds, $L$ ends up with an element $v$ with $J_1(v)\geq\ell$, so
$C(L)\geq\ell$. Thus every $z$ with $L$ a leaf of $S_1(z)$ has
$I(z)\geq100\ell\geq s$.

If the exchange fails, step~3 sets the bits of $L$ and its ancestors
through level $\ell-1$ to $1$. We claim that, for every $z$ with $L$ a leaf of $S_1(z)$,
$K_1(z)\geq\ell$. Otherwise, the root of
$S_1(z)$ is the level-$K_1(z)$ ancestor of $L$, whose bit is now $1$.
But the definition of $K_1(z)$ requires this root to have its bit set to $0$, a contradiction.
Hence $K_1(z)\geq\ell$, and
$I(z)\geq100\ell\geq s$. This proves \eqref{eq:certify}.

To see that $\min\{I(z),s\}$ does not decrease for any element $z$,
observe that the only leaf of $T_1$ whose contents change is $L=L_1(x)$.
If $S_1(z)$ does not contain $L$ after the insertion, none of its leaf
values $C$ has changed, and $K_1(z)$ can only increase, so $I(z)$ cannot
decrease. Otherwise, \eqref{eq:certify} gives $I(z)\geq s$ afterward,
so $\min\{I(z),s\}=s$ cannot be a decrease either.
\end{proof}

We now prove, by induction on the insertions,
that every element in the table is always safe.

\begin{lemma}\label{lem:invariant}
After every insertion during an epoch $s$, every element $x$ in the table
is safe, even with $I(x)$ replaced by $\min\{I(x),s\}$.
\end{lemma}
\begin{proof}
We induct over the insertions; initially, the table is empty. Consider
an insertion of $x$ during epoch $s$, and say that an element $z$ is
\defn{$s$-safe} if it is safe with $I(z)$ replaced by
$\min\{I(z),s\}$. Increasing $s$ can only increase this quantity, so every
element is $s$-safe before the insertion. Let $L=L_1(x)$, and, if
$s>100$, let $\ell=\ceil{s/100}$, $V$, $v$, and $w$ be as in step~3.
If an $s$-safe element $z$ lies in a leaf of $T_1$, that leaf belongs
to $S_1(z)$.

\emph{Elements that the insertion does not move.} Their $J_i$ values
are fixed, their $K_i$ values never decrease, and
\Cref{lem:certificate-stability} shows that their bounds
$\min\{I(z),s\}$ never decrease. Thus they remain $s$-safe.

\emph{Elements that the insertion places or moves.} Since the algorithm
of \Cref{sec:stable} never moves an element that is already placed,
it suffices to check the following cases for an element placed or
moved by the insertion:
\begin{itemize}
\item \emph{The element $x$, when $s\geq3$.} It lands in $L$, which lies
in $S_1(x)$.
\item \emph{The element $u$ placed by \Cref{sec:stable}.}
Let $T_i$ be its destination tree, where $i\in\{s-1,s\}$. Either $u$
lies in an ancestor of $L_i(u)$ at a level at most $J_i(u)$, or step~2
sets the bits of its ancestors at levels $1$ through $j-1$ to $1$,
where $j$ is the smallest level of a home subtree containing its new
location, as in step~2. In the
latter case, the definition of $K_i(u)$ gives $K_i(u)\geq j$, so its
new location lies in $S_i(u)$.
It remains to check that $i\leq\min\{I(u),s\}$. If $s\leq2$, this
follows from $i\leq s$ and $I(u)\geq100$. Otherwise, $u=y$, which was
safe in $L$ before the insertion. Hence $L$ belonged to $S_1(y)$ and
still does, since $K_1(y)$ never decreases. By \eqref{eq:certify},
$I(y)\geq s\geq i$.
\item \emph{The element $v$ moved down from $V$ to $L$ in step~3.} This
swap requires $J_1(v)\geq\ell$. If $V$ is not an ancestor of $L_1(v)$,
$v$'s prior safety implies that $S_1(v)$ already contains $V$ (and thus
$L$). Otherwise, step~3 sets the bit of $v$'s level-$(\ell-1)$ ancestor
to $1$. Since $\ell-1<J_1(v)$, the definition of $K_1(v)$ gives
$K_1(v)\geq\ell$, so $S_1(v)$ contains the subtree of $V$, including $L$.
\item \emph{The element $w$ moved up from $L$ to $V$ in step~3.} Since $w$
was safe in $L$ before the insertion, $L$ is a leaf of $S_1(w)$ and \eqref{eq:certify} gives $I(w)\geq s$.
If $V$ lies outside $S_1(w)$, it is above that subtree's root and hence
is also the level-$\ell$ ancestor of $L_1(w)$, where
$\ell\leq s=\min\{I(w),s\}$. Thus $w$ is $s$-safe in either case.
\end{itemize}
\end{proof}

\begin{lemma}[Correctness]\label{lem:final-correct}
At every time, a query never reports a present element as absent.
\end{lemma}
\begin{proof}
The set of nodes that a query for $x$ waits for only grows with $I(x)$,
so by \Cref{lem:invariant}, every element in the table is always safe.
\end{proof}

\subsection{\texorpdfstring{Proving~\Cref{thm:intro-full}}{Proving Theorem 1.1}}
\label{subsec:expensive-node-analysis}

We now show that all operations (insertions, positive queries, and
negative queries) take $O(1)$ expected time. For our analysis, we
lump together many of the events that can make insertions or queries
expensive, and bound their cost very loosely: whenever one occurs at a
node, we assume that a query for any element residing in that node's
subtree may have to search the entire subtree. In some cases this bound is
exponentially loose, but these events are also exponentially unlikely, so
the loose bound still suffices.

\paragraph{Expensive nodes.}
A node $V$ is \defn{expensive} at a given time if, before then, any of
the following has occurred:
\begin{enumerate}
\item a current-array insertion runs $\Fill$ with target node $V$;
\item a cleanup attempt in a nonfinal phase has target node $V$ and
inserts its element below $V$;
\item $V$ is the root of its tree and a cleanup attempt in the final
phase succeeds in that tree;
\item $V$ is in $T_1$ and step~\ref{step:neg-record} of an insertion in
some epoch $s>100$ selects $V$.
\end{enumerate}
For our analysis, we assume that, if $V$ is expensive, then an insertion or query for an element in the subtree rooted at $V$ must search through the entire subtree.
Accordingly, for a leaf $L$, let $R(L)$ be the level of the highest
expensive node on the path from $L$ to the root, or $1$ if there is none. The following lemma shows that it is
unlikely for nodes to be expensive.

\begin{lemma}\label{lem:expensive-nodes}
A fixed node $V$ at level $\ell$ of any tree is expensive at any fixed
time with probability $O(2^{-20\ell})$.
\end{lemma}
\begin{proof}
We bound the probability of each of the four cases in the definition in turn.
The first case has probability $e^{-\Omega(2^\ell)}$ by \Cref{lem:early}.
For the second case, a nonfinal cleanup attempt with target node $V$ is made in
phase $\ell$, and it inserts its element below $V$ only if a child of $V$
is still nonfull after phase $\ell-1$. For the third case, a final-phase
cleanup attempt succeeds only if the tree is still nonfull after its
last nonfinal phase. By \Cref{lem:full,lem:full-tail}, each of these
has probability $e^{-2^{\Omega(\ell)}}$.

For the fourth case, step~\ref{step:neg-record} of an insertion of $x$
in epoch $s$ selects $V$ only if $\ceil{s/100}=\ell$ and
$L_1(x)$ lies below $V$. Since $T_1$ has $\Omega(n)$ leaves and the size of the subtree rooted at $V$ is $2^\ell$, $L_1(x)$ lies below $V$ with probability $O(2^\ell/n)$. By
\eqref{eq:epoch-length}, there are
$O(n(3/4)^{100(\ell-1)})$ insertions in total in epochs $s$ with $\ell = \lceil s/ 100\rceil$. Each of them selects
$V$ with probability $O(2^\ell/n)$, so by a union bound, the probability $V$ is ever
selected is at most
\[
 O\bigl(n(3/4)^{100(\ell-1)}\bigr)\cdot O(2^\ell/n)=O(2^{-20\ell}).
\]
\end{proof}

Note that, for any fixed leaf $L$, a union bound over the levels $r\geq\ell$ gives
\begin{equation}\label{eq:R-tail}
 \Prb(R(L)\geq\ell)\leq\sum_{r\geq\ell}O(2^{-20r})=O(2^{-20\ell}).
\end{equation}

\paragraph{Insertion time.}
An insertion takes $O(1)$ time unless it triggers one of the first three
cases at a level-$\ell$ node, or
step~\ref{step:neg-record} fails at a level-$\ell$ node; the latter
requires $R(L_1(x))\geq\ell$, since otherwise every element in that node
was placed there by a phase-$\ell$ cleanup attempt, and so has
$J_1\geq\ell$. Either way it
costs $O(\ell)$, including the bits written by
step~\ref{step:neg-trail}. The node involved is determined by the fresh
hashes of the element being placed, so by \Cref{lem:early} and
\eqref{eq:R-tail} this happens with probability $O(2^{-20\ell})$, and
the expected insertion time is $O(1)+\sum_{\ell\geq1}O(\ell)2^{-20\ell}=O(1)$.

\paragraph{Positive queries.}
Fix an element $x$. By the same
reasoning as in the proof of \Cref{thm:intro-reorder}, $x$ has $O(1)$
expected search bound immediately after its insertion (the recording
step of its own insertion never moves $x$), and its possible later eviction
from $T_1$ adds $O(1)$ in expectation. The only other way $x$'s search bound can increase
is through a later exchange in step~\ref{step:neg-record} of the insertion algorithm, so it remains
to bound the probability and cost of such exchanges.

An exchange at a node $V$ may only move $x$ between $V$ and a leaf below
$V$. If $V$ is not above $L_1(x)$, this keeps $x$ in the same home
subtree, so $q(x)$ does not increase. If $V$ is above $L_1(x)$ at level $\ell$, then $x$
ends up in its level-$\ell$ home subtree, so $q(x)=O(4^\ell)$. So
exchanges add at most $O(4^\ell)$ to $q(x)$ for each level-$\ell$ node
above $L_1(x)$ that step~\ref{step:neg-record} of a later insertion
selects. For each later insertion of an element $y$, $L_1(y)$ is
independent of $L_1(x)$, so the analysis of the fourth case in
\Cref{lem:expensive-nodes} still applies, and this has probability
$O(2^{-20\ell})$. Hence
\[
 \E[q(x)]\leq O(1)+\sum_{\ell\geq1}O(4^\ell)\,O(2^{-20\ell})=O(1).
\]

\paragraph{Negative queries.}
To analyze negative queries, we fix some element $x$ and a time
when $x$ is not present. The table at this time does not depend on later
insertions, so we may assume that $x$ is never inserted; then $x$'s
hashes are independent of the table at all times.
A node's bit is $1$ only if that node or one of its ancestors is
expensive, since every step that sets a bit does so below a node that
it makes expensive. So every bit above level $R(L_i(x))$ is $0$, and
$K_i(x)\leq R(L_i(x))+1$. Since $L_i(x)$ is independent of the table,
\eqref{eq:R-tail} gives $\Prb(K_i(x)>\ell)=O(2^{-20\ell})$ at any fixed
time, and hence
\begin{equation}\label{eq:K-moment}
                         \E[4^{4K_i(x)}]=O(1).
\end{equation}

We now bound $I(x)$.
By definition, $I(x)/100$ is the maximum of $K_1(x)$ and the values
$J_1(y)$ of the elements $y$ in the leaves of $S_1(x)$.
$S_1(x)$ depends on the
table, so, to avoid issues arising from this dependence, we directly
bound the probability that $I(x) \geq 100b$ for any value
$b \geq 2$. Let $\ell=\ceil{\log_2b}<b$, and let $P$
be $x$'s level-$\ell$ home subtree in $T_1$
(all of $T_1$ if $\ell$ exceeds its height). Then $I(x)\geq100b$ only if
one of the following three events occurs:
\begin{enumerate}
\item $K_1(x) > \ell$;
\item some element $y$ with $L_1(y)\notin P$ has been moved into $P$;
\item some inserted element $y$ has $L_1(y)\in P$ and $J_1(y)\geq b$.
\end{enumerate}
Each of the first two events implies $R(L_1(x))\geq\ell$, which has
probability $O(2^{-20\ell})$ by \eqref{eq:R-tail}. For the first event, observe that
$K_1(x)\leq R(L_1(x))+1$, because a node's bit is $1$ only if that node or one of its ancestors is
expensive, and thus every bit on $x$'s path above level $R(L_1(x))$ is $0$.
For the second, such an element can only have entered $P$ by first
going through a node above $P$: either the target of a call of $\Fill$, or
a node at which step~\ref{step:neg-record} performs an exchange. Either
way, that node is expensive and lies on $x$'s path at a level above
$\ell$.

We now bound the third event. Each of the at most $n$ inserted elements has
$L_1(y)\in P$ with probability $O(2^\ell/n)$ and, independently,
$J_1(y)\geq b$ with probability $b^{-4}$, so a union bound gives
$O(2^\ell b^{-4})$. Together,
\begin{equation}\label{eq:I-tail}
 \Prb(I(x)\geq100b)\leq O(2^{-20\ell})+O(2^\ell b^{-4})=O(b^{-3}).
\end{equation}
In particular, $\E[I(x)^2]=O(1)$.

The nodes waited for in $T_i$ occur within the first
$O(J_i(x)+2^{K_i(x)})$ probes of its local list, with an additional
$O(I(x))$ probes in $T_1$. By \eqref{eq:weights}, the total query cost
is at most
\[
 O\!\left(I(x)^2+
       \sum_{i\leq\min\{I(x),\alpha\}}
                    i\bigl(J_i(x)^2+4^{K_i(x)}\bigr)\right).
\]
The $I(x)^2$ and $i=1$ terms have constant expectation,
by
\eqref{eq:K-moment} and the fact that
$\E[J_i(x)^2]=\sum_{k\geq1}(2k-1)k^{-4}<\infty$.
For $i\geq2$, $J_i(x)$ depends only on $g_i(x)$, so it is independent
of $I(x)$, and hence
\[
 \E[\1\{I(x)\geq i\}J_i(x)^2]=O(i^{-3}).
\]
By H\"older's inequality,
\eqref{eq:K-moment}, and \eqref{eq:I-tail} we have that
\[
 \E[\1\{I(x)\geq i\}4^{K_i(x)}]
 \leq\Prb(I(x)\geq i)^{3/4}
           \bigl(\E[4^{4K_i(x)}]\bigr)^{1/4}=O(i^{-9/4}).
\]
The expected negative-query cost is therefore
$O(1)+\sum_{i\geq2}O(i^{-2}+i^{-5/4})=O(1)$.

Putting the pieces together, we obtain our final theorem.

\thmintrofull*
\begin{proof}
Insertions succeed by \Cref{lem:trail}, and queries are correct by
\Cref{lem:final-correct}. Insertions, positive queries, and negative
queries take $O(1)$ expected time by the bounds above.
\end{proof}

\section*{Acknowledgments}

Mart\'{\i}n Farach-Colton was supported in part by the Leonard J. Shustek
Professorship and by NSF grants  2551614, 2420942, 2247576, and 2230742. Andrew Krapivin and William
Kuszmaul were supported in part by NSF grant 2504471, by NSF CAREER award
2542165, and by a Jane Street Research Grant. Andrew Krapivin was also
partially supported by the Jeanne~B.\ and Richard~F.\ Berdik ARCS Pittsburgh Endowed
Scholar Award, the David
and Lucile Packard Foundation (Award \#202071730), and gifts from Nikolai Mushegian and Microsoft.

\section*{AI Methodology}
All of the major technical contributions in this paper were made by the authors. AI tools (Claude Opus 5.5, GPT 6 Astra, Gemini 4 Argon) were used to help structure the paper and produce initial drafts of the main sections of the paper; those sections were then almost entirely rewritten by the authors. The AI tools were also used to help clean-up and tighten the prose in the final editing pass.

The $\Omega(\log \log \epsilon^{-1})$ lower bound in Appendix \ref{app:resizing}, which itself is a natural extension of the existing dynamic-hash-table lower bound in \cite{DBLP:conf/focs/BenderKZ24}, was produced with guidance from the authors by Claude Opus 5.5 and GPT 6 Astra. The authors have independently verified the correctness of the proof.

\bibliographystyle{alpha}
\bibliography{refs}

\newcommand{\etalchar}[1]{$^{#1}$}
\begin{thebibliography}{DadHPP06}

\bibitem[ABH{\etalchar{+}}24]{DBLP:journals/corr/abs-2411-19394}
Anders Aamand, Ioana~O. Bercea, Jakob B{\ae}k~Tejs Houen, Jonas Klausen, and
  Mikkel Thorup.
\newblock Hashing for sampling-based estimation.
\newblock {\em CoRR}, abs/2411.19394, 2024.

\bibitem[AK74]{DBLP:journals/cj/AmbleK74}
Ole Amble and Donald~E. Knuth.
\newblock Ordered hash tables.
\newblock {\em Comput. J.}, 17(2):135--142, 1974.

\bibitem[AKK{\etalchar{+}}20]{DBLP:conf/stoc/AamandKKRT20}
Anders Aamand, Jakob B{\ae}k~Tejs Knudsen, Mathias B{\ae}k~Tejs Knudsen, Peter
  Michael~Reichstein Rasmussen, and Mikkel Thorup.
\newblock Fast hashing with strong concentration bounds.
\newblock In {\em Proceedings of the 52nd Annual {ACM} {SIGACT} Symposium on
  Theory of Computing, {STOC} 2020}, 2020.

\bibitem[AKS78]{DBLP:journals/ipl/AjtaiKS78}
Mikl{\'{o}}s Ajtai, J{\'{a}}nos Koml{\'{o}}s, and Endre Szemer{\'{e}}di.
\newblock There is no fast single hashing algorithm.
\newblock {\em Inf. Process. Lett.}, 7(6):270--273, 1978.

\bibitem[ANS09]{DBLP:conf/icalp/ArbitmanNS09}
Yuriy Arbitman, Moni Naor, and Gil Segev.
\newblock De-amortized cuckoo hashing: Provable worst-case performance and
  experimental results.
\newblock In {\em Automata, Languages and Programming, 36th International
  Colloquium, {ICALP} 2009}, 2009.

\bibitem[ANS10]{DBLP:conf/focs/ArbitmanNS10}
Yuriy Arbitman, Moni Naor, and Gil Segev.
\newblock Backyard cuckoo hashing: Constant worst-case operations with a
  succinct representation.
\newblock In {\em 51st Annual {IEEE} Symposium on Foundations of Computer
  Science, {FOCS} 2010}, 2010.

\bibitem[Bat75]{DBLP:journals/cacm/Batagelj75}
Vladimir Batagelj.
\newblock The quadratic hash method when the table size is not a prime number.
\newblock {\em Commun. {ACM}}, 18(4):216--217, 1975.

\bibitem[BBK{\etalchar{+}}23]{DBLP:conf/focs/Bercea0KHT23}
Ioana~O. Bercea, Lorenzo Beretta, Jonas Klausen, Jakob B{\ae}k~Tejs Houen, and
  Mikkel Thorup.
\newblock Locally uniform hashing.
\newblock In {\em 64th {IEEE} Annual Symposium on Foundations of Computer
  Science, {FOCS} 2023}, 2023.

\bibitem[BCF{\etalchar{+}}23]{DBLP:journals/jacm/BenderCFKT23}
Michael~A. Bender, Alex Conway, Mart{\'{\i}}n Farach{-}Colton, William
  Kuszmaul, and Guido Tagliavini.
\newblock Iceberg hashing: Optimizing many hash-table criteria at once.
\newblock {\em J. {ACM}}, 70(6):40:1--40:51, 2023.

\bibitem[BCF{\etalchar{+}}25]{DBLP:journals/talg/BenderCFKT25}
Michael~A. Bender, Alexander Conway, Mart{\'{\i}}n Farach{-}Colton, William
  Kuszmaul, and Guido Tagliavini.
\newblock Tiny pointers.
\newblock {\em {ACM} Trans. Algorithms}, 21(4):38:1--38:43, 2025.

\bibitem[BE23]{DBLP:journals/algorithmica/BerceaE23}
Ioana~O. Bercea and Guy Even.
\newblock Dynamic dictionaries for multisets and counting filters with constant
  time operations.
\newblock {\em Algorithmica}, 85(6):1786--1804, 2023.

\bibitem[BF24]{DBLP:conf/focs/BellF24}
Tolson Bell and Alan~M. Frieze.
\newblock {$O(1)$} insertion for random walk $d$-ary cuckoo hashing up to the
  load threshold.
\newblock In {\em 65th {IEEE} Annual Symposium on Foundations of Computer
  Science, {FOCS} 2024}, 2024.

\bibitem[BFK{\etalchar{+}}22]{DBLP:conf/stoc/BenderFKKL22}
Michael~A. Bender, Mart{\'{\i}}n Farach{-}Colton, John Kuszmaul, William
  Kuszmaul, and Mingmou Liu.
\newblock On the optimal time/space tradeoff for hash tables.
\newblock In {\em 54th Annual {ACM} {SIGACT} Symposium on Theory of Computing,
  {STOC} 2022}, 2022.

\bibitem[BFKK24]{DBLP:conf/sosa/BenderFKK24}
Michael~A. Bender, Mart{\'{\i}}n Farach{-}Colton, John Kuszmaul, and William
  Kuszmaul.
\newblock Modern hashing made simple.
\newblock In {\em 2024 Symposium on Simplicity in Algorithms, {SOSA} 2024},
  2024.

\bibitem[BG07]{DBLP:conf/focs/BlellochG07}
Guy~E. Blelloch and Daniel Golovin.
\newblock Strongly history-independent hashing with applications.
\newblock In {\em 48th Annual {IEEE} Symposium on Foundations of Computer
  Science, {FOCS} 2007}, 2007.

\bibitem[BK24]{DBLP:conf/focs/BravermanK24}
Mark Braverman and William Kuszmaul.
\newblock Tight analyses of ordered and unordered linear probing.
\newblock In {\em 65th {IEEE} Annual Symposium on Foundations of Computer
  Science, {FOCS} 2024}, 2024.

\bibitem[BK26]{DBLP:journals/corr/abs-2607-24545}
Tolson Bell and William Kuszmaul.
\newblock Fast insertion for bucketized cuckoo hashing.
\newblock {\em CoRR}, abs/2607.24545, 2026.

\bibitem[BKK21]{DBLP:conf/focs/BenderKK21}
Michael~A. Bender, Bradley~C. Kuszmaul, and William Kuszmaul.
\newblock Linear probing revisited: Tombstones mark the demise of primary
  clustering.
\newblock In {\em 62nd {IEEE} Annual Symposium on Foundations of Computer
  Science, {FOCS} 2021}, 2021.

\bibitem[BKZ24]{DBLP:conf/focs/BenderKZ24}
Michael~A. Bender, William Kuszmaul, and Renfei Zhou.
\newblock Tight bounds for classical open addressing.
\newblock In {\em 65th {IEEE} Annual Symposium on Foundations of Computer
  Science, {FOCS} 2024}, 2024.

\bibitem[BKZ25]{DBLP:conf/stoc/BenderKZ25}
Michael~A. Bender, William Kuszmaul, and Renfei Zhou.
\newblock Optimal non-oblivious open addressing.
\newblock In {\em Proceedings of the 57th Annual {ACM} Symposium on Theory of
  Computing, {STOC} 2025}, 2025.

\bibitem[Bre73]{DBLP:journals/cacm/Brent73}
Richard~P. Brent.
\newblock Reducing the retrieval time of scatter storage techniques.
\newblock {\em Commun. {ACM}}, 16(2):105--109, 1973.

\bibitem[CFS18]{DBLP:conf/icalp/ConwayFS18}
Alexander Conway, Mart{\'{\i}}n Farach{-}Colton, and Philip Shilane.
\newblock Optimal hashing in external memory.
\newblock In {\em 45th International Colloquium on Automata, Languages, and
  Programming, {ICALP} 2018}, 2018.

\bibitem[CLM85]{DBLP:conf/focs/CelisLM85}
Pedro Celis, Per{-}{\AA}ke Larson, and J.~Ian Munro.
\newblock {Robin Hood} hashing (preliminary report).
\newblock In {\em 26th Annual Symposium on Foundations of Computer Science,
  {FOCS} 1985}, 1985.

\bibitem[CPT15]{DBLP:conf/stoc/ChristianiPT15}
Tobias Christiani, Rasmus Pagh, and Mikkel Thorup.
\newblock From independence to expansion and back again.
\newblock In {\em Proceedings of the Forty-Seventh Annual {ACM} on Symposium on
  Theory of Computing, {STOC} 2015}, 2015.

\bibitem[CSW07]{DBLP:conf/soda/CainSW07}
Julie~Anne Cain, Peter Sanders, and Nicholas~C. Wormald.
\newblock The random graph threshold for $k$-orientability and a fast algorithm
  for optimal multiple-choice allocation.
\newblock In {\em Proceedings of the Eighteenth Annual {ACM-SIAM} Symposium on
  Discrete Algorithms, {SODA} 2007}, 2007.

\bibitem[CW79]{DBLP:journals/jcss/CarterW79}
Larry Carter and Mark~N. Wegman.
\newblock Universal classes of hash functions.
\newblock {\em J. Comput. Syst. Sci.}, 18(2):143--154, 1979.

\bibitem[DadHPP06]{DBLP:conf/latin/DemaineHPP06}
Erik~D. Demaine, Friedhelm~Meyer auf~der Heide, Rasmus Pagh, and Mihai
  P{\u{a}}tra{\c{s}}cu.
\newblock De dictionariis dynamicis pauco spatio utentibus.
\newblock In {\em {LATIN} 2006}, 2006.

\bibitem[DGM{\etalchar{+}}10]{DBLP:conf/icalp/DietzfelbingerGMMPR10}
Martin Dietzfelbinger, Andreas Goerdt, Michael Mitzenmacher, Andrea Montanari,
  Rasmus Pagh, and Michael Rink.
\newblock Tight thresholds for cuckoo hashing via {XORSAT}.
\newblock In {\em Automata, Languages and Programming, 37th International
  Colloquium, {ICALP} 2010}, 2010.

\bibitem[DKRT15]{DBLP:conf/focs/DahlgaardKRT15}
S{\o}ren Dahlgaard, Mathias B{\ae}k~Tejs Knudsen, Eva Rotenberg, and Mikkel
  Thorup.
\newblock Hashing for statistics over {$k$}-partitions.
\newblock In {\em {IEEE} 56th Annual Symposium on Foundations of Computer
  Science, {FOCS} 2015}, 2015.

\bibitem[DR98]{DBLP:journals/rsa/DubhashiR98}
Devdatt~P. Dubhashi and Desh Ranjan.
\newblock Balls and bins: {A} study in negative dependence.
\newblock {\em Random Struct. Algorithms}, 13(2):99--124, 1998.

\bibitem[DR09]{DBLP:conf/icalp/DietzfelbingerR09}
Martin Dietzfelbinger and Michael Rink.
\newblock Applications of a splitting trick.
\newblock In {\em Automata, Languages and Programming, 36th International
  Colloquium, {ICALP} 2009}, 2009.

\bibitem[DW03]{DBLP:conf/stoc/DietzfelbingerW03}
Martin Dietzfelbinger and Philipp Woelfel.
\newblock Almost random graphs with simple hash functions.
\newblock In {\em Proceedings of the 35th Annual {ACM} Symposium on Theory of
  Computing, {STOC} 2003}, 2003.

\bibitem[DW07]{DBLP:journals/tcs/DietzfelbingerW07}
Martin Dietzfelbinger and Christoph Weidling.
\newblock Balanced allocation and dictionaries with tightly packed constant
  size bins.
\newblock {\em Theor. Comput. Sci.}, 380(1-2):47--68, 2007.

\bibitem[Eck74]{DBLP:journals/cj/Ecker74}
A.~Ecker.
\newblock The period of search for the quadratic and related hash methods.
\newblock {\em Comput. J.}, 17(4):340--343, 1974.

\bibitem[FJ19]{DBLP:journals/rsa/FriezeJ19}
Alan~M. Frieze and Tony Johansson.
\newblock On the insertion time of random walk cuckoo hashing.
\newblock {\em Random Struct. Algorithms}, 54(4):721--729, 2019.

\bibitem[FKK24]{DBLP:conf/focs/Farach-ColtonKK24}
Mart{\'{\i}}n Farach{-}Colton, Andrew Krapivin, and William Kuszmaul.
\newblock Optimal bounds for open addressing without reordering.
\newblock In {\em 65th {IEEE} Annual Symposium on Foundations of Computer
  Science, {FOCS} 2024}, 2024.

\bibitem[FKK26]{DBLP:conf/stoc/Farach-ColtonKK26}
Mart{\'{\i}}n Farach{-}Colton, Andrew Krapivin, and William Kuszmaul.
\newblock Greedy open addressing revisited: Beyond {Yao's} lower bound.
\newblock In {\em Proceedings of the 58th Annual {ACM} Symposium on Theory of
  Computing, {STOC} 2026}, 2026.

\bibitem[FKP16]{DBLP:journals/cpc/FountoulakisKP16}
Nikolaos Fountoulakis, Megha Khosla, and Konstantinos Panagiotou.
\newblock The multiple-orientability thresholds for random hypergraphs.
\newblock {\em Comb. Probab. Comput.}, 25(6):870--908, 2016.

\bibitem[FM12]{DBLP:journals/rsa/FriezeM12}
Alan~M. Frieze and P{\'{a}}ll Melsted.
\newblock Maximum matchings in random bipartite graphs and the space
  utilization of cuckoo hash tables.
\newblock {\em Random Struct. Algorithms}, 41(3):334--364, 2012.

\bibitem[FMM11]{DBLP:journals/siamcomp/FriezeMM11}
Alan~M. Frieze, P{\'{a}}ll Melsted, and Michael Mitzenmacher.
\newblock An analysis of random-walk cuckoo hashing.
\newblock {\em {SIAM} J. Comput.}, 40(2):291--308, 2011.

\bibitem[FN93]{DBLP:journals/siamcomp/FiatN93}
Amos Fiat and Moni Naor.
\newblock Implicit {$O(1)$} probe search.
\newblock {\em {SIAM} J. Comput.}, 22(1):1--10, 1993.

\bibitem[FNSS92]{DBLP:journals/jacm/FiatNSS92}
Amos Fiat, Moni Naor, Jeanette~P. Schmidt, and Alan Siegel.
\newblock Nonoblivious hashing.
\newblock {\em J. {ACM}}, 39(4):764--782, 1992.

\bibitem[FP10]{DBLP:conf/icalp/FountoulakisP10}
Nikolaos Fountoulakis and Konstantinos Panagiotou.
\newblock Orientability of random hypergraphs and the power of multiple
  choices.
\newblock In {\em Automata, Languages and Programming, 37th International
  Colloquium, {ICALP} 2010}, 2010.

\bibitem[FP18]{DBLP:journals/ipl/FriezeP18}
Alan~M. Frieze and Samantha Petti.
\newblock Balanced allocation through random walk.
\newblock {\em Inf. Process. Lett.}, 131:39--43, 2018.

\bibitem[FPS13]{DBLP:journals/siamcomp/FountoulakisPS13}
Nikolaos Fountoulakis, Konstantinos Panagiotou, and Angelika Steger.
\newblock On the insertion time of cuckoo hashing.
\newblock {\em {SIAM} J. Comput.}, 42(6):2156--2181, 2013.

\bibitem[FPSS05]{DBLP:journals/mst/FotakisPSS05}
Dimitris Fotakis, Rasmus Pagh, Peter Sanders, and Paul~G. Spirakis.
\newblock Space efficient hash tables with worst case constant access time.
\newblock {\em Theory Comput. Syst.}, 38(2):229--248, 2005.

\bibitem[FR07]{DBLP:conf/soda/FernholzR07}
Daniel Fernholz and Vijaya Ramachandran.
\newblock The $k$-orientability thresholds for {$G_{n,p}$}.
\newblock In {\em Proceedings of the Eighteenth Annual {ACM-SIAM} Symposium on
  Discrete Algorithms, {SODA} 2007}, 2007.

\bibitem[GHMT12]{DBLP:conf/medalg/GoodrichHMT12}
Michael~T. Goodrich, Daniel~S. Hirschberg, Michael Mitzenmacher, and Justin
  Thaler.
\newblock Cache-oblivious dictionaries and multimaps with negligible failure
  probability.
\newblock In {\em Design and Analysis of Algorithms -- First Mediterranean
  Conference on Algorithms, {MedAlg} 2012}, 2012.

\bibitem[GM79]{DBLP:journals/siamcomp/GonnetM79}
Gaston~H. Gonnet and J.~Ian Munro.
\newblock Efficient ordering of hash tables.
\newblock {\em {SIAM} J. Comput.}, 8(3):463--478, 1979.

\bibitem[GPW26a]{GPW26b}
Yuhao Guo, Seth Pettie, and Chengzhang Wan.
\newblock A sharp analysis of fixed offset hash tables.
\newblock arXiv preprint arXiv:2610.31363, 2026.

\bibitem[GPW26b]{DBLP:journals/corr/abs-2608-08013}
Yuhao Guo, Seth Pettie, and Chengzhang Wan.
\newblock A simple analysis of quadratic probing and other open addressing
  schemes.
\newblock {\em CoRR}, abs/2608.08013, 2026.

\bibitem[GS78]{DBLP:journals/jcss/GuibasS78}
Leonidas~J. Guibas and Endre Szemer{\'{e}}di.
\newblock The analysis of double hashing.
\newblock {\em J. Comput. Syst. Sci.}, 16(2):226--274, 1978.

\bibitem[GW10]{DBLP:conf/stoc/GaoW10}
Pu~Gao and Nicholas~C. Wormald.
\newblock Load balancing and orientability thresholds for random hypergraphs.
\newblock In {\em Proceedings of the 42nd {ACM} Symposium on Theory of
  Computing, {STOC} 2010}, 2010.

\bibitem[HD72]{DBLP:journals/cj/HopgoodD72}
F.~Robert~A. Hopgood and James~H. Davenport.
\newblock The quadratic hash method when the table size is a power of 2.
\newblock {\em Comput. J.}, 15(4):314--315, 1972.

\bibitem[HKL{\etalchar{+}}26a]{DBLP:journals/corr/abs-2608-28512}
Yang Hu, William Kuszmaul, Jingxun Liang, Stefan Walzer, Huacheng Yu, and
  Renfei Zhou.
\newblock Quadratic probing insertions are $\epsilon^{-(1+o(1))}$ time.
\newblock {\em CoRR}, abs/2608.28512, 2026.

\bibitem[HKL{\etalchar{+}}26b]{DBLP:journals/corr/abs-2607-13247}
Yang Hu, William Kuszmaul, Jingxun Liang, Stefan Walzer, Huacheng Yu, and
  Renfei Zhou.
\newblock Quadratic probing revisited: Smoothed analysis and the fall of {Robin
  Hood}.
\newblock {\em CoRR}, abs/2607.13247, 2026.

\bibitem[HLY{\etalchar{+}}25]{DBLP:conf/stoc/HuLYZZ25}
Yang Hu, Jingxun Liang, Huacheng Yu, Junkai Zhang, and Renfei Zhou.
\newblock Optimal static dictionary with worst-case constant query time.
\newblock In {\em Proceedings of the 57th Annual {ACM} Symposium on Theory of
  Computing, {STOC} 2025}, 2025.

\bibitem[HMP01]{DBLP:journals/jal/HagerupMP01}
Torben Hagerup, Peter~Bro Miltersen, and Rasmus Pagh.
\newblock Deterministic dictionaries.
\newblock {\em J. Algorithms}, 41(1):69--85, 2001.

\bibitem[IP12]{DBLP:conf/soda/IaconoP12}
John Iacono and Mihai P{\u{a}}tra{\c{s}}cu.
\newblock Using hashing to solve the dictionary problem.
\newblock In {\em Proceedings of the Twenty-Third Annual {ACM-SIAM} Symposium
  on Discrete Algorithms, {SODA} 2012}, 2012.

\bibitem[JP08]{DBLP:journals/algorithmica/JensenP08}
Morten~Skaarup Jensen and Rasmus Pagh.
\newblock Optimality in external memory hashing.
\newblock {\em Algorithmica}, 52(3):403--411, 2008.

\bibitem[KA19]{DBLP:journals/algorithmica/KhoslaA19}
Megha Khosla and Avishek Anand.
\newblock A faster algorithm for cuckoo insertion and bipartite matching in
  large graphs.
\newblock {\em Algorithmica}, 81(9):3707--3724, 2019.

\bibitem[KKW26]{DBLP:journals/corr/abs-2607-17494}
Andrew Krapivin, William Kuszmaul, and Jolyne Wang.
\newblock Linear probing with non-greedy insertions.
\newblock {\em CoRR}, abs/2607.17494, 2026.

\bibitem[KM25]{DBLP:conf/soda/KuszmaulM25}
William Kuszmaul and Michael Mitzenmacher.
\newblock Efficient $d$-ary cuckoo hashing at high load factors by bubbling up.
\newblock In {\em Proceedings of the 2025 Annual {ACM-SIAM} Symposium on
  Discrete Algorithms, {SODA} 2025}, 2025.

\bibitem[KMW09]{DBLP:journals/siamcomp/KirschMW09}
Adam Kirsch, Michael Mitzenmacher, and Udi Wieder.
\newblock More robust hashing: Cuckoo hashing with a stash.
\newblock {\em {SIAM} J. Comput.}, 39(4):1543--1561, 2009.

\bibitem[Knu63]{Knu63}
Donald~E. Knuth.
\newblock Notes on ``open'' addressing.
\newblock Unpublished memorandum, 1963.

\bibitem[Knu74]{Knu74}
Donald~E. Knuth.
\newblock Computer science and its relation to mathematics.
\newblock {\em The American Mathematical Monthly}, 81(4):323--343, 1974.

\bibitem[Knu98]{Knu98}
Donald~E. Knuth.
\newblock {\em The Art of Computer Programming, Volume 3: Sorting and
  Searching}.
\newblock Addison-Wesley, 2nd edition, 1998.

\bibitem[KPX{\etalchar{+}}25]{DBLP:conf/soda/KuszmaulPXZZ25}
William Kuszmaul, Aaron Putterman, Tingqiang Xu, Hangrui Zhou, and Renfei Zhou.
\newblock Tight bounds and phase transitions for incremental and dynamic
  retrieval.
\newblock In {\em Proceedings of the 2025 Annual {ACM-SIAM} Symposium on
  Discrete Algorithms, {SODA} 2025}, 2025.

\bibitem[KPX{\etalchar{+}}26]{DBLP:journals/corr/abs-2606-15944}
William Kuszmaul, Aaron Putterman, Tingqiang Xu, Hangrui Zhou, and Renfei Zhou.
\newblock Resizable retrieval.
\newblock {\em CoRR}, abs/2606.15944, 2026.

\bibitem[Kus22]{DBLP:conf/focs/Kuszmaul22}
William Kuszmaul.
\newblock A hash table without hash functions, and how to get the most out of
  your random bits.
\newblock In {\em 63rd {IEEE} Annual Symposium on Foundations of Computer
  Science, {FOCS} 2022}, 2022.

\bibitem[Kus23]{DBLP:conf/focs/Kuszmaul23}
William Kuszmaul.
\newblock Strongly history-independent storage allocation: New upper and lower
  bounds.
\newblock In {\em 64th {IEEE} Annual Symposium on Foundations of Computer
  Science, {FOCS} 2023}, 2023.

\bibitem[Kus26]{DBLP:journals/corr/abs-2608-00762}
William Kuszmaul.
\newblock An analysis of {Brent's} insertion method for hash tables.
\newblock {\em CoRR}, abs/2608.00762, 2026.

\bibitem[KW66]{KW66}
Alan~G. Konheim and Benjamin Weiss.
\newblock An occupancy discipline and applications.
\newblock {\em {SIAM} J. Appl. Math.}, 14(6):1266--1274, 1966.

\bibitem[KX24]{DBLP:conf/icalp/KuszmaulX24}
William Kuszmaul and Zoe Xi.
\newblock Towards an analysis of quadratic probing.
\newblock In {\em 51st International Colloquium on Automata, Languages, and
  Programming, {ICALP} 2024}, 2024.

\bibitem[Lel12]{DBLP:conf/soda/Lelarge12}
Marc Lelarge.
\newblock A new approach to the orientation of random hypergraphs.
\newblock In {\em Proceedings of the Twenty-Third Annual {ACM-SIAM} Symposium
  on Discrete Algorithms, {SODA} 2012}, 2012.

\bibitem[LLYZ23a]{DBLP:conf/focs/LiLYZ23}
Tianxiao Li, Jingxun Liang, Huacheng Yu, and Renfei Zhou.
\newblock Dynamic ``succincter''.
\newblock In {\em 64th {IEEE} Annual Symposium on Foundations of Computer
  Science, {FOCS} 2023}, 2023.

\bibitem[LLYZ23b]{DBLP:conf/focs/LiLYZ23a}
Tianxiao Li, Jingxun Liang, Huacheng Yu, and Renfei Zhou.
\newblock Tight cell-probe lower bounds for dynamic succinct dictionaries.
\newblock In {\em 64th {IEEE} Annual Symposium on Foundations of Computer
  Science, {FOCS} 2023}, 2023.

\bibitem[LLYZ24]{DBLP:conf/soda/LiLYZ24}
Tianxiao Li, Jingxun Liang, Huacheng Yu, and Renfei Zhou.
\newblock Dynamic dictionary with subconstant wasted bits per key.
\newblock In {\em Proceedings of the 2024 {ACM-SIAM} Symposium on Discrete
  Algorithms, {SODA} 2024}, 2024.

\bibitem[LM93]{DBLP:journals/combinatorica/LuekerM93}
George~S. Lueker and Mariko Molodowitch.
\newblock More analysis of double hashing.
\newblock {\em Comb.}, 13(1):83--96, 1993.

\bibitem[LP09]{DBLP:conf/esa/LehmanP09}
Eric~P. Lehman and Rina Panigrahy.
\newblock 3.5-way cuckoo hashing for the price of 2-and-a-bit.
\newblock In {\em 17th Annual European Symposium on Algorithms, {ESA} 2009},
  2009.

\bibitem[LPP{\etalchar{+}}24]{DBLP:conf/icalp/LarsenPPPYZ24}
Kasper~Green Larsen, Rasmus Pagh, Giuseppe Persiano, Toniann Pitassi, Kevin
  Yeo, and Or~Zamir.
\newblock Optimal non-adaptive cell probe dictionaries and hashing.
\newblock In {\em 51st International Colloquium on Automata, Languages, and
  Programming, {ICALP} 2024}, 2024.

\bibitem[Mal77]{DBLP:journals/cj/Mallach77}
Efrem~G. Mallach.
\newblock Scatter storage techniques: {A} unifying viewpoint and a method for
  reducing retrieval times.
\newblock {\em Comput. J.}, 20(2):137--140, 1977.

\bibitem[Mau83]{DBLP:journals/cacm/Maurer83}
W.~Douglas Maurer.
\newblock An improved hash code for scatter storage (reprint).
\newblock {\em Commun. {ACM}}, 26(1):36--38, 1983.

\bibitem[Mit14]{DBLP:conf/spaa/Mitzenmacher14}
Michael Mitzenmacher.
\newblock Balanced allocations and double hashing.
\newblock In {\em 26th {ACM} Symposium on Parallelism in Algorithms and
  Architectures, {SPAA} '14}, 2014.

\bibitem[Mit16]{DBLP:conf/analco/Mitzenmacher16}
Michael Mitzenmacher.
\newblock More analysis of double hashing for balanced allocations.
\newblock In {\em Proceedings of the Thirteenth Workshop on Analytic
  Algorithmics and Combinatorics, {ANALCO} 2016}, 2016.

\bibitem[MS80]{DBLP:journals/jcss/MunroS80}
J.~Ian Munro and Hendra Suwanda.
\newblock Implicit data structures for fast search and update.
\newblock {\em J. Comput. Syst. Sci.}, 21(2):236--250, 1980.

\bibitem[MU17]{MU17}
Michael Mitzenmacher and Eli Upfal.
\newblock {\em Probability and Computing: Randomization and Probabilistic
  Techniques in Algorithms and Data Analysis}.
\newblock Cambridge University Press, 2nd edition, 2017.

\bibitem[NSW08]{DBLP:conf/icalp/NaorSW08}
Moni Naor, Gil Segev, and Udi Wieder.
\newblock History-independent cuckoo hashing.
\newblock In {\em Automata, Languages and Programming, 35th International
  Colloquium, {ICALP} 2008}, 2008.

\bibitem[NT01]{DBLP:journals/iacr/NaorT01}
Moni Naor and Vanessa Teague.
\newblock Anti-persistence: History independent data structures.
\newblock {\em {IACR} Cryptol. ePrint Arch.}, 2001:36, 2001.

\bibitem[{\"{O}}P03]{DBLP:conf/stoc/OstlinP03}
Anna {\"{O}}stlin and Rasmus Pagh.
\newblock Uniform hashing in constant time and linear space.
\newblock In {\em Proceedings of the 35th Annual {ACM} Symposium on Theory of
  Computing, {STOC} 2003}, 2003.

\bibitem[Pag01]{DBLP:journals/siamcomp/Pagh01}
Rasmus Pagh.
\newblock Low redundancy in static dictionaries with constant query time.
\newblock {\em {SIAM} J. Comput.}, 31(2):353--363, 2001.

\bibitem[P{\u{a}}t08]{DBLP:conf/focs/Patrascu08}
Mihai P{\u{a}}tra{\c{s}}cu.
\newblock Succincter.
\newblock In {\em 49th Annual {IEEE} Symposium on Foundations of Computer
  Science, {FOCS} 2008}, 2008.

\bibitem[Pet57]{DBLP:journals/ibmrd/Peterson57}
W.~Wesley Peterson.
\newblock Addressing for random-access storage.
\newblock {\em {IBM} J. Res. Dev.}, 1(2):130--146, 1957.

\bibitem[PM89]{DBLP:journals/jal/PobleteM89}
Patricio~V. Poblete and J.~Ian Munro.
\newblock Last-come-first-served hashing.
\newblock {\em J. Algorithms}, 10(2):228--248, 1989.

\bibitem[PP08]{DBLP:journals/siamcomp/PaghP08}
Anna Pagh and Rasmus Pagh.
\newblock Uniform hashing in constant time and optimal space.
\newblock {\em {SIAM} J. Comput.}, 38(1):85--96, 2008.

\bibitem[PPR09]{DBLP:journals/siamcomp/PaghPR09}
Anna Pagh, Rasmus Pagh, and Milan Ru{\v{z}}i{\'{c}}.
\newblock Linear probing with constant independence.
\newblock {\em {SIAM} J. Comput.}, 39(3):1107--1120, 2009.

\bibitem[PR04]{DBLP:journals/jal/PaghR04}
Rasmus Pagh and Flemming~Friche Rodler.
\newblock Cuckoo hashing.
\newblock {\em J. Algorithms}, 51(2):122--144, 2004.

\bibitem[PT12]{DBLP:journals/jacm/PatrascuT12}
Mihai P{\u{a}}tra{\c{s}}cu and Mikkel Thorup.
\newblock The power of simple tabulation hashing.
\newblock {\em J. {ACM}}, 59(3):14:1--14:50, 2012.

\bibitem[PT13]{DBLP:conf/soda/PatrascuT13}
Mihai P{\u{a}}tra{\c{s}}cu and Mikkel Thorup.
\newblock Twisted tabulation hashing.
\newblock In {\em Proceedings of the Twenty-Fourth Annual {ACM-SIAM} Symposium
  on Discrete Algorithms, {SODA} 2013}, 2013.

\bibitem[PT16]{DBLP:journals/talg/PatrascuT16}
Mihai P{\u{a}}tra{\c{s}}cu and Mikkel Thorup.
\newblock On the $k$-independence required by linear probing and minwise
  independence.
\newblock {\em {ACM} Trans. Algorithms}, 12(1):8:1--8:27, 2016.

\bibitem[Rad70]{DBLP:journals/cacm/Radke70}
Charles~E. Radke.
\newblock The use of quadratic residue research.
\newblock {\em Commun. {ACM}}, 13(2):103--105, 1970.

\bibitem[Riv78]{DBLP:journals/jacm/Rivest78}
Ronald~L. Rivest.
\newblock Optimal arrangement of keys in a hash table.
\newblock {\em J. {ACM}}, 25(2):200--209, 1978.

\bibitem[RR03]{DBLP:conf/icalp/RamanR03}
Rajeev Raman and S.~Srinivasa Rao.
\newblock Succinct dynamic dictionaries and trees.
\newblock In {\em Automata, Languages and Programming, 30th International
  Colloquium, {ICALP} 2003}, 2003.

\bibitem[Ru{\v{z}}08]{DBLP:journals/talg/Ruzic08}
Milan Ru{\v{z}}i{\'{c}}.
\newblock Uniform deterministic dictionaries.
\newblock {\em {ACM} Trans. Algorithms}, 4(1):1:1--1:23, 2008.

\bibitem[Sie04]{DBLP:journals/siamcomp/Siegel04}
Alan Siegel.
\newblock On universal classes of extremely random constant-time hash
  functions.
\newblock {\em {SIAM} J. Comput.}, 33(3):505--543, 2004.

\bibitem[SS90]{DBLP:conf/stoc/SchmidtS90}
Jeanette~P. Schmidt and Alan Siegel.
\newblock The analysis of closed hashing under limited randomness.
\newblock In {\em Proceedings of the 22nd Annual {ACM} Symposium on Theory of
  Computing, {STOC} 1990}, 1990.

\bibitem[Sun91]{DBLP:conf/focs/Sundar91}
Rajamani Sundar.
\newblock A lower bound for the dictionary problem under a hashing model.
\newblock In {\em 32nd Annual Symposium on Foundations of Computer Science,
  {FOCS} 1991}, 1991.

\bibitem[Tho13]{DBLP:conf/focs/Thorup13}
Mikkel Thorup.
\newblock Simple tabulation, fast expanders, double tabulation, and high
  independence.
\newblock In {\em 54th Annual {IEEE} Symposium on Foundations of Computer
  Science, {FOCS} 2013}, 2013.

\bibitem[Tho17]{DBLP:journals/cacm/Thorup17}
Mikkel Thorup.
\newblock Fast and powerful hashing using tabulation.
\newblock {\em Commun. {ACM}}, 60(7):94--101, 2017.

\bibitem[Ull72]{DBLP:journals/jacm/Ullman72}
Jeffrey~D. Ullman.
\newblock A note on the efficiency of hashing functions.
\newblock {\em J. {ACM}}, 19(3):569--575, 1972.

\bibitem[VZ13]{DBLP:journals/siamcomp/VerbinZ13}
Elad Verbin and Qin Zhang.
\newblock The limits of buffering: {A} tight lower bound for dynamic membership
  in the external memory model.
\newblock {\em {SIAM} J. Comput.}, 42(1):212--229, 2013.

\bibitem[Waj17]{Waj17}
David Wajc.
\newblock Negative association: Definition, properties, and applications, 2017.
\newblock Manuscript.

\bibitem[Wal22]{DBLP:conf/esa/Walzer22}
Stefan Walzer.
\newblock Insertion time of random walk cuckoo hashing below the peeling
  threshold.
\newblock In {\em 30th Annual European Symposium on Algorithms, {ESA} 2022},
  2022.

\bibitem[Wal23]{DBLP:journals/talg/Walzer23}
Stefan Walzer.
\newblock Load thresholds for cuckoo hashing with overlapping blocks.
\newblock {\em {ACM} Trans. Algorithms}, 19(3):24:1--24:22, 2023.

\bibitem[WC81]{DBLP:journals/jcss/WegmanC81}
Mark~N. Wegman and Larry Carter.
\newblock New hash functions and their use in authentication and set equality.
\newblock {\em J. Comput. Syst. Sci.}, 22(3):265--279, 1981.

\bibitem[Yao81]{DBLP:journals/jacm/Yao81a}
Andrew~Chi{-}Chih Yao.
\newblock Should tables be sorted?
\newblock {\em J. {ACM}}, 28(3):615--628, 1981.

\bibitem[Yao85]{DBLP:journals/jacm/Yao85}
Andrew~Chi{-}Chih Yao.
\newblock Uniform hashing is optimal.
\newblock {\em J. {ACM}}, 32(3):687--693, 1985.

\bibitem[Yu22]{DBLP:journals/siamcomp/Yu22}
Huacheng Yu.
\newblock Nearly optimal static {Las Vegas} succinct dictionary.
\newblock {\em {SIAM} J. Comput.}, 51(3):20--174, 2022.

\bibitem[Zam26]{DBLP:journals/corr/abs-2607-16390}
Or~Zamir.
\newblock Locality in open addressing hash tables.
\newblock {\em CoRR}, abs/2607.16390, 2026.

\bibitem[Zob70]{Zob70}
Albert~L. Zobrist.
\newblock A new hashing method with application for game playing.
\newblock Technical Report~88, Computer Sciences Department, University of
  Wisconsin, 1970.
\newblock Reprinted in \emph{ICGA Journal} 13(2):69--73, 1990.

\end{thebibliography}

\appendix

\section{Explicit Calculations of Addresses and Probe Sequences}
\label{app:explicit}

\Cref{app:addresses} removes the assumption that $n$ is a power of two
and shows how to compute, in constant time from $n$ and the number of
insertions performed so far, the address of every slot and the current
epoch and phase (which is straightforward when $n$ is a power of two).
\Cref{app:probes} gives the fixed probe schedule used by queries,
shows that it satisfies \eqref{eq:weights}, its next address is
also computable in constant time, and the bounds necessary
for negative queries can be tracked with $O(1)$ words of scratch space.

\paragraph{Computational model.}
We use words of $\Theta(\log n)$ bits, with unit-cost integer
arithmetic (including multiplication and division), bitwise operations,
shifts, population count, and the highest-set-bit operation.
The word length is large enough for the polynomially bounded counters
below. Hash evaluations and priority threshold comparisons are assumed to take constant time.

\subsection{Arbitrary Table Sizes, and Exact Addresses}
\label{app:addresses}

Without loss of generality, we assume $n \geq 8$ throughout this subsection.

\paragraph{Assumptions used in the proofs.}
Throughout the paper, the proofs use only the following property of the
sizes of the trees: they shrink geometrically, but not too quickly,
\begin{equation}\label{eq:sizes}
 |T_{i+1}|\geq|T_i|/4,\qquad
 \sum_{r>i}|T_r|\leq3|T_i|.
\end{equation}
For $n$ a power of two, both bounds hold trivially; below, we choose
the sizes for arbitrary $n$ so that they still hold.
Writing $S_i=\sum_{r\geq i}|T_r|$, the second bound gives
$|T_i|\geq S_i/4$, so $S_{i+1}\leq\frac34S_i$ and
$|T_i|\leq S_i\leq n(3/4)^{i-1}$; together with \eqref{eq:stages},
this gives \eqref{eq:epoch-length}. It also gives $|T_1|\geq n/4$.

\paragraph{Layout within a tree.}
Each tree occupies a contiguous part of the array; for
concreteness we use the implicit layout of a binary heap, as in
\Cref{fig:forest}. That is, in a tree of height $h$, node
$[j,k]$ sits at depth $d=h-j$ and has heap index $p=2^d+k$. The root, with
$p=1$, occupies the first four slots of the tree, and node $p\geq2$
occupies the two slots at offsets $2p$ and $2p+1$. Conversely, from a heap
index $p$ we recover $d=\floor{\log_2p}$, $j=h-d$, and $k=p-2^d$. The
children of heap index $p$ are $2p$ and $2p+1$, and the level-$j$ ancestor
of the leaf with heap index $p$ has index $p\gg(j-1)$.

\paragraph{Arbitrary $n$.}
Let $L=\floor{\log_2n}$ and write $n=2^L+m$ with $0\leq m<2^L$. Start
from the forest for $2^L$, which has one tree of each size
$2^{L-1},\ldots,8,4$ and a second tree of size $4$. For each set bit
$b\geq2$ of $m$, add one more tree of size $2^b$. If $n$ is not a
multiple of four, the remaining $n\bmod4$ slots form one final
single-node tree. Trees are laid out in order of non-increasing
size (i.e. we maintain $|T_i| \leq |T_{i-1}|$).

Every size above four occurs once or twice, size four occurs two or three
times, and the final small node, if any, has at least one slot, so
$|T_{i+1}|\geq |T_i|/4$. After a tree of size $2^b$ come at most one more tree
of size $2^b$, at most two of each smaller size, one extra tree of size
four, and at most three further slots. For $b\geq3$ these total at most
$2^b+2(2^b-4)+4+3<3\cdot2^b$, and for $b=2$ at most $4+4+3<12$. So
$\sum_{r>i}|T_r|\leq3|T_i|$, and \eqref{eq:sizes} holds.

\paragraph{Addresses.}
The trees larger than $2^b$ are the trees of sizes $2^{b+1},\ldots,
2^{L-1}$ from the forest for $2^L$, together with one extra tree for each
set bit of $m$ above position $b$. So the first tree of size $2^b$ begins
at offset
\[
            2^L-2^{b+1}+\bigl((m\gg(b+1))\ll(b+1)\bigr),
\]
and each later tree of the same size begins $2^b$ further on. The final
small node, if any, begins at offset $n-(n\bmod4)$. All of this is a
constant number of shifts, masks, and additions. For $n=2^L$ it places
$T_i$ at offset $n-2n/2^i$.

\paragraph{The current epoch.}
Insertions find their epoch from $c$ alone, the number $0\leq c<n$ of
elements present, by first finding the tree $T_k$ that contains array
position $c$, counting from zero. Since $2^L+m=n$, the offset formula
above says that the first tree of size $2^b$ begins at
$s_b=(\floor{n/2^{b+1}}-1)\,2^{b+1}$, so the trees of size at most $2^b$,
together with the final small node, occupy exactly the positions from
$s_b$ on. Hence, if $c\geq n-(n\bmod4)$, then $T_k$ is the final small
node and $k=L+\mathrm{popcount}(m\gg2)$. Otherwise $T_k$ has size $2^b$
for the smallest $b\geq2$ with $c\geq s_b$, and it is copy
$j=\floor{(c-s_b)/2^b}$ of that size, so
\[
 k=L-b+\mathrm{popcount}(m\gg(b+1))+j.
\]
To compute $b$, let $p=\floor{\log_2(n-c)}$, set $b=\max\{2,p-1\}$, and
increase $b$ by one if $c<s_b$. This is correct because
$2^{b+1}\leq n-s_b<2^{b+2}$, so the condition $c\geq s_b$ fails when
$n-c\geq2^{b+2}$ and holds when $n-c<2^{b+1}$; hence the smallest valid
$b$ is $\max\{2,p-1\}$ or $\max\{2,p\}$.

Let $o_k$ be the offset of $T_k$. By \eqref{eq:stages},
$t_{k-1}\leq o_k\leq c$, and $c<t_{k+1}$ if $k<\alpha$, so the next insertion belongs to epoch
$k$ if $c<t_k$ and to epoch $k+1$ otherwise, where
$t_k=o_k+\floor{(1-\eta)|T_k|}$, or $t_k=n$ if $T_k$ is the last tree.
Given the epoch $i$, the phase follows from the formula in
\Cref{sec:stable} with $r=c-t_{i-1}$. If $i=k$, this needs the tree
before $T_k$, which has offset $o_k-|T_{k-1}|$, where $|T_{k-1}|$ is
$2^b$ if $j>0$, $2^{b+1}$ if $j=0$, and $4$ if $T_k$ is the final small
node. So insertions need no stored state besides $c$.

\subsection{An Explicit Fixed Probe Schedule}
\label{app:probes}

We now describe the probe schedule queries follow and
show that the next probe location
can always be computed in constant time.
The schedule is designed so that negative queries, as in
\Cref{sec:full}, need only $O(1)$ words of scratch space (with a naive use of the
injection of~\cite{DBLP:conf/focs/Farach-ColtonKK24}, a query may need to remember its progress toward the stopping rule in
each of the $\Theta(\log n)$ trees, using $\Theta(\log n)$ words).
Our schedule instead returns to each tree repeatedly, reading
progressively longer prefixes of its local list, and rereads each prefix
from scratch, so this progress can be discarded between visits.
For simplicity, the schedule spends exactly four probes on every node it
reads, rereading the first slot of nodes with fewer than four slots.

\paragraph{The local lists.}
The local list of $T_i$ alternates between the ancestor list and the
expanding-subtree list, with the latter reading each home subtree in
breadth-first order. Calculating the node to read at each probe of
the local list then takes constant time: since every visit reads the
list from its start, the query can keep the current level $h$ and
breadth-first index $v$ as counters, and the $v$th node in breadth-first
order of a subtree rooted at $[h,w]$ is $[h-d,\,2^dw+v-2^d]$, where
$d=\floor{\log_2v}$. The
level-$j$ ancestor is read within the first $8j$ probes of the local
list, and the whole level-$h$ home subtree within the first $2^{h+4}$.

\paragraph{Merging across trees.}
In round $r=0,1,\ldots$, for each $a=0,\ldots,r$ in turn, the schedule
\emph{visits} the trees $T_{2^a},\ldots,T_{2^{a+1}-1}$ in order, reading the
first $2^{\floor{(r-a)/2}}$ probes of each one's local list; visits
to trees $T_i$ with $i>\alpha$ are spent on probes of some arbitrary fixed address.
Since each round visits $T_1,T_2,\ldots$ consecutively, the query can
keep the current tree's offset and size as counters and step to the next
tree in constant time using \Cref{app:addresses}, so the schedule needs
only a constant number of counters, each updated in constant time per
probe.
Every local list
has length $O(n)$, so by round $O(\log n)$ the query has read every
address it can ever read, and each counter fits in a word.

Round $r$ takes
$\sum_{a=0}^r2^{a+\floor{(r-a)/2}}<3\cdot2^r$ probes. Now fix a tree
$T_i$ and a position $k$ in its local list, and let $a=\floor{\log_2i}$
and $e=\ceil{\log_2k}$. Round $r=a+2e$ reads the first $k$ probes of
this local list in a single visit. The earlier rounds take fewer than
$3\cdot2^{a+2e}$ probes, and the earlier visits in this round fewer than
$(i+2^a)2^e$, so the $k$th probe is reached within
\[
 3\cdot2^{a+2e}+(i+2^a)2^e+k\leq16ik^2
\]
probes, as required by \eqref{eq:weights}. For $k\geq2$, this uses
$2^a\leq i$ and $2^e<2k\leq k^2$; for $k=1$, the left side is at most
$6i$.

Combined with the bounds for the local lists, the level-$j$ ancestor in
$T_i$ is read within $16i(8j)^2=2^{10}ij^2$ probes and the level-$h$
home subtree within $16i(2^{h+4})^2=2^{12}i4^h$, so $D=2^{12}$ suffices
in \Cref{def:search-bound}.

\paragraph{Negative queries.}
Since each visit rereads its prefix from scratch, a negative query
can recompute the bounds of \Cref{tab:bounds} during each visit using
$O(1)$ words. The only
subtlety is $I(x)$, which depends on the values $J_1(y)$ of other
elements, that is, on fourth roots of priorities. Instead of computing
a fourth root to calculate $I(x)$, the query tracks the minimum priority $\mu$ of the elements in the
leaves of $T_1$ that it waits for: for an integer $q\geq1$, writing
$z=\ceil{q/100}$, we have
$I(x)\geq q$ exactly when $z\leq K_1(x)$ or $\mu<1/z^4$. The query can
compute $\mu$ during its scan of the level-$K_1(x)$ home subtree, since
whether $K_1(x)\geq h$ depends only on ancestors that are read before
the expanding-subtree list reaches level $h$.

Our stopping condition is that the query stops at the end of the first
round in which, for every tree $T_i$ with $i\leq\min\{I(x),\alpha\}$,
that round's visit to $T_i$ read every node the query waits for in
$T_i$. Since, for every $k$, round $\floor{\log_2i}+2\ceil{\log_2k}$ reads the first
$k$ probes of $T_i$'s local list in a single visit, \eqref{eq:weights}
bounds when this condition is first met, up to a constant factor.

\section{Growing Tables: An \texorpdfstring{$\Omega(\log\log \epsilon^{-1})$}{Omega(log log 1/eps)} Lower Bound}
\label{app:resizing}

\providecommand{\pc}{\operatorname{pc}}

Every table in this paper has a fixed capacity: it is told $n$ in advance
and fills an array of exactly $n$ slots. A natural follow-up question is
whether such a table can instead \emph{grow}, keeping its array completely
full at every moment as elements arrive, while still spending constant time
per operation. Bender, Kuszmaul, and Zhou~\cite{DBLP:conf/focs/BenderKZ24} showed that a
classical open-addressed table can be resized so as to keep load factor at
least $1-\epsilon$ at all times, with $O(1)$ expected-time queries and
$O(\log\log\epsilon^{-1})$ expected-time insertions and deletions; at load
factor one this is $O(\log\log n)$. They also showed that
$\Omega(\log\log\epsilon^{-1})$ is optimal for fixed-capacity tables that
support deletions. This appendix shows that for \emph{growing} tables the
$\log\log$ is unavoidable even without deletions, however the table
decides when and by how much to grow. The constant bounds of
\Cref{thm:intro-stable,thm:intro-reorder,thm:intro-full} are therefore a genuinely
fixed-capacity phenomenon.

The proof adapts the potential-function argument of
\cite[Section~7]{DBLP:conf/focs/BenderKZ24}. The new difficulty is that a growing table may
change its probe sequences every time it gains a slot, so there is no
single probe function against which costs can be measured. We handle this
with reference costs that stay frozen on nested time windows, with one
window length per tier, combined across tiers by a running minimum. A
table may also choose its size after seeing the elements, which makes these
references depend on the elements. Each window, however, admits only
polynomially many references, and the windows are long enough to absorb a
union bound over all of them.

The proof in this appendix was generated with guidance from the authors by Claude Opus 5.5 and GPT 6 Astra. The authors take full responsibility for its correctness.

\subsection{Model and Result}

\paragraph{Growing tables.}
Elements arrive one at a time and are never deleted. After $t$ insertions the
table stores $n_t=n_0+t$ elements in the first $N_t$ slots of an unbounded
array, where $n_t\le N_t\le n_t/(1-\epsilon)$. The table chooses $N_t$
itself, as an arbitrary function of the elements it has seen. It may grow one
slot at a time or in large chunks, at moments of its choosing, and it may
even shrink when its top slots are empty. At load factor one, $N_t=n_t$
and there is nothing to choose. For each size $N$ there is a probe function $h_N$, fixed in
advance, that maps each element to a sequence over $[N]\cup\{\mathrm{Null}\}$.
A query issued while the table has $N$ slots follows $h_N$. We write
$\pc_N(x,s)$ for the first index of slot $s$ in $h_N(x)$ (or $\infty$),
the probe complexity of element $x$ in slot $s$. We require that at every
moment between operations the expected average probe complexity of the
stored elements is at most $A$.

\paragraph{Cost.}
Insertions may rearrange the stored elements arbitrarily. Following
\cite{DBLP:conf/focs/BenderKZ24}, each new element first sits in a \defn{special slot}, outside the
array, and every relocation of an element counts as one move; the
\defn{switching cost} $S$ is the total number of moves, including each new
element's placement into the array. It is a lower bound on running time. For
the analysis we allow any number of special slots, so that every
rearrangement decomposes into \defn{half-moves}, each of which carries one
element between a special slot and an empty slot of the array. A move is at
most two half-moves. During insertion $t$ the array consists of the first
$\max(N_{t-1},N_t)$ slots. An element that the table parks anywhere else in the
middle of an insertion counts as sitting in a special slot.

\paragraph{Hard input.}
All elements, including the $n_0$ initial ones, are distinct and uniformly
random from a universe of size polynomial in $n$. By Yao's principle we may
assume that the table is deterministic. We may also assume that
$n_0\ge n/2$. Otherwise, we start the analysis after insertion
$\ceil{n/2}-n_0$ and treat the elements present at that moment as the initial
elements. They are uniformly random, and at least $n/2\ge H/2$ insertions
remain. Every size then lies in $[n/2,2n]$.

\begin{theorem}\label{thm:resizing}
Let $1/n\le\epsilon\le\epsilon_0$ for a sufficiently small constant
$\epsilon_0$, and let $\lambda=\ceil{\frac12\log_2\log_2\epsilon^{-1}}$.
Consider a growing classical open-addressed table that chooses its size
adaptively, keeps load factor at least $1-\epsilon$ at all times, and has
expected average probe complexity at most $A=2^{o(2^\lambda)}$ at every
moment; for instance $A=O(1)$. Insert $H$ uniformly random elements into it,
starting from $n_0$ elements, where $H\ge\delta n$ for $n:=n_0+H$ and a
constant $\delta>0$. Then the expected switching cost is
$\Omega(H\log\log\epsilon^{-1})$. In particular, at load factor one, a
growing table with constant average query cost makes
$\Omega(\log\log n)$ amortized expected moves per insertion.
\end{theorem}

Three remarks. First, the bound is tight: the resizable construction of
\cite{DBLP:conf/focs/BenderKZ24} achieves it with $O(1)$ expected queries. Second, only the \emph{average}
query cost needs to be small, so the theorem applies even to tables like
the one in \Cref{thm:intro-stable}. Third, there is no contradiction with
\Cref{thm:intro-reorder}, which achieves constant insertion time at load factor
one with fixed capacity. The argument below needs the table to stay within
$\epsilon$ of full for many more than $\epsilon n$ insertions. A
fixed-capacity table that fills from empty is that close to full only
during its last $\epsilon n$ insertions, and over such a short stretch the
budget can absorb every late element sitting at a high tier.

\subsection{Tools from \texorpdfstring{\cite{DBLP:conf/focs/BenderKZ24}}{BKZ24}}
\label{subsec:bkz-tools}

\paragraph{Near-uniformity.}
A probe function $h$ on $N$ slots is \defn{nearly uniform} if
$N\cdot\Prb_x[\pc_h(x,s)\le m]\le\kappa_0m^{10}$ for every slot $s$ and
every $m\ge1$, where $\kappa_0$ is an absolute constant. Bender et al.\
show that every probe function $h$ on $N$ slots can be replaced by a
nearly uniform probe function $h'$ on the same $N$ slots such that every
assignment of at most $N$ elements to slots has total probe complexity under
$h'$ at most $\kappa_0$ times its total under $h$, plus $\kappa_0N$
\cite[Section~7]{DBLP:conf/focs/BenderKZ24}. We apply this to every $h_N$ separately and write
$\pc'_N$ for the resulting probe complexities. Every size we use lies in
$[n/2,2n]$, so, after doubling $\kappa_0$,
$n\cdot\Prb_x[\pc'_N(x,s)\le m]\le\kappa_0m^{10}$ for all of them. The
table then meets the budget with $A'=\kappa_0(A+1)$ under $\pc'$. Since
$h'_N$ is again a probe sequence over $[N]$,
$\#\{s:\pc'_N(x,s)\le m\}\le m$ for every element $x$, and
$\pc'_N(x,s)=\infty$ for $s\ge N$.

\paragraph{Tiers.}
Bender et al.\ call these \emph{levels}; we say \defn{tiers} to avoid a
clash with the tree levels of \Cref{sec:stable}. Given any cost function
$c$ on element--slot pairs, the tier $\tau(x,s)$ is $0$ if
$c(x,s)\le2^{2^\lambda}$; it is $i$ if
$c(x,s)\in(2^{2^{\lambda+i-1}},2^{2^{\lambda+i}}]$ for $1\le i<\lambda$;
and it is $\lambda$ if $c(x,s)>2^{2^{2\lambda-1}}$. Empty slots and
special slots have tier $\lambda$. An element has tier at least $k$ exactly when
its cost exceeds
\begin{equation}\label{eq:tier-threshold}
                \theta_k:=2^{2^{\lambda+k-1}}.
\end{equation}

\paragraph{Stanzas and the potential.}
Fix a tier $10\le k\le\lambda$ and a tier function $\tau$. A
\defn{$k$-stanza} is a sequence of slots $s_1,\ldots,s_j$ with $j\ge2$,
such that:
\begin{itemize}
\item $s_1$ is nonempty with tier at least $k$;
\item $s_j$ has tier at least $k$, and may be nonempty or empty;
\item $s_2,\ldots,s_{j-1}$ are distinct and nonempty, with tier at most
      $k-10$;
\item writing $x_r$ for the element in $s_r$, $\tau(x_r,s_{r+1})\le k-10$ for
      every $r<j$.
\end{itemize}
Its potential is
\[
 1-\sum_{r=2}^{j-1}\sqrt2^{\,\tau(x_r,s_r)-k+10}
  -\sum_{r=1}^{j-1}\sqrt2^{\,\tau(x_r,s_{r+1})-k+10}.
\]
$\Phi_k$ is the maximum total potential of a collection of stanzas in
which every slot of tier at least $k$ serves at most once as a start and
at most once as an end, and every other slot serves at most once as an
internal slot. Special slots are never ends, since no probe sequence
contains them. Let $J_k$ be the number of elements whose tier at their current
position is at least $k$, counting elements in special slots, and let
\[
                       Q_k:=J_k-\Phi_k .
\]

\begin{lemma}\label{lem:q-range}
$0\le Q_k\le J_k$.
\end{lemma}
\begin{proof}
The empty collection gives $\Phi_k\ge0$. Every stanza has potential at
most $1$, and a collection of positive-potential stanzas uses distinct
starts, each of which is a nonempty slot of tier at least $k$; hence
$\Phi_k\le J_k$.
\end{proof}

The next lemma is the second property of the potential in
\cite[Section~7]{DBLP:conf/focs/BenderKZ24}, written one tier at a time. Let
\[
 f(\delta):=0\ \ (\delta\le0),\qquad
 f(\delta):=1\ \ (1\le\delta\le9),\qquad
 f(\delta):=\kappa_1 2^{-(\delta-10)/2}\ \ (\delta\ge10),
\]
for a suitable absolute constant $\kappa_1$, and let
$F:=\sum_{\delta\in\mathbb Z}f(\delta)=O(1)$.

\begin{lemma}[\cite{DBLP:conf/focs/BenderKZ24}]\label{lem:half-move}
Fix $\tau$ and $k$. Suppose an element $x$ moves from a special slot to
an empty slot $b$ of the array, or from $b$ to an empty special slot. Let
$j=\tau(x,b)$. Then $|\Delta Q_k|\le f(k-j)$.
\end{lemma}
\begin{proof}
For the move into $b$, the case analysis of \cite{DBLP:conf/focs/BenderKZ24} gives:
\begin{itemize}
\item if $j\ge k$, then $\Delta J_k=\Delta\Phi_k=0$;
\item if $k-10<j<k$, then $\Delta J_k=-1$ and $\Delta\Phi_k\in[-2,0]$;
\item if $j\le k-10$, then $\Delta J_k=-1$ and
      $\Delta\Phi_k=-1\pm\kappa_1\sqrt2^{\,j-k+10}$.
\end{itemize}
The move out of $b$ is the reverse of a move into $b$ between the same
two states. Bender et al.\ state this for the tiers of a single probe
function. Their proof, however, examines one tier at a time, and it uses
only:
\begin{itemize}
\item the tier of each pair;
\item that empty and special slots have tier $\lambda$;
\item that special slots are never ends;
\item that $b$ is empty before the move.
\end{itemize}
It therefore applies verbatim to any tier function, and to an array of
any length, including one padded with empty slots that the table never
uses.
\end{proof}

\paragraph{Short paths.}
For a cost function $c$, let $G$ be the bipartite graph whose left
vertices are a set $\mathcal A$ of $2n$ uniformly random elements and whose
right vertices are slots; it has an edge between $x$ and $s$ whenever
$\tau(x,s)<\lambda$, labelled by that tier. For $0\le i\le\lambda-10$, an
\defn{$i$-short path} uses no edge of tier above $i$, and, for every
$\delta\ge0$, at most $\sqrt2^{\,\delta}$ edges of tier $i-\delta$. If an
$(i+10)$-stanza $s_1,\ldots,s_j$ has positive potential, then
$x_1,s_2,x_2,\ldots,x_{j-1},s_j$ is an $i$-short path
\cite[Section~7]{DBLP:conf/focs/BenderKZ24}, since every edge it uses contributes a term to
the potential. Let
\[
 \beta_i:=2^{\lambda+i},\qquad a:=\frac{32}{1-2^{-1/2}}<109.3,\qquad
 R_i:=2^{(a+o(1))\beta_i}.
\]
A pair has tier at most $\ell$ exactly when its cost is at most
$2^{\beta_\ell}$. For a slot $s$, let $\deg_\ell(s)$ be the number of elements
$x\in\mathcal A$ with $\tau(x,s)\le\ell$. Call $s$ \defn{heavy} for $i$ if
$\deg_\ell(s)>2^{32\beta_\ell}$ for some $\ell\le i$, and let
$\mathcal H_i$ consist of the heavy slots for $i$ together with the elements
joined to them by edges of tier at most $i$. Call the cost function $c$
\defn{admissible} if it is finite only on slots below $2n$ and, for every
element $x$, slot $s$, and $m\ge1$,
\[
 \#\{s':c(x,s')\le m\}\le\lambda m
 \qquad\text{and}\qquad
 n\cdot\Prb_x[c(x,s)\le m]\le\lambda\kappa_0m^{10}.
\]

\begin{lemma}[\cite{DBLP:conf/focs/BenderKZ24}]\label{lem:short-paths}
Let $c$ be admissible. Then for every vertex set $X$, the vertices
reachable from $X\cup\mathcal H_i$ by $i$-short paths number at most
$(|X|+|\mathcal H_i|)R_i$.
\end{lemma}
\begin{proof}
Admissibility bounds every left degree at tier $\ell$ by
$\lambda2^{\beta_\ell}\le2^{32\beta_\ell}$. Outside the heavy slots,
every right degree at tier $\ell\le i$ is at most $2^{32\beta_\ell}$ as
well. These are the low-degree thresholds of \cite[Section~7]{DBLP:conf/focs/BenderKZ24}, and
their path-encoding count shows that from any vertex, at most
$2^{o(2^\lambda)}\cdot2^{a\beta_i}$ vertices are reachable by $i$-short
paths that avoid heavy slots. Now take an $i$-short path from
$X\cup\mathcal H_i$ to a vertex $v$. If it meets a heavy slot, cut it
just after the last one. The remaining path starts at an element joined to
that slot by an edge of tier at most $i$, which lies in $\mathcal H_i$,
and it is again $i$-short. (If the last heavy slot is $v$ itself, then
$v\in\mathcal H_i$.)
\end{proof}

Admissibility also makes heavy slots extremely rare. We need this in a
strong form, because we shall apply it to polynomially many cost
functions at once.

\begin{lemma}\label{lem:heavy}
Let $\epsilon_0$ be small enough. Fix $0\le i\le\lambda-10$ and an
admissible cost function $c$. Then
$\Prb\bigl[|\mathcal H_i|>n2^{-227\beta_i}\bigr]\le\exp(-n^{1/17})$.
\end{lemma}
\begin{proof}
We use the Chernoff bound $\Prb[\Sigma\ge r]\le(e\mu/r)^{r/D}\le e^{-r/D}$
for $r\ge e^2\mu$, where $\Sigma$ is a sum of independent $[0,D]$-valued
variables with mean at most $\mu$. By Hoeffding's comparison theorem, it
also holds when the summands are indexed by a sample drawn without
replacement, such as $\mathcal A$. Recall that $\epsilon\ge1/n$, so
$n\ge\epsilon_0^{-1}$ is large.

Let $m:=\ceil{n2^{-240\beta_i}}$. Since $2^{2\lambda}<4\log_2\epsilon^{-1}$,
we have $\beta_i\le2^{2\lambda-10}<\frac1{256}\log_2\epsilon^{-1}$, and
hence $m\ge n\epsilon^{15/16}\ge n^{1/16}$. Fix a tier $\ell\le i$, write
$\varphi=2^{\beta_\ell}$, and fix a set $S$ of $m$ slots below $2n$.
Then
\[
 \sum_{s\in S}\deg_\ell(s)=\sum_{x\in\mathcal A}
     \#\{s\in S:c(x,s)\le\varphi\}.
\]
By admissibility, each summand is at most $\lambda\varphi$ and the mean is
at most $2m\lambda\kappa_0\varphi^{10}$. So the sum reaches
$m\varphi^{12}$ with probability at most $e^{-m\varphi^{11}/\lambda}$.
There are at most $(2en/m)^m\le(2e\cdot2^{240\beta_i})^m$ such sets $S$,
and $\varphi^{11}/\lambda\ge2^{11\cdot2^\lambda}/\lambda$ far exceeds
$240\beta_i+3$. A union bound over the sets and over the tiers $\ell$
therefore shows that, except with probability $\lambda e^{-m}$, every
set of at most $m$ slots has total tier-$\ell$ degree below
$m2^{12\beta_\ell}$, for every $\ell\le i$.

On this event, fewer than $m$ slots are heavy at each tier $\ell$, since
$m$ of them would have total degree above $m2^{32\beta_\ell}$. So fewer
than $\lambda m$ slots are heavy for $i$. Split them into at most
$\lambda$ groups of at most $m$ slots each. Their total tier-$i$ degree,
which bounds the number of elements joined to them by edges of tier at most
$i$, is then below $\lambda m2^{12\beta_i}$. Hence
$|\mathcal H_i|<\lambda m(1+2^{12\beta_i})\le4\lambda n2^{-228\beta_i}
\le n2^{-227\beta_i}$, and $\lambda e^{-m}\le\exp(-n^{1/17})$.
\end{proof}

\subsection{Frozen References on Nested Windows}

\paragraph{Constants.}
Throughout, $k=i+10$ ranges over $[10,\lambda]$. Fix
\[
                        b:=512,\qquad\rho:=230,\qquad\nu:=113.
\]
These satisfy $a<\nu$ and $a+\nu<\rho<b$, and also $a+\nu<227$, the
exponent in \Cref{lem:heavy}. Since
$2^{2\lambda}<4\log_2\epsilon^{-1}\le4\log_2n$, every
$\beta_i\le2^{2\lambda-10}$ is below $\frac1{256}\log_2\epsilon^{-1}$, and
hence
\[
 2^{\rho\beta_i}<n^{0.9},\qquad
 2^{a\beta_i}<\epsilon^{-0.43}.
\]
Since $n\ge\epsilon^{-1}\ge\epsilon_0^{-1}$, choosing $\epsilon_0$ small
also makes $n$ as large as we need.
By \eqref{eq:tier-threshold}, tier at least $k$ means cost above
$\theta_k=2^{b\beta_i}$. Let $B_k(t)$ be the number of elements whose
\emph{true} tier, measured with $\pc'_{N_t}$, is at least $k$ at time $t$.
Markov's inequality then gives, at every time,
\begin{equation}\label{eq:true-budget}
           \E[B_k(t)]\le nA'2^{-b\beta_i}=n\,2^{-(b-o(1))\beta_i}.
\end{equation}
Here time $t$ means the state after the $t$th insertion has been fully
processed. At such a moment, every element is in the array.

\paragraph{Windows.}
Let $M_k:=2^{\floor{\log_2n-\rho\beta_i}}$. This is a power of two, at
least $n^{0.1}/2$, and non-increasing in $k$. A \defn{$k$-window} is a
block $(rM_k,(r+1)M_k]$ of consecutive insertions. Since $M_{k+1}$ divides
$M_k$, every $k$-window is a union of $(k+1)$-windows. We write $e_k(t)$
for the last insertion of the $k$-window containing $t$.

\paragraph{References.}
For each insertion $t$ and tier $k$, define the reference cost
\[
     c^{(k)}_t(x,s):=\min_{10\le k'\le k}\pc'_{N(e_{k'}(t))}(x,s),
\]
and let $\tau^{(k)}_t$ be its tier function. It has four properties.
\begin{enumerate}[label=(R\arabic*)]
\item\label{R:frozen} $c^{(k)}_t$ depends on $t$ only through the
      $k$-window containing $t$, because every coarser window containing
      it is the same.
\item\label{R:monotone} $c^{(k+1)}_t\le c^{(k)}_t$ pointwise, hence
      $\tau^{(k+1)}_t\le\tau^{(k)}_t$.
\item\label{R:end} $c^{(k)}_t\le\pc'_{N(e_k(t))}$. In particular, at time
      $e=e_k(t)$, an element whose $\tau^{(k)}_t$-tier at its slot is at least
      $k$ has true tier at least $k$.
\item\label{R:fixed} $c^{(k)}_t$ lies in the family $\mathcal C^{(k)}_t$
      of all functions $\min_{10\le k'\le k}\pc'_{\nu_{k'}}$ with
      $n_{e_{k'}(t)}\le\nu_{k'}\le n_{e_{k'}(t)}/(1-\epsilon)$. This
      family is fixed in advance and depends on $t$ only through its
      $k$-window. It has at most $(2\epsilon n+2)^\lambda\le(3n)^\lambda$
      members, and each is admissible, being a minimum of at most
      $\lambda$ nearly uniform probe sequences over at most $2n$ slots.
\end{enumerate}
The reference looks into the future, toward the ends of the windows
containing $t$. This is harmless: it is never used to bound the cost of a
query at time $t$, only as a yardstick that holds still while the
potential is tracked. Because the table picks its sizes after seeing the
elements, the reference also depends on the elements, including elements that arrive
during the window itself. Property~\ref{R:fixed} limits the damage: a
union bound over $\mathcal C^{(k)}_t$ costs $O(\lambda\log n)$ per window,
while windows have length $M_k\ge n^{0.1}/2$.

\subsection{The Windowed Potential}

\paragraph{The analysis array.}
The potential never looks at the table's actual size. Let
$L_t:=\floor{n_t/(1-\epsilon)}$. This is the largest size the table may
have after $t$ insertions, so $N_t\le L_t$. Unlike $N_t$, it is fixed in
advance; it is non-decreasing, and $L_t-L_{t-1}\in\{1,2\}$. For each tier
$k$, draw $D_k$ uniformly from $[d_k,2d_k)$, where
$d_k:=\ceil{n2^{-\nu\beta_i}}$, independently of everything else. This
choice exists only in the analysis.

For tier $k$, the analysis works with an array of $L_t+D_k$ slots. Every
slot of it that holds no element counts as empty, including every slot from
$\max(N_{t-1},N_t)$ on, which the table cannot use. While insertion $t$ is processed,
every element that is not in a special slot lies below
$\max(N_{t-1},N_t)\le L_t$. After insertion $t-1$, all empty slots of the
array lie below $L_t+2d_k$, and at most
$L_t-n_{t-1}\le2\epsilon n+1$ of them lie below $L_t$. The $D_k$ empty
slots at the top play the role of the lookahead slots of
\cite{DBLP:conf/focs/BenderKZ24}: because $D_k$ is random, the table cannot aim a chain of
displacements at the next slot to appear.

\paragraph{The potential.}
While insertion $t$ is being processed, we compute $J_k$, $\Phi_k$, and
$Q_k$ as in \Cref{subsec:bkz-tools}, using the tier function
$\tau^{(k)}_t$ and the array of $L_t+D_k$ slots. The processing of
insertion $t$ consists of the following events:
\begin{enumerate}[label=(\roman*)]
\item the array grows from $L_{t-1}+D_k$ to $L_t+D_k$ slots, one
      \defn{allocation} at a time;
\item the new element enters a special slot;
\item the elements are rearranged by half-moves;
\item if $t$ is the last insertion of a $k$-window, the reference for
      tier $k$ switches to the next window, a \defn{switch}.
\end{enumerate}

\begin{lemma}\label{lem:events}
For every tier $k$:
\begin{enumerate}[label=(\alph*)]
\item\label{ev:insert} when the new element enters a special slot,
      $\Delta Q_k=1-g_k$ for some $g_k\in[0,1]$;
\item\label{ev:alloc} each allocation has $\Delta Q_k=-u_k$ for some
      $u_k\in[0,1]$;
\item\label{ev:half} each half-move of an element $x$ to or from slot $b$ has
      $|\Delta Q_k|\le f\bigl(k-\tau^{(k)}_t(x,b)\bigr)$;
\item\label{ev:switch} each switch, at the end $e$ of a $k$-window, has
      $\Delta Q_k\ge-B_k(e)$;
\item no other event changes $Q_k$.
\end{enumerate}
\end{lemma}
\begin{proof}
\ref{ev:insert} The new element adds $1$ to $J_k$ and creates one new possible
start. Any collection afterwards uses that start at most once, so
$\Phi_k$ rises by at most $1$.

\ref{ev:alloc} An allocation adds one empty slot. Every old collection
stays valid, and the new slot can only end a stanza, at most one per
collection; so $\Phi_k$ rises by between $0$ and $1$. $J_k$ is
unchanged, because tiers are frozen within a window by \ref{R:frozen}.

\ref{ev:half} This is \Cref{lem:half-move}.

\ref{ev:switch} After the switch, $Q_k\ge0$ by \Cref{lem:q-range}.
Before it, $Q_k\le J_k\le B_k(e)$, by \Cref{lem:q-range} and \ref{R:end};
no element is in a special slot at time $e$.

Finally, by \ref{R:frozen}, tiers change only at switches, and queries do
not change the state. Nor do the table's own changes of size: a slot
between $N_t$ and $L_t+D_k$ counts as an empty slot of the array either
way.
\end{proof}

\subsection{Three Estimates}

\begin{lemma}[slack is summable]\label{lem:slack-sum}
For a half-move of element $x$ to or from slot $b$ during insertion $t$,
$\sum_{k=10}^{\lambda}f\bigl(k-\tau^{(k)}_t(x,b)\bigr)\le F$.
\end{lemma}
\begin{proof}
By \ref{R:monotone}, $\tau^{(k)}_t(x,b)$ is non-increasing in $k$, so
$k-\tau^{(k)}_t(x,b)$ is strictly increasing in $k$. The sum therefore
evaluates $f$ at distinct integers.
\end{proof}

Without the running minimum this step can fail. With plain window-end
references, one can build nearly uniform families whose pairs become
cheaper over time, for instance
$\pc_N(x,s)\approx r(x,s)\,(n/(N-s))^{\alpha}$ with an element-dependent
factor $r$. In such a family a typical destination lands a few tiers
below $k$ for $\Theta(\lambda)$ tiers at once, costing $\Theta(\lambda)$
slack per move.

\begin{lemma}[expensive elements are transient]\label{lem:transient}
Let $W$ be a $k$-window with last insertion $e$, and let $S_W$ be the
number of moves made while processing the insertions of $W$. Let
$J_k(t^-)$ denote $J_k$ at the start of insertion $t$, before any
allocation. Then for every $t\in W$,
\begin{align*}
 J_k(t^-)&\le\#\{\text{elements moved during insertions }t,\ldots,e\}+B_k(e)
          \le S_W+B_k(e),\\
 \sum_{t\in W}J_k(t^-)&\le M_k\bigl(S_W+B_k(e)\bigr).
\end{align*}
\end{lemma}
\begin{proof}
Let $x$ be an element counted in $J_k(t^-)$, sitting in slot $s$. By
\ref{R:frozen} and \ref{R:end}, the pair $(x,s)$ has true tier at least
$k$ at time $e$. So either $x$ moves during the processing of insertions
$t,\ldots,e$, or $x$ is still in $s$ at time $e$ and is counted in
$B_k(e)$. This proves the pointwise bound. For the sum, each move is
counted for at most $M_k$ values of $t$.
\end{proof}

\begin{lemma}[gains are rare]\label{lem:gains}
Let $\epsilon_0$ be small enough, and let $W$ be a $k$-window with last
insertion $e$. Let $G_W$ and $U_W$ be the totals of $g_k$ over the
insertions of $W$ and of $u_k$ over its allocations, and let $S_W$ be as
in \Cref{lem:transient}. Then
\[
 \E\bigl[G_W+U_W+B_k(e)\bigr]\le2^{-2\beta_i}\,\E[S_W]
   +M_k\bigl(2^{-2\beta_i}+\epsilon^{1/2}+n^{-0.05}\bigr).
\]
\end{lemma}
\begin{proof}
Let $\mathcal A$ consist of the $n$ elements of the input together with $n$
further uniformly random elements. Then $\mathcal A$ is a uniformly random set
of $2n$ elements, and the input is a uniformly random sequence of distinct
elements of $\mathcal A$. Let $\mathcal F_t$ be generated by
$\mathcal A$, by $D_k$, and by the first $t$ inserted elements, together with
the initial elements. The table is deterministic, so its state after
insertion $t$, size included, is determined by $\mathcal F_t$. Given
$\mathcal F_{t-1}$, the new element $y_t$ is uniform over the at least $n$
elements of $\mathcal A$ that have not been inserted yet.

The reference $c$ on $W$ is \emph{not} determined by $\mathcal F_{t-1}$:
it depends on sizes at the ends of windows, which the table chooses after
seeing later elements. By \ref{R:fixed}, however, $c$ lies in a fixed family
$\mathcal C$ of at most $(3n)^\lambda$ admissible cost functions. Let
$\mathcal G$ be the event that $|\mathcal H_i(c')|\le n2^{-227\beta_i}$
for every $c'\in\mathcal C$, where $\mathcal H_i(c')$ is the set
$\mathcal H_i$ computed for $c'$. By \Cref{lem:heavy},
$\Prb[\mathcal G^c]\le(3n)^\lambda\exp(-n^{1/17})$. The event
$\mathcal G$ depends only on $\mathcal A$. Write
$J_W:=\sum_{t\in W}J_k(t^-)$.

\emph{Insertion gains.} For $c'\in\mathcal C$ and $t\in W$, let
$X_t(c')$ consist of:
\begin{itemize}
\item the slots that are nonempty after insertion $t-1$ and have
      $c'$-tier at least $k$;
\item the slots below $L_t+2d_k$ that are empty after insertion $t-1$.
\end{itemize}
It is determined by $\mathcal F_{t-1}$ and $c'$, and
$|X_t(c)|\le J_k(t^-)+2\epsilon n+2d_k+1$. Let $E_t(c')$ be the event
that $y_t$ is reachable from $X_t(c')\cup\mathcal H_i(c')$ by an
$i$-short path in the graph of $c'$. By \Cref{lem:short-paths},
\[
 \Prb[E_t(c')\mid\mathcal F_{t-1}]\le p_t(c'):=
   \frac{R_i}{n}\bigl(|X_t(c')|+|\mathcal H_i(c')|\bigr).
\]
A gain at insertion $t$ requires a positive stanza that starts at $y_t$'s
special slot. Its end lies in $X_t(c)$: between the end of insertion $t-1$
and the arrival of $y_t$, only allocations have happened, and they add
empty slots below $L_t+2d_k$. The stanza gives an $i$-short path from
$y_t$ to that end. Hence $g_k(t)\le\1[E_t(c)]$.

Fix $c'\in\mathcal C$. Since
$\E[e^{\1[E]}\mid\mathcal F]\le1+(e-1)p\le e^{2p}$ whenever
$\Prb[E\mid\mathcal F]\le p$, the process
$\exp\sum_{t'\le t,\,t'\in W}(\1[E_{t'}(c')]-2p_{t'}(c'))$ is a
supermartingale. So
$\sum_{t\in W}\1[E_t(c')]\ge2\sum_{t\in W}p_t(c')+s$ has probability at
most $e^{-s}$. Take $s=3\lambda\ln(3n)$ and take a union bound over
$\mathcal C$. Then, except with probability $(3n)^{-2\lambda}$, this
bound fails for no $c'\in\mathcal C$, and in particular not for the
reference $c$. On this event and on $\mathcal G$,
\[
 G_W\le\frac{2R_i}{n}\Bigl(J_W+M_k\bigl(2\epsilon n+2d_k+1
        +n2^{-227\beta_i}\bigr)\Bigr)+3\lambda\ln(3n).
\]
Otherwise $G_W\le M_k$, and the exceptional events have probability at
most $1/M_k$. So
\[
 \E[G_W]\le\frac{2R_i}{n}\Bigl(\E[J_W]+M_k\bigl(2\epsilon n+2d_k+1
        +n2^{-227\beta_i}\bigr)\Bigr)+3\lambda\ln(3n)+1.
\]

\emph{Allocation gains.} Consider an allocation during insertion
$t\in W$; it adds the empty slot $L+D_k$, where $L<L_t$. A gain requires
a positive stanza ending at this slot, which gives an $i$-short path to
it from an element in a nonempty slot of tier at least $k$. By
\Cref{lem:short-paths}, applied with $X$ the set of these $J_k(t^-)$
elements, the vertices reachable from $X\cup\mathcal H_i(c)$ number at most
$(J_k(t^-)+|\mathcal H_i(c)|)R_i$. This set does not depend on $D_k$,
since internal slots of a stanza are nonempty. It is determined by the elements and
the behaviour of the table, which never sees $D_k$. Neither does
$\mathcal G$, which depends only on $\mathcal A$. Given all of these,
$L+D_k$ is still uniform over $d_k$ slots. Hence
$\E[u_k\1[\mathcal G]]\le(R_i/d_k)\,\E[J_k(t^-)+n2^{-227\beta_i}]$.
There are at most two allocations per insertion, and
$2M_k\Prb[\mathcal G^c]\le1$, so
\[
 \E[U_W]\le\frac{2R_i}{d_k}\,\E[J_W]+2M_k\,\frac{R_in}{d_k}\,
     2^{-227\beta_i}+1.
\]

\emph{Combining.} By \Cref{lem:transient},
$J_W\le M_k(S_W+B_k(e))$. By \eqref{eq:true-budget},
$\E[B_k(e)]\le n2^{-(b-o(1))\beta_i}\le M_k2^{-(b-\rho-o(1))\beta_i}$.
Add the two bounds above to $\E[B_k(e)]$, and use $2R_i/n\le2R_i/d_k$.
The coefficient of $\E[S_W]$ is then at most
$4R_iM_k/d_k\le2^{-(\rho-a-\nu-o(1))\beta_i}$. The remaining terms are
$M_k$ times:
\begin{itemize}
\item $4\epsilon R_i\le\epsilon^{0.57-o(1)}\le\epsilon^{1/2}$;
\item $(3\lambda\ln(3n)+2)/M_k\le n^{-0.05}$, since $M_k\ge n^{0.1}/2$;
\item $2R_i(2d_k+1)/n\le2^{-(\nu-a-o(1))\beta_i}$, and
      $2R_i(1+n/d_k)2^{-227\beta_i}\le2^{-(227-a-\nu-o(1))\beta_i}$;
\item $(4R_iM_k/d_k+1)\,\E[B_k(e)]/M_k\le2^{-(b-\rho-1)\beta_i}$.
\end{itemize}
The differences $\rho-a-\nu$, $\nu-a$, $227-a-\nu$, and $b-\rho$ all
exceed $2.7$. For $\epsilon_0$ small, $\beta_i\ge2^\lambda$ is large, so
the coefficient of $\E[S_W]$ and the last two items are at most
$2^{-2\beta_i}$ in total.
\end{proof}

\subsection{Proof of \texorpdfstring{\Cref{thm:resizing}}{Theorem}}

Let $H'$ be the largest multiple of $M_{10}$ that is at most $H$. Since
$M_{10}\le n2^{-\rho2^\lambda}=o(n)$, we have $H-H'=o(H)$. We work with
insertions $1,\ldots,H'$, and so every $k$-window lies inside this range.

\paragraph{One tier.}
Fix $k$ and add up $\Delta Q_k$ over all events, using
\Cref{lem:events}. By \Cref{lem:q-range}, $Q_k\ge0$ at the start. At the
end, just before the final switch, $Q_k\le J_k\le B_k(H')$, by
\Cref{lem:q-range} and \ref{R:end}. Therefore
\begin{equation}\label{eq:one-tier}
 \sum_{\text{half-moves}}f\bigl(k-\tau^{(k)}(x,b)\bigr)
 \;\ge\;H'-\sum_W\bigl(G_W+U_W+B_k(e_W)\bigr),
\end{equation}
where $W$ ranges over the $H'/M_k$ windows of tier $k$, and $e_W$ is the
last insertion of $W$.

\paragraph{All tiers.}
Sum \eqref{eq:one-tier} over the $\lambda-9$ tiers and take
expectations. By \Cref{lem:slack-sum}, the left side is at most $F$
times the number of half-moves, which is at most $2F\,S$. On the right,
apply \Cref{lem:gains} to every window. The windows of one tier
partition the insertions, so $\sum_WS_W\le S$ and $\sum_WM_k=H'$. Since
$\sum_i2^{-2\beta_i}\le1$,
\[
 2F\,\E[S]\;\ge\;(\lambda-9)H'-\E[S]
   -H'\bigl(1+\lambda\epsilon^{1/2}+\lambda n^{-0.05}\bigr),
\]
and hence
\[
 (2F+1)\,\E[S]\;\ge\;\bigl(\lambda-10-\lambda\epsilon_0^{1/2}
   -\lambda n^{-0.05}\bigr)H'\;=\;\Omega(\lambda H),
\]
because $\epsilon_0$ is small and $\lambda=O(\log\log n)$. Finally,
$\lambda=\Theta(\log\log\epsilon^{-1})$.
\qed

\subsection{Discussion}

\paragraph{Why changing probe sequences does not help.}
A growing table may redefine its probe sequences every time it gains a
slot, and one might hope that this churn substitutes for moves. The proof
shows why it cannot. Inside a $k$-window the reference is frozen, so
churn never enters the accounting. All of it is paid at window ends, and
each payment is at most the number of elements that are \emph{genuinely}
expensive at that moment, which the query budget makes negligible
compared with $M_k$.

An element that is cheap now but expensive under the reference is
\emph{transient}. It must move before its window closes, or else be
genuinely expensive when it does. So the time such elements spend counted as
expensive is at most $M_k$ times the number of moves. And because
$M_kR_i/n\ll1$, the potential that random new elements can extract from them
is a vanishing multiple of the move count. This is the precise sense in
which an element whose probe sequence keeps changing must pay for it in moves.
The windows can be made short enough because of the gap between the
budget exponent $b=512$ and the reach exponent $a\approx109$, and that gap
is exactly what the separation of ten tiers inside a stanza buys.

\paragraph{Why choosing when to grow does not help.}
At load factor one the size is forced. At load factor $1-\epsilon$ the
table may grow or shrink whenever it likes, by any amount, after seeing
the elements. Two devices neutralize this freedom.
\begin{itemize}
\item The potential never looks at the actual size. It works with an
      array of $L_t+D_k$ slots, whose length is fixed in advance up to the
      hidden offset $D_k$, and treats unused slots as empty. So
      allocations, and the fresh empty slots they create, happen at fixed
      times, no more than two per insertion.
\item Adaptivity therefore enters only through the references, and only
      through the at most $\lambda$ sizes at window ends. That leaves at most
      $(3n)^\lambda$ candidates per window. Both random ingredients of
      \Cref{lem:gains}, the new element and the heavy-slot set, can be union
      bounded over all of them, at a cost of $O(\lambda\log n)$ per
      window of length $M_k\ge n^{0.1}/2$.
\end{itemize}
Windows stay defined by insertion counts, so the switch bound only ever
uses the query budget at fixed times, never at stopping times.

The argument never uses that the probe function is indexed by the size.
Suppose that at each time $t$, queries follow one of at most $2^w$ probe
functions over $[L_t]$, fixed in advance for that time and chosen by the
table after seeing the elements. Choosing the size is the case
$2^w\le3n$. Near-uniformize each of these functions on $L_t\ge n/2$
slots. Then the families of \ref{R:fixed} have at most $2^{w\lambda}$
members, and the proof goes through with $s$ in \Cref{lem:gains}
increased by $w\lambda$, as long as $w\le n^{1/20}$. The term
$n^{-0.05}$ in \Cref{lem:gains} then becomes $O(\lambda n^{-0.05})$, and
the theorem survives because $\lambda^2n^{-0.05}=o(1)$. In particular, \Cref{thm:resizing} still
holds if the table, besides choosing its size, keeps $n^{1/20}/2$ bits of
metadata outside the array, which queries read for free.

\end{document}